\documentclass[aps,prl,twocolumn,superscriptaddress,nofootinbib]{revtex4-1}

\usepackage{graphicx,color,amsmath,amssymb,bm}
\usepackage{slashed}

\usepackage{array, multirow}
\usepackage{float}
\usepackage{booktabs}

\newcommand{\nocontentsline}[3]{}
\newcommand{\toclesslab}[3]{\bgroup\let\addcontentsline=\nocontentsline#1{#2\label{#3}}\egroup}
\newcommand{\tocless}[2]{\bgroup\let\addcontentsline=\nocontentsline#1{#2}\egroup}
\newcommand{\epar}{E_\parallel}

\newcommand{\bfE}{{\bf E}}
\newcommand{\bfB}{{\bf B}}
\newcommand{\edb}{{\bf E} \cdot {\bf B}}

\newcommand{\bfx}{{\bf x}}
\newcommand{\gagg}{g_{a\gamma\gamma}}
\newcommand{\gan}{g_{aNN}}

\newcommand{\nablabold}{\boldsymbol{\nabla}}

\newcommand{\boldnabla}{\boldsymbol{\nabla}}

\newcommand{\br}{\mathbf{r}}

\newcommand{\bp}{\mathbf{p}}
\newcommand{\bq}{\mathbf{q}}

\newcommand{\bpi}{\boldsymbol{\Pi}}
\newcommand{\moddist}{|\br - \br'|}

\newcommand{\expom}{e^{-i\Omega t}}
\newcommand{\expomp}{e^{i\Omega t}}

\newcommand{\expompm}{e^{\pm i\Omega t}}

\usepackage[export]{adjustbox}
\usepackage{tikz}
\usepackage[normalem]{ulem}
\usepackage{tkz-euclide}
\usetikzlibrary{decorations.pathmorphing}	
\tikzset{
    v/.style={decorate, decoration={snake, segment length=3mm, amplitude=0.75mm}, draw},
    f/.style={draw=black, postaction={decorate},
        decoration={markings,mark=at position .6 with {\arrow[very thick]{latex}}}},
    fb/.style={draw=black, postaction={decorate},
        decoration={markings,mark=at position .4 with {\arrowreversed[very thick]{latex}}}},
    fnar/.style={draw=black},
    g/.style={decorate, draw=black,
        decoration={coil,amplitude=3pt, segment length=3.5pt}},
    s/.style={dashed,draw=black, postaction={decorate},
        decoration={markings,mark=at position .55 with {\arrow[very thick]{latex}}}},
    sb/.style={dashed,draw=black, postaction={decorate},
        decoration={markings,mark=at position .55 with {\arrowreversed[draw=black,very thick]{latex}}}},
    snar/.style={dashed,draw=black,line width =1.25pt},
}

\usepackage{hyperref} 
\usepackage{amsmath}
\hypersetup{
    colorlinks=true,
    citecolor=blue,
    linkcolor=magenta,
    filecolor=magenta,
    urlcolor=blue,
}

\newcommand{\es}[2] {\begin{equation} \label{#1} \begin{split} #2 \end{split} \end{equation}}

\newcommand{\rpc}{r_{\scriptscriptstyle{\rm PC}}}
\newcommand{\rns}{R_{\rm NS}}
\newcommand{\rhodm}{\rho_{_{\rm DM}}}
\newcommand{\rhops}{\rho_{_{\rm PS}}}
\newcommand{\bfmu}{\boldsymbol{\mu}}
\newcommand{\bcrab}{B_{\rm Crab}}
\newcommand{\rcrab}{R_{\rm Crab}}

\definecolor{mypurple}{RGB}{164,64,214}

\begin{document}

\title{An Axion Pulsarscope
}

\author{Mariia Khelashvili}
\email{khelashvili@fias.uni-frankfurt.de}
\affiliation{Department of Physics, Princeton University, Princeton, NJ 08544, USA}
\affiliation{Bogolyubov Institute for Theoretical Physics of the NAS of Ukraine, Metrolohichna Str. 14-b, Kyiv, 03143, Ukraine}
\affiliation{ Goethe Universität, Max-von-Laue Str. 1, Frankfurt am Main, 60438, Germany}
\author{Mariangela Lisanti}
\email{mlisanti@princeton.edu}
\affiliation{Department of Physics, Princeton University, Princeton, NJ 08544, USA}
\affiliation{Center for Computational Astrophysics, Flatiron Institute, New York, NY 10010, USA}
\author{Anirudh Prabhu}
\email{prabhu@princeton.edu}
\affiliation{Princeton Center for Theoretical Science, Princeton University, Princeton, NJ 08544, USA}
\author{Benjamin R. Safdi}
\email{brsafdi@berkeley.edu}
\affiliation{Berkeley Center for Theoretical Physics, University of California, Berkeley, CA 94720, USA}
\affiliation{Theoretical Physics Group, Lawrence Berkeley National Laboratory, Berkeley, CA 94720, USA}

\date{\today}

\begin{abstract}
Electromagnetic fields surrounding pulsars may source coherent ultralight axion signals at the known rotational frequencies of the neutron stars, which can be detected by laboratory experiments ({\it e.g.}, pulsarscopes).  As a promising {case study}, we model axion emission from the well-studied Crab pulsar, which would yield a prominent signal at $f \approx 29.6$~Hz regardless of whether the axion contributes to the dark matter abundance. We estimate the relevant sensitivity of future axion dark matter detection experiments such as DMRadio-GUT, Dark~SRF, and CASPEr, assuming different magnetosphere models to bracket the uncertainty in astrophysical modeling.  {For example, depending on final experimental parameters, the Dark~SRF experiment could probe axions with any mass $m_a \ll 10^{-13}$~eV down to $g_{a\gamma\gamma} \sim 3 \times 10^{-13}$ GeV$^{-1}$ with one year of data and assuming the vacuum magnetosphere model.}  These projected sensitivities may be degraded depending on the extent to which the magnetosphere is screened by charge-filled plasma. The promise of pulsar-sourced axions as a clean target for direct detection experiments motivates dedicated simulations of axion production in pulsar magnetospheres.

\end{abstract}

\maketitle

{\hypersetup{linkcolor=blue}

\noindent \textbf{\textit{Introduction.---}} The existence of axions is predicted by numerous well-motivated extensions to the Standard Model~\cite{Svrcek_2006, Axiverse2010}. While interesting in their own right, these ultralight pseudoscalar bosons can potentially account for the dark matter~(DM), as is the case for the quantum chromodynamics~(QCD)  axion~\cite{PRESKILL1983127,Abbott1982,Fischler1982,Planck2018, PQ1, PQ2, WeinbergAxion, WilczekAxion}.  Given their ubiquity in theoretical models, a broad laboratory program exists to search for axions (see~\cite{Graham:2015ouw,Adams:2022pbo,Safdi:2022xkm}).  This includes haloscope experiments, which search for axion DM in the Milky Way~\cite{Hagmann:1990, Hagmann:1998cb, Asztalos:2001tf, Asztalos:2009yp, Du:2018uak, Braine2020, Bradley2003, Bradley2004, Shokair2014, HAYSTAC, Zhong2018, Backes_2021, Choi_2021, QUAX:2020adt, mcallister2017organ}, helioscope experiments, which search for axions produced by the Sun~\cite{Sikivie:1983ip,CAST:2017uph}, and ``light shining through walls'' experiments, which aim to produce axions directly in the lab~\cite{Ehret:2010mh,DIAZORTIZ2022100968}.  This Letter proposes a fourth alternative---the \emph{pulsarscope}---and demonstrates that planned experiments can successfully search for coherent axion signals emitted by nearby pulsars. 

Axions are generically expected to couple to Standard Model fields through dimension-five operators, motivating efforts to detect them in the laboratory. For example, the axion field, $a(x)$, couples to electromagnetism through the operator $\mathcal{L} \supset - \gagg a F \tilde{F}/4 = \gagg a \edb  $, where $\gagg$ is the coupling constant, $F$ is the quantum electrodynamics~(QED) field strength, and ${\bf E}$ (${\bf B}$) is the electric (magnetic) field.  The axion may also couple to Standard Model fermions $f$ 
through the operator $\mathcal{L} \supset (g_{aff} / 2 m_f ) \partial_\mu a \bar{f} \gamma^\mu \gamma^5 f = -(g_{aff} / m_f )\nablabold a \cdot {\bf S}$ (in the non-relativistic limit), where ${\bf S}$ is the fermion spin operator, $m_f$ is the fermion mass, and $g_{aff}$ is the coupling constant.  
These operators may induce observable signatures in laboratory experiments, such as axion-to-photon conversion ({\it e.g.},~\cite{Sikivie:1983ip,CAST:2017uph,ADMX:2018gho,ADMX:2019uok,ADMX:2020hay,PhysRevLett.127.261803,Ouellet:2018beu,Salemi:2021gck}) or spin precession ({\it e.g.},~\cite{Graham:2013gfa,Casper2013, JacksonKimball:2017elr,Garcon2019, Wu2019, Gao:2022nuq, Foster:2023bxl}).

Rapidly-rotating neutron stars~(NSs), or pulsars, can serve as axion factories because their magnetospheres may possess regions of large, un-screened accelerating electric fields ($\edb \ne 0$)~\cite{Garbrecht:2018akc, Prabhu2021, noordhuis2023novel}. In non-axisymmetric pulsar magnetospheres, ultralight axions are efficiently radiated at the rotation frequency of the pulsar, $\Omega$~\cite{Garbrecht:2018akc}.  As illustrated in Fig.~\ref{fig:ill}, this relativistic axion signal travels towards Earth, where it may be detected by a ground-based experiment.  While their densities are generally much smaller than the DM density near Earth~\cite{Adams:2022pbo}, pulsar-sourced axions benefit from large coherence times and have known frequencies. 

We estimate the sensitivity of proposed axion DM detection experiments to pulsar-sourced axions.  We consider two extreme models of the pulsar magnetosphere to bracket the uncertainty on the expected signal. The leading sensitivity to low-mass axions is obtained from the CASPEr-wind~\cite{Casper2013, JacksonKimball:2017elr,Garcon2019, Wu2019}, 
Dark~SRF~\cite{Berlin2020, Berlin2021}, and DMRadio-GUT~\cite{DMRadio:2022pkf,DMRadio:2022jfv}
experiments. 
Most optimistically, these experiments could probe previously unexplored regions of axion parameter space for axion masses $m_a \ll 10^{-13}$ eV; however, if the magnetospheres are heavily screened, then the sensitivities may be subdominant relative to current constraints.

\begin{figure}
    \centering
    \includegraphics[width=\columnwidth]{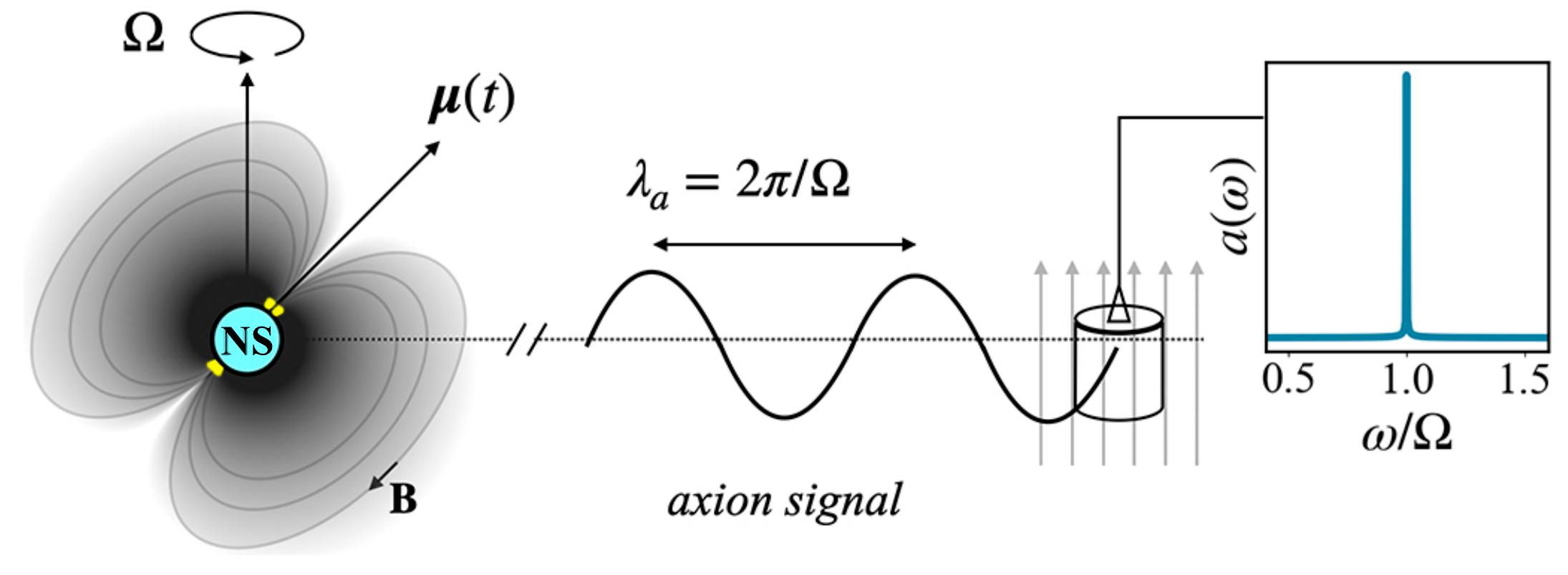}
    \caption{ 
    A neutron star~(NS) rotating with frequency $\Omega$ emits axions from gap regions where $\textbf{E}\cdot \textbf{B} \neq 0$ (shown in grey for the VDM model and in yellow for the PC model, not to scale). Relativistic axions emitted by the NS may convert to detectable signals in laboratory experiments ({\it pulsarscopes}). The resulting signal is highly coherent, with a prominent peak in frequency at the pulsar rotation frequency, $\Omega$.
    }
    \label{fig:ill}
\end{figure}

\noindent
\textbf{\textit{Axion Radiation from Pulsars.---}} 
The axion field sourced by a pulsar is determined by the distribution of $\edb$ in the magnetosphere according to the Klein-Gordon equation, $(\Box + m_a^2) a(x) = \gagg \edb$.
In the limit where the axion mass $m_a \to 0$, this equation is analogous to that for the electric potential, interpreted as $a(x)$, in Lorenz gauge with charge density $\rho_{\rm eff} = g_{a\gamma\gamma} \edb$.  Thus, one can calculate the radiated axion signal given a model for $\edb$ exterior to a NS. 

We consider two limiting scenarios for $\edb$. The first is the Vacuum Dipole Model (VDM)~\cite{Deutsch1955,Hoyle1964,Pacini1967, Pacini1968, OstrikerGunn1969,Melrose2012}.  In this case, the exterior of the NS is assumed to be vacuum and the magnetic field is described by a rotating dipole with magnetic moment  $\bfmu(t) = B_0 \rns^3  {\hat {\mu}}(t)$, where $B_0$ is the surface magnetic field and $\rns$ the NS radius. The unit vector $\hat {\mu}$ is misaligned from the rotation axis by an angle $\theta_m$ and rotates around the axis with frequency $\Omega$. The NS is well-approximated as a perfect conductor, therefore $\edb = 0$ in the interior. The exterior $\edb$ is determined by solving Laplace's equation with conducting boundary conditions imposed on the NS surface~\cite{Deutsch1955}.  

In the VDM, it follows that
\es{eqn:edbvac}{
    \edb = {|\bfmu|^2 \Omega \over 4 r^5} &\left[\cos\theta \sin^2 \theta_m + {1 \over 2} \sin\theta \sin{(2\theta_m)} \times \right. \\
    &\left. \big(\Omega r \cos(\phi - \Omega t) - \sin(\phi - \Omega t)\big)\right] \,,
}
with $(\theta,\phi)$ the spherical coordinates in the frame in which $\hat{z}$ is aligned with the NS's rotation axis. The first term on the right-hand-side of~\eqref{eqn:edbvac} does not have time dependence and thus does not radiate axions, while the second term gives sub-dominant radiated power relative to the third by an amount proportional to $\Omega R_{\rm NS} \ll 1$. The third term in~\eqref{eqn:edbvac} shows that the radiation will be at angular frequency $\Omega$ with dipole moment  
\es{eq:p_dipole}{
| {\bf p}_{\rm eff}| = \left|\int d^3x\, {\bf x} \rho_{\rm eff}(\bfx)\right| \approx { \pi \over 6} g_{a\gamma\gamma} {|\bfmu|^2 \Omega \over R_{\rm NS}} \sin(2 \theta_m) \,.
}
Using the dipole radiation formula, this suggests a radiated power in axions of the form 
\es{eqn:vacuumdipolepower}{
P_a &\approx {\pi \over 432 } g_{a\gamma\gamma}^2 B_0^4 \Omega^6 R_{\rm NS}^{10} \sin^2(2 \theta_m) \\
&\approx 1.4 \times 10^{35} \, \, {{\rm erg} \over {\rm s}} \left( {g_{a\gamma\gamma} \over 10^{-12} \, \, {\rm GeV}^{-1}} \right)^2 \left( {\Omega \over 2\pi \times 30 \, \, {\rm Hz}} \right)^6 \\ &\times \left( {{R_{\rm NS}} \over 14 \, \, {\rm km}} \right)^{10} \left( {B_0 \over 8.5\times 10^{12} \, \, {\rm G}} \right)^4 \sin^2(2 \theta_m) \,.
}
The fiducial pulsar parameters above correspond to those of the Crab pulsar, as motivated below. The Supplementary Material~(SM)  generalizes (\ref{eqn:vacuumdipolepower}) to include the non-zero axion mass (see also~\cite{Garbrecht:2018akc}).

For sufficiently rapid pulsar rotation, $\edb$ is strong enough to liberate charges from the NS surface, {which populate the magnetosphere and screen the electric field in the process.} 
However, even in active pulsars, $\edb$ is not expected to be screened everywhere.  An example of a partially-screened magnetosphere is the disk-dome or electrosphere solution found in slowly-rotating NSs~\cite{2002A&A...384..414P,Philippov2022Review,Cruz:2023vne}.  These solutions have screened magnetospheres extending out to distances of order $R_{\rm NS}$ from the NS surface, beyond which the VDM applies. 
Referring to~\eqref{eq:p_dipole}, excluding distances {up to} $x R_{\rm NS}$ from the NS center, with $x \sim$ few, reduces the radiated power by a factor of $x$. 

A lower bound on the axion emission power may be computed assuming the magnetosphere is mostly screened. The production of screening $e^\pm$ pairs occurs in gap regions with non-vanishing $\edb$, such as those that arise near the magnetic poles of the pulsar. Pulsar rotation defines a boundary, called the light cylinder (at radius $R_{\rm LC} = 1/\Omega$ in natural units), beyond which particles cannot co-rotate with the NS. Charged particles escape the magnetosphere along open field lines (defined as those that do not close within the light cylinder), leading to charge starvation above the polar cap (PC), defined as the region on the NS surface that contains the footpoints of all open field lines. The dearth of plasma above the PC opens a gap with $\edb \ne 0$. As the gap grows, it becomes unstable to runaway $e^\pm$ pair production, the dynamics of which have been modeled semi-analytically in~\cite{Tolman:2022unu, noordhuis2023novel} and from first-principles kinetic plasma simulations in~\cite{Cruz2021, Cruz:2022zyp}. All timescales associated with screening of the electric field are much smaller than the rotational period of the pulsar. Thus, when considering axion radiation at the rotational frequency, the gap may be treated as a point particle with ``axion charge''
\begin{align}
    Q_a = \gagg \displaystyle\int_{\rm gap} d^3x \left\langle \edb ({\bf x}, t) \right\rangle_t \, , 
\end{align} 
where $\left\langle . \right \rangle_t$ represents the time average. We assume that the gap is screened for one light crossing time and model the gap as a cylinder with radius $\rpc = \rns \sqrt{\Omega \rns}$ and height $h$.  Note that the gap radius may be derived by equating the magnetic flux through the PC region, $\pi B_0 \rpc^2$, and the flux through the light cylinder~\cite{GoldreichJulian1969}. At low altitude, $\ell \ll \rns$ with $\ell$ the height above the NS surface, the electric field in the gap is $E_\parallel(\ell) = 2 \Omega B_0 \ell$, which corresponds to a voltage drop across the gap of $\Phi_{{\scriptscriptstyle{\rm PC}}} = \Omega B_0 h^2$~\cite{RudermanSutherland1975}. With this geometry, the axion charge is $Q_a = \pi \gagg B_0^2 \Omega \rpc^2 h^2$.

The height of the gap is limited by the vacuum breakdown; as $h$ increases, the voltage drop increases and the gap becomes unstable to pair production. Most conservatively, we may set $h$ to be the mean free path of a pair-producing photon~\cite{Timokhin:2015dua,Timokhin:2018vdn,Caputo:2023cpv}, which is approximated by~\cite{RudermanSutherland1975}
\es{}{
h \sim  7 \, \, {\rm m} \left( {\Omega \over 2\pi \times 30 \, \,\, {\rm Hz}} {B_0 \over 8.5 \times 10^{12} \, \, {\rm G}} \right)^{-4/7}  \,. \label{eqn:hgap}
}
The true gap height may be larger than that in (\ref{eqn:hgap}) by a factor $N_{\rm gen}$, which represents the number of generations of pair production required to screen the gap. Assuming the charge density grows exponentially during the cascade, $N_{\rm gen}$ depends logarithmically on the ratio between the current supplied by the magnetosphere and the Goldreich-Julian current density $j_{\rm GJ} \sim 2 \Omega B_0$, giving $N_{\rm gen} \sim$ few. Modeling the current supply of the magnetosphere requires global simulations, and so we assume $N_{\rm gen} = 1$ in this work to be conservative. {\color{black}
Alternative models of the PC and their implications for axion emission power are discussed in Supplementary Material~\ref{sec:PG-models-sm}.}

A NS with a dipole magnetic field has two antipodal PCs with opposite ``axion charge.'' Taking the model of two oppositely charged point masses rotating rigidly and inclined at an angle $\theta_m$ relative to the rotation axis, the radiated power in the dipole approximation is 
\es{eq:P_a_PC}{
P_a &\approx { {1 \over 3\pi}} R_{\rm NS}^2 \Omega^4 Q_a^2 \sin^2 \theta_m \\
&\approx { 10^{20}} \, \, {{\rm erg} \over {\rm s}}\left( {g_{a\gamma\gamma} \over 10^{-12} \, \, {\rm GeV}^{-1}} \right)^2 \left( {\Omega \over 2\pi \times 30 \, \, {\rm Hz}} \right)^{ {40 \over 7}} \\ &\times \left( {{R_{\rm NS}} \over 14 \, \, {\rm km}} \right)^{8} \left( {B_0 \over 8.5 \times 10^{12} \, \, {\rm G}} \right)^{12/7} \sin^2(\theta_m) \,,
}
for $m_a \ll \Omega$. See SM for {the} generalization to non-zero axion mass.  Equations~\eqref{eqn:vacuumdipolepower} and~\eqref{eq:P_a_PC} fully bracket the current uncertainties on the emitted power in axions.

Among pulsars in the ATNF catalog~\cite{ATNF}, the Crab pulsar provides the highest axion power, by a factor of $\approx 20$ in the VDM, and is within a factor of two of the most efficient axion-emitting pulsar in the PC model.  
The spin-down parameters of the Crab are measured directly from timing of the radio beam: $P = 0.0338238880741$~s ($\Omega / 2\pi = 29.5649038871$ Hz) and $\dot{P} = 4.1958812 \times 10^{-13}$ (on 15/1/2024)~\cite{Lyne1993,ATNF}. Its radius, $\rcrab$, may be determined by observations of the Crab nebula~\cite{Bejger2002}: $\rcrab = 14$--$15$~km, where the spread arises due to uncertainty in the NS equation of state. Self-consistently incorporating the measurement of the pulsar radius, mass, and moment of inertia with the equation for vacuum-filled pulsar spin down gives $\bcrab > 8.5~(6.9) \times 10^{12}$~G for $\rcrab = 14$~($15$)~km~\cite{Philippov2014}.
These inferred magnetic field values are roughly consistent with those from analytic spin-down models studied in~\cite{Kou2013}, $B_{\rm Crab} = (5.0$--$8.3) \times 10^{12}$~G.  Motivated by these analyses, we fix the relevant Crab pulsar parameters to be $R_{\rm NS}= 14$ km and $B_0 = 8.5 \times 10^{12}$~G, which are conservative within the aforementioned observational constraints.

\vspace{0.1in}
\noindent \textbf{\textit{Detecting Pulsar-Sourced Axions.---}}
The pulsar-induced axion field $a({\bf x},t)$ on Earth
is analogous to that of DM (see, {\it e.g.},~\cite{Foster:2017hbq}) with a few exceptions.  First, the local energy density in axions due to the pulsar is given by $\rhops = P_a/(4\pi D^2)$, with $D$ the distance to the pulsar. 
The terrestrial energy density of pulsar-sourced axions, $\rhops$, is in general much less than the local DM density, $\rhodm$. For example, if the full spin-down power of the Crab pulsar were emitted in axions, the ratio of energy densities would be $\rhops/\rhodm \sim 10^{-13}$. Despite the orders-of-magnitude suppression in the local axion density, pulsar-sourced axions possess several key differences relative to DM axions that facilitate their detection.

While DM axions are non-relativistic, with characteristic momenta $k \sim 10^{-3} \omega$ and $\omega \sim m_a$, pulsar-induced axions are generically relativistic, with $k = \sqrt{\omega^2 - m_a^2} \sim \omega$ for $m_a \ll \omega \approx \Omega$. (Note that axions are only produced by the pulsar if $m_a < \Omega$.)  This aspect of the pulsar-induced axion signal is useful {for} experiments such as CASPEr-wind that rely on the axion field's {spatial} gradient.

DM axions are expected to be ultra-narrow in frequency space, with the axion field only having non-trivial support over a spread in frequencies $\delta \omega / \omega \sim 10^{-6}$, but the pulsar-induced axion signals may be even narrower.  
Assuming that the un-screened parts of magnetosphere co-rotate with the NS, then the axion signal has a natural {line width} set by the pulsar spin-down parameters: $\delta \omega / \omega = \dot P$, where $\dot P$ is the observed rate of change of the period. Most observed pulsars have $\dot{P} \ll 10^{-10}$, giving nearly monochromatic axion signals.  For example, the Crab pulsar has $\dot P \approx 4.2 \times 10^{-13}$ as mentioned previously, making the Crab-induced axion signal millions of times narrower than the expected DM signal. (Note that every few years the Crab undergoes small sporadic ``glitches,'' resulting in changes in the frequency and amplitude  of the signal; we assume the observing window does not contain any such major timing anomalies.)  By causality, deviations from co-rotation are expected near the light cylinder. On the other hand, in both the VDM model and the PC model, along with the electrosphere model, the axion emission is predominantly produced near the NS surface and far from the light cylinder. {Understanding} the extent to which magnetosphere drag may affect the signal {line width} requires dedicated simulations that include general relativistic effects, which are beyond the scope of this work.

Another key aspect of the pulsar-induced axion signal is that it is generated at known frequencies, unlike {in} the case of DM. Given that the mass of the DM axion is currently unknown, experiments must scan over millions of different possibilities. On the other hand, the Crab signal is at the known frequency $f \approx 29.6$ Hz. {The frequency of the signal is modulated by known effects ({\it e.g.} pulsar spin down and motion of the Earth), requiring real-time adjustment of the observing frequency.}

We now follow the formalism in~\cite{Foster:2017hbq,Dror2021} to project the sensitivity of key upcoming axion DM experiments to the signal from the Crab pulsar.  We concentrate on {experiments like} Dark~SRF and CASPEr-wind, 
though in the SM we discuss DMRadio-GUT, which may yield comparable sensitivity relative to a Dark~SRF experiment. 

{Superconducting radio frequency} (SRF) cavities have demonstrated quality factors $Q\gtrsim 10^{11}$~\cite{Romanenko2014, Posen_2019}, making them ideal resonators to look for highly monochromatic axion signals.  The Dark~SRF approach (called ``heterodyne detection'') to low-frequency axion detection involves the axion driving a transition between two nearly degenerate modes of an SRF cavity~\cite{Berlin2020, Berlin2021}. In this approach, the axion field drives a transition from a cavity ``pump mode'' of frequency $\omega_0 \sim$ GHz to a nearly-degenerate ``signal mode'' of frequency $\omega_1 \approx \omega_0 + \omega_a$, where $\omega_a$ is the frequency of the axion field. The signal is measured by a readout waveguide coupled to the signal mode~\cite{Berlin2020,Berlin2021}. 

We envision measuring the power in the signal mode at equal time intervals over a total integration time $t$ and then computing the power spectral density~(PSD) of the data in frequency space.
Since the axion coherence time ($\tau_a \gtrsim$ kyr) is much larger than $t$, the axion signal is expected to lie entirely in a single frequency bin, labeled $k^*$.
The power in bin $k^*$, obtained by integrating the PSD over the bin (see SM for details), is 
\begin{align}
    S_{k^*} \simeq \pi^2 \left(\gagg \eta_{10} B_p \right)^2 V_{\rm cav} \left( {Q_1 \over \omega_1} \right) {\rhops}\, ,
\end{align}
where $\eta_{10}$ is a dimensionless mode overlap factor, $B_p$ is the magnetic field amplitude of the pump mode, $V_{\rm cav}$ is the cavity volume, and $Q_1$ is the loaded quality factor of the signal mode. At the Crab pulsar frequency, the dominant noise source is thermal. The noise power in bin $k^*$ is (see SM)
\begin{align}
    B_{k^*} &\simeq 4 \pi T_{\rm cav} \left( {Q_1 \over Q_{\rm int}} \right) \Delta \omega \, ,
\end{align}
where $T_{\rm cav}$ is the cavity temperature, $Q_{\rm int}$ is the intrinsic quality factor of the cavity, and $\Delta \omega = 2\pi/t$ is the frequency bin width. 
The probability of observing a given PSD in bin $k^*$ is exponentially distributed with mean $S_{k^*}+ B_{k^*}$ for fixed signal parameters; this implies that the expected 95\% upper limit under the null hypothesis may be determined by setting $S_{k^*}/B_{k^*} \approx 8.48$~\cite{Dror2022}. 
Figure~\ref{fig:results}~(left panel) shows this projection, adopting fiducial parameters for the Crab pulsar as described previously and fiducial detector parameters: $\eta_{10} = 1$, $B_p = 0.2$~T, $V_{\rm cav} = 1 \ {\rm m}^3$, $\omega_0 \approx \omega_1 = 2\pi \times 100$ MHz, $Q_{\rm int} \approx Q_1 \approx 10^{12}$, $T_{\rm cav} = 1.8$ K, and $t = 1$ year as in~\cite{Berlin2020, Berlin2021}.

\begin{figure*}
    \centering
    \includegraphics[width=\linewidth]{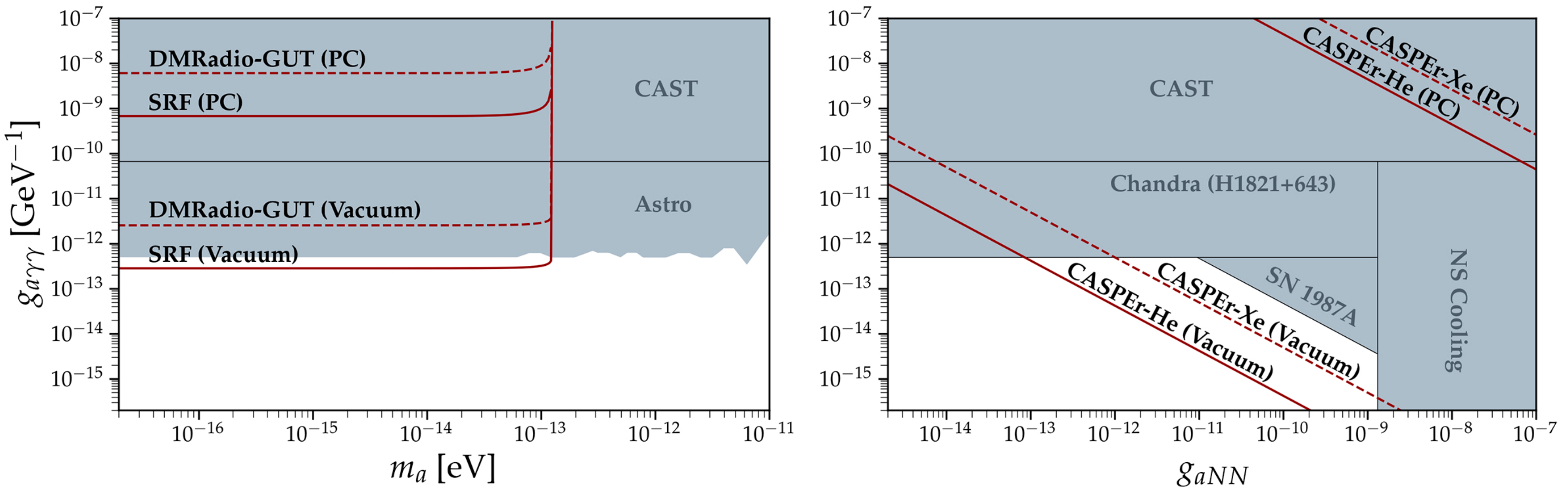}
    \caption{\textbf{Left:}~The projected sensitivity to the axion-photon coupling $g_{a\gamma \gamma}$ as a function of the axion mass, $m_a$, from one-year dedicated searches with the proposed Dark~SRF axion experiment and the DMRadio-GUT experiment for the axion-induced signal from the Crab pulsar at $f \approx 29.6$~Hz. 
    We illustrate projected 95\% upper limits under the null hypothesis assuming both the VDM magnetosphere (`Vacuum') and for the PC model, where the axions are only produced in small gaps near the NS poles.  
    The grey shaded regions represent current constraints on this parameter space, as summarized in the main text. \textbf{Right:}~As in the left panel, but for the future CASPER-wind experiment, which searches for the axion-nucleon coupling $g_{aNN}$. This figure assumes $m_a \ll 10^{-13}$~eV and illustrates the projected 95\% upper limits in the $g_{a\gamma\gamma}-g_{aNN}$ plane, for two different target materials. 
    }
    \label{fig:results}
\end{figure*}

Figure~\ref{fig:results} (left panel) compares the expected sensitivity to existing constraints from the CAST experiment~\cite{CAST2015}, which looks for axions produced in the Sun.
The shaded grey region
also illustrates the current constraints from a number of astrophysical axion probes, including searches for spectral distortions of astrophysical sources with
 Chandra X-ray data~\cite{Reynes:2021bpe, Reynolds:2019uqt, Marsh:2017yvc, Wouters:2013hua} and super star cluster searches for axions with NuSTAR X-ray data~\cite{Dessert:2020lil} (see also~\cite{Eby:2024mhd, Song:2024rru}).  Note that some of the astrophysical limits shown may also be subject to theoretical uncertainties ({\it e.g.},~\cite{Libanov:2019fzq,Matthews:2022gqi}).  Assuming the VDM (labeled `Vacuum'), the Dark~SRF projected sensitivity would be world leading, {improving upon the CAST bounds on $g_{a\gamma \gamma}$ by more than two orders of magnitude}. In the electrosphere model, the axion power is lower than predicted by the VDM by a factor of $x\sim$ few, which leads to a modest drop in sensitivity of $x^{1/4}$. Assuming the conservative PC model, with radiated power given in~\eqref{eq:P_a_PC}, the projected sensitivity does not pass that of the CAST experiment. {This motivates more detailed modeling of the Crab pulsar magnetosphere to determine precisely where the sensitivity lies.} 

Figure~\ref{fig:results}~(left panel) also illustrates the projected sensitivity of the planned DMRadio-GUT experiment~\cite{DMRadio:2022jfv}, assuming the experiment takes data with a quality factor of $2\times 10^7$~\cite{DMRadio:2022jfv} for one year with a resonant readout at the Crab frequency. See the SM for further details.

The right panel of Fig.~\ref{fig:results} shows the projected sensitivity of the CASPEr-wind experiment~\cite{Casper2013, JacksonKimball:2017elr,Garcon2019, Wu2019} under both the VDM and PC magnetosphere models, assuming $m_a \ll 10^{-13}$ eV.  CASPEr-wind searches for the axion through {its coupling to the nucleon}:  $H_{\rm int} = \gamma_N {\bm B}_a \cdot {\bm S}_N$, with $\gamma_N$ the nucleon gyromagnetic ratio and ${\bm B}_a$ an effective magnetic field that is equal to ${\bm B}_a = - g_{aNN} {\bm \nabla} a / \gamma_N$.  CASPEr-wind, and also co-magnetometer experiments such as~\cite{Lee2023, Bloch2022}, use nuclear magnetic resonance techniques with a transverse magnetometer to search for spin precession of a magnetized sample due to the axion interaction Hamiltonian; the effect is resonantly enhanced when the axion frequency matches the Larmor frequency $\omega_0 = \gamma_N B_E$, with $B_E$ the static external magnetic field that aligns the spins.

 Measuring the transverse magnetic field over $t= 1 \, \, {\rm year}$, the axion signal contributes to a single frequency bin because $t \ll \tau_a$.  On the other hand, the width of the resonance is limited by the transverse spin relaxation time $T_2$, which is expected to be much less than $t$.  As in the SRF case, one may compute the expected signal contribution within the frequency bin $k^*$  and compare it to the expected background noise, which arises predominantly from spin projection and magnetometer noise for CASPEr-wind~\cite{Dror2022}. (See the SM for details.)

 Figure~\ref{fig:results}~(right panel) shows the projected 95\% upper limits in the space of $|g_{a\gamma\gamma}|$ and $|g_{aNN}|$.
 (Note that the observed signal is proportional to the combination $g_{a\gamma\gamma}^2 g_{aNN}^2$.) The projected sensitivity is shown for two different assumptions regarding the CASPEr-wind type experiment (see, {\it e.g.},~\cite{Dror2022} for the justification of these choices): the more conservative case takes a Xe-129 target, which has a nuclear magnetic moment of $0.78 \mu_N$, with $\mu_N$ the nuclear magneton, along with the number density $n \approx 1.3 \times 10^{22}$ cm$^{-3}$. The more optimistic scenario assumes a Helium-3 target with a nuclear magnetic moment $\sim$$2.1 \mu_N$ and $n \approx 2.8 \times 10^{22}$ cm$^{-3}$.  For both, the volume is $1$ m$^3$, spin relaxation time is $T_2 = 100$~sec, and the effective SQUID loop area is $0.3 \ {\rm cm}^2$.

 We compare the expected upper limits from CASPEr-wind to the existing constraints on this parameter space on $g_{a\gamma\gamma}$ alone
 along with the NS cooling constraints on $g_{aNN}$~\cite{Buschmann:2021juv}.  Additionally, the non-observation of gamma-rays from SN1987A that would have been produced by nucleon bremsstrahlung (with $g_{aNN}$) and then converted to gamma-rays in the Galactic magnetic fields (with $g_{a\gamma\gamma}$) gives a constraint on the combination $g_{a\gamma\gamma} \times g_{aNN}$~\cite{Payez:2014xsa,safdi2024}, which is also illustrated.  Both the Xe and He targets could probe previously unexplored regions of parameter space for the VDM magnetosphere.

\vspace{0.1in}

\noindent \textbf{\textit{Discussion.---}} We proposed a new target for axion DM detection experiments: pulsar-sourced axions.  
These relativistic axions are produced at known frequencies; for example, the signal from the Crab pulsar is around $f\approx 29.6$ Hz.  
Upcoming searches for axions from the Crab pulsar may probe previously-unexplored regions of low-mass axion parameter space.  Global particle-in-cell~(PIC) plasma simulations of the Crab magnetosphere along the lines of~\cite{Spitkovsky2006, Kalapotharakos2009, Petri2012} are needed to better model the distribution of $\edb$ in the magnetosphere and compute the axion luminosity and line width more accurately. 

It is also possible that better pulsar candidates exist that are not known at present. These could include high-field, high-period pulsars that are not beamed towards Earth or are otherwise obscured to electromagnetic radiation. Young pulsars born after supernova, including possible future supernova, may also be superior targets.  Transient axion signals could also arise from gamma-ray bursts and magnetar glitches.

The axion emission from pulsars may also directly affect the observed pulsar quantities such as the period and magnetic field.  For example, the spin-down luminosity of the Crab pulsar is $L_{\rm spin-down} = 4.5 \times 10^{38}$ erg/s~\cite{ATNF}.  
The axion-induced emission formula in~\eqref{eqn:vacuumdipolepower}  surpasses the spin-down luminosity for axion-photon couplings larger than $g_{a\gamma\gamma} \sim 5 \times 10^{-11}$ GeV$^{-1}$. This strongly suggests a conservative upper limit of $|g_{a\gamma\gamma}| \lesssim 5 \times 10^{-11}$ GeV$^{-1}$, assuming the VDM model, for $m_a \lesssim 10^{-13}$ eV.  On the other hand, the limit can likely be substantially improved by studying the back-reaction of the axion emission more carefully on the pulsar dynamics. 

Another class of astrophysically-produced low-frequency axions that could be detectable with terrestrial DM detectors are signals narrow in the time domain.  For example, we estimate that NS-NS inspirals and NS-black-hole inspirals may yield detectable but short-lived axion signals, which would be coincident in time with gravitational wave signals, depending on the magnetic properties of the NSs and the distance to the event. We leave detailed estimates of this signal to future work, which could most optimistically add axions as an additional messenger to the growing field of multi-messenger astronomy. 

\vspace{0.1in}

\noindent {\it Acknowledgments}---
The authors acknowledge A.~Berlin, S.~Chaudhuri, J.~Foster, H.~ Hakobyan, Y.~Kahn,  N.~Rodd, A.~Sushkov, L.~Winslow, and S.~Witte for helpful discussions. 
ML and MK are supported by the Department of Energy~(DOE) under Award Number DE-SC0007968.  ML also acknowledges support from the Simons Investigator in Physics Award.  AP acknowledges support from the Princeton Center for Theoretical Science.  BRS is supported in part by the DOE Early Career Grant DESC0019225 and in part by a Sloan Research Fellowship. 
MK is grateful to Princeton University for their hospitality.
The work presented in this paper was performed on computational resources managed and supported by Princeton Research Computing.
This research also made extensive use of the publicly available codes \texttt{IPython}~\citep{PER-GRA:2007}, \texttt{matplotlib}~\citep{Hunter:2007},
\texttt{Jupyter}~\citep{Kluyver2016jupyter},
\texttt{NumPy}~\citep{harris2020array}, and 
\texttt{SciPy}~\citep{2020SciPy-NMeth}.


\bibliography{pulsarscope}

\begin{thebibliography}{123}%
\makeatletter
\providecommand \@ifxundefined [1]{%
 \@ifx{#1\undefined}
}%
\providecommand \@ifnum [1]{%
 \ifnum #1\expandafter \@firstoftwo
 \else \expandafter \@secondoftwo
 \fi
}%
\providecommand \@ifx [1]{%
 \ifx #1\expandafter \@firstoftwo
 \else \expandafter \@secondoftwo
 \fi
}%
\providecommand \natexlab [1]{#1}%
\providecommand \enquote  [1]{``#1''}%
\providecommand \bibnamefont  [1]{#1}%
\providecommand \bibfnamefont [1]{#1}%
\providecommand \citenamefont [1]{#1}%
\providecommand \href@noop [0]{\@secondoftwo}%
\providecommand \href [0]{\begingroup \@sanitize@url \@href}%
\providecommand \@href[1]{\@@startlink{#1}\@@href}%
\providecommand \@@href[1]{\endgroup#1\@@endlink}%
\providecommand \@sanitize@url [0]{\catcode `\\12\catcode `\$12\catcode
  `\&12\catcode `\#12\catcode `\^12\catcode `\_12\catcode `\%12\relax}%
\providecommand \@@startlink[1]{}%
\providecommand \@@endlink[0]{}%
\providecommand \url  [0]{\begingroup\@sanitize@url \@url }%
\providecommand \@url [1]{\endgroup\@href {#1}{\urlprefix }}%
\providecommand \urlprefix  [0]{URL }%
\providecommand \Eprint [0]{\href }%
\providecommand \doibase [0]{http://dx.doi.org/}%
\providecommand \selectlanguage [0]{\@gobble}%
\providecommand \bibinfo  [0]{\@secondoftwo}%
\providecommand \bibfield  [0]{\@secondoftwo}%
\providecommand \translation [1]{[#1]}%
\providecommand \BibitemOpen [0]{}%
\providecommand \bibitemStop [0]{}%
\providecommand \bibitemNoStop [0]{.\EOS\space}%
\providecommand \EOS [0]{\spacefactor3000\relax}%
\providecommand \BibitemShut  [1]{\csname bibitem#1\endcsname}%
\let\auto@bib@innerbib\@empty
\bibitem [{\citenamefont {Svrcek}\ and\ \citenamefont
  {Witten}(2006)}]{Svrcek_2006}%
  \BibitemOpen
  \bibfield  {author} {\bibinfo {author} {\bibfnamefont {P.}~\bibnamefont
  {Svrcek}}\ and\ \bibinfo {author} {\bibfnamefont {E.}~\bibnamefont
  {Witten}},\ }\href {\doibase 10.1088/1126-6708/2006/06/051} {\bibfield
  {journal} {\bibinfo  {journal} {JHEP}\ }\textbf {\bibinfo {volume} {2006}},\
  \bibinfo {pages} {051–051} (\bibinfo {year} {2006})}\BibitemShut {NoStop}%
\bibitem [{\citenamefont {Arvanitaki}\ \emph {et~al.}(2010)\citenamefont
  {Arvanitaki}, \citenamefont {Dimopoulos}, \citenamefont {Dubovsky},
  \citenamefont {Kaloper},\ and\ \citenamefont {March-Russell}}]{Axiverse2010}%
  \BibitemOpen
  \bibfield  {author} {\bibinfo {author} {\bibfnamefont {A.}~\bibnamefont
  {Arvanitaki}}, \bibinfo {author} {\bibfnamefont {S.}~\bibnamefont
  {Dimopoulos}}, \bibinfo {author} {\bibfnamefont {S.}~\bibnamefont
  {Dubovsky}}, \bibinfo {author} {\bibfnamefont {N.}~\bibnamefont {Kaloper}}, \
  and\ \bibinfo {author} {\bibfnamefont {J.}~\bibnamefont {March-Russell}},\
  }\href {\doibase 10.1103/PhysRevD.81.123530} {\bibfield  {journal} {\bibinfo
  {journal} {Phys. Rev. D}\ }\textbf {\bibinfo {volume} {81}},\ \bibinfo
  {pages} {123530} (\bibinfo {year} {2010})}\BibitemShut {NoStop}%
\bibitem [{\citenamefont {Preskill}\ \emph {et~al.}(1983)\citenamefont
  {Preskill}, \citenamefont {Wise},\ and\ \citenamefont
  {Wilczek}}]{PRESKILL1983127}%
  \BibitemOpen
  \bibfield  {author} {\bibinfo {author} {\bibfnamefont {J.}~\bibnamefont
  {Preskill}}, \bibinfo {author} {\bibfnamefont {M.~B.}\ \bibnamefont {Wise}},
  \ and\ \bibinfo {author} {\bibfnamefont {F.}~\bibnamefont {Wilczek}},\ }\href
  {\doibase https://doi.org/10.1016/0370-2693(83)90637-8} {\bibfield  {journal}
  {\bibinfo  {journal} {Physics Letters B}\ }\textbf {\bibinfo {volume}
  {120}},\ \bibinfo {pages} {127 } (\bibinfo {year} {1983})}\BibitemShut
  {NoStop}%
\bibitem [{\citenamefont {Abbott}\ and\ \citenamefont
  {Sikivie}(1983)}]{Abbott1982}%
  \BibitemOpen
  \bibfield  {author} {\bibinfo {author} {\bibfnamefont {L.}~\bibnamefont
  {Abbott}}\ and\ \bibinfo {author} {\bibfnamefont {P.}~\bibnamefont
  {Sikivie}},\ }\href {\doibase 10.1016/0370-2693(83)90638-X} {\bibfield
  {journal} {\bibinfo  {journal} {Phys. Lett. B}\ }\textbf {\bibinfo {volume}
  {120}},\ \bibinfo {pages} {133} (\bibinfo {year} {1983})}\BibitemShut
  {NoStop}%
\bibitem [{\citenamefont {Dine}\ and\ \citenamefont
  {Fischler}(1983)}]{Fischler1982}%
  \BibitemOpen
  \bibfield  {author} {\bibinfo {author} {\bibfnamefont {M.}~\bibnamefont
  {Dine}}\ and\ \bibinfo {author} {\bibfnamefont {W.}~\bibnamefont
  {Fischler}},\ }\href {\doibase 10.1016/0370-2693(83)90639-1} {\bibfield
  {journal} {\bibinfo  {journal} {Phys. Lett. B}\ }\textbf {\bibinfo {volume}
  {120}},\ \bibinfo {pages} {137} (\bibinfo {year} {1983})}\BibitemShut
  {NoStop}%
\bibitem [{\citenamefont {{Planck Collaboration}}\ \emph
  {et~al.}(2020)\citenamefont {{Planck Collaboration}}, \citenamefont
  {{Aghanim, N.}} \emph {et~al.}}]{Planck2018}%
  \BibitemOpen
  \bibfield  {author} {\bibinfo {author} {\bibnamefont {{Planck
  Collaboration}}}, \bibinfo {author} {\bibnamefont {{Aghanim, N.}}},  \emph
  {et~al.},\ }\href {\doibase 10.1051/0004-6361/201833910} {\bibfield
  {journal} {\bibinfo  {journal} {A\&A}\ }\textbf {\bibinfo {volume} {641}},\
  \bibinfo {pages} {A6} (\bibinfo {year} {2020})}\BibitemShut {NoStop}%
\bibitem [{\citenamefont {Peccei}\ and\ \citenamefont
  {Quinn}(1977{\natexlab{a}})}]{PQ1}%
  \BibitemOpen
  \bibfield  {author} {\bibinfo {author} {\bibfnamefont {R.~D.}\ \bibnamefont
  {Peccei}}\ and\ \bibinfo {author} {\bibfnamefont {H.~R.}\ \bibnamefont
  {Quinn}},\ }\href {\doibase 10.1103/PhysRevLett.38.1440} {\bibfield
  {journal} {\bibinfo  {journal} {Phys. Rev. Lett.}\ }\textbf {\bibinfo
  {volume} {38}},\ \bibinfo {pages} {1440} (\bibinfo {year}
  {1977}{\natexlab{a}})}\BibitemShut {NoStop}%
\bibitem [{\citenamefont {Peccei}\ and\ \citenamefont
  {Quinn}(1977{\natexlab{b}})}]{PQ2}%
  \BibitemOpen
  \bibfield  {author} {\bibinfo {author} {\bibfnamefont {R.~D.}\ \bibnamefont
  {Peccei}}\ and\ \bibinfo {author} {\bibfnamefont {H.~R.}\ \bibnamefont
  {Quinn}},\ }\href {\doibase 10.1103/PhysRevD.16.1791} {\bibfield  {journal}
  {\bibinfo  {journal} {Phys. Rev. D}\ }\textbf {\bibinfo {volume} {16}},\
  \bibinfo {pages} {1791} (\bibinfo {year} {1977}{\natexlab{b}})}\BibitemShut
  {NoStop}%
\bibitem [{\citenamefont {Weinberg}(1978)}]{WeinbergAxion}%
  \BibitemOpen
  \bibfield  {author} {\bibinfo {author} {\bibfnamefont {S.}~\bibnamefont
  {Weinberg}},\ }\href {\doibase 10.1103/PhysRevLett.40.223} {\bibfield
  {journal} {\bibinfo  {journal} {Phys. Rev. Lett.}\ }\textbf {\bibinfo
  {volume} {40}},\ \bibinfo {pages} {223} (\bibinfo {year} {1978})}\BibitemShut
  {NoStop}%
\bibitem [{\citenamefont {Wilczek}(1978)}]{WilczekAxion}%
  \BibitemOpen
  \bibfield  {author} {\bibinfo {author} {\bibfnamefont {F.}~\bibnamefont
  {Wilczek}},\ }\href {\doibase 10.1103/PhysRevLett.40.279} {\bibfield
  {journal} {\bibinfo  {journal} {Phys. Rev. Lett.}\ }\textbf {\bibinfo
  {volume} {40}},\ \bibinfo {pages} {279} (\bibinfo {year} {1978})}\BibitemShut
  {NoStop}%
\bibitem [{\citenamefont {Graham}\ \emph {et~al.}(2015)\citenamefont {Graham},
  \citenamefont {Irastorza}, \citenamefont {Lamoreaux}, \citenamefont
  {Lindner},\ and\ \citenamefont {van Bibber}}]{Graham:2015ouw}%
  \BibitemOpen
  \bibfield  {author} {\bibinfo {author} {\bibfnamefont {P.~W.}\ \bibnamefont
  {Graham}}, \bibinfo {author} {\bibfnamefont {I.~G.}\ \bibnamefont
  {Irastorza}}, \bibinfo {author} {\bibfnamefont {S.~K.}\ \bibnamefont
  {Lamoreaux}}, \bibinfo {author} {\bibfnamefont {A.}~\bibnamefont {Lindner}},
  \ and\ \bibinfo {author} {\bibfnamefont {K.~A.}\ \bibnamefont {van Bibber}},\
  }\href {\doibase 10.1146/annurev-nucl-102014-022120} {\bibfield  {journal}
  {\bibinfo  {journal} {Ann. Rev. Nucl. Part. Sci.}\ }\textbf {\bibinfo
  {volume} {65}},\ \bibinfo {pages} {485} (\bibinfo {year} {2015})},\ \Eprint
  {http://arxiv.org/abs/1602.00039} {arXiv:1602.00039 [hep-ex]} \BibitemShut
  {NoStop}%
\bibitem [{\citenamefont {Adams}\ \emph {et~al.}(2022)\citenamefont {Adams}
  \emph {et~al.}}]{Adams:2022pbo}%
  \BibitemOpen
  \bibfield  {author} {\bibinfo {author} {\bibfnamefont {C.~B.}\ \bibnamefont
  {Adams}} \emph {et~al.},\ }in\ \href@noop {} {\emph {\bibinfo {booktitle}
  {{Snowmass 2021}}}}\ (\bibinfo {year} {2022})\ \Eprint
  {http://arxiv.org/abs/2203.14923} {arXiv:2203.14923 [hep-ex]} \BibitemShut
  {NoStop}%
\bibitem [{\citenamefont {Safdi}(2024)}]{Safdi:2022xkm}%
  \BibitemOpen
  \bibfield  {author} {\bibinfo {author} {\bibfnamefont {B.~R.}\ \bibnamefont
  {Safdi}},\ }\href {\doibase 10.22323/1.439.0009} {\bibfield  {journal}
  {\bibinfo  {journal} {PoS}\ }\textbf {\bibinfo {volume} {TASI2022}},\
  \bibinfo {pages} {009} (\bibinfo {year} {2024})},\ \Eprint
  {http://arxiv.org/abs/2303.02169} {arXiv:2303.02169 [hep-ph]} \BibitemShut
  {NoStop}%
\bibitem [{\citenamefont {Hagmann}\ \emph {et~al.}(1990)\citenamefont
  {Hagmann}, \citenamefont {Sikivie}, \citenamefont {Sullivan},\ and\
  \citenamefont {Tanner}}]{Hagmann:1990}%
  \BibitemOpen
  \bibfield  {author} {\bibinfo {author} {\bibfnamefont {C.}~\bibnamefont
  {Hagmann}}, \bibinfo {author} {\bibfnamefont {P.}~\bibnamefont {Sikivie}},
  \bibinfo {author} {\bibfnamefont {N.~S.}\ \bibnamefont {Sullivan}}, \ and\
  \bibinfo {author} {\bibfnamefont {D.~B.}\ \bibnamefont {Tanner}},\ }\href
  {\doibase 10.1103/PhysRevD.42.1297} {\bibfield  {journal} {\bibinfo
  {journal} {Phys. Rev. D}\ }\textbf {\bibinfo {volume} {42}},\ \bibinfo
  {pages} {1297} (\bibinfo {year} {1990})}\BibitemShut {NoStop}%
\bibitem [{\citenamefont {Hagmann}\ \emph {et~al.}(1998)\citenamefont {Hagmann}
  \emph {et~al.}}]{Hagmann:1998cb}%
  \BibitemOpen
  \bibfield  {author} {\bibinfo {author} {\bibfnamefont {C.}~\bibnamefont
  {Hagmann}} \emph {et~al.} (\bibinfo {collaboration} {ADMX}),\ }\href
  {\doibase 10.1103/PhysRevLett.80.2043} {\bibfield  {journal} {\bibinfo
  {journal} {Phys. Rev. Lett.}\ }\textbf {\bibinfo {volume} {80}},\ \bibinfo
  {pages} {2043} (\bibinfo {year} {1998})},\ \Eprint
  {http://arxiv.org/abs/astro-ph/9801286} {arXiv:astro-ph/9801286} \BibitemShut
  {NoStop}%
\bibitem [{\citenamefont {Asztalos}\ \emph {et~al.}(2001)\citenamefont
  {Asztalos} \emph {et~al.}}]{Asztalos:2001tf}%
  \BibitemOpen
  \bibfield  {author} {\bibinfo {author} {\bibfnamefont {S.~J.}\ \bibnamefont
  {Asztalos}} \emph {et~al.} (\bibinfo {collaboration} {ADMX}),\ }\href
  {\doibase 10.1103/PhysRevD.64.092003} {\bibfield  {journal} {\bibinfo
  {journal} {Phys. Rev. D}\ }\textbf {\bibinfo {volume} {64}},\ \bibinfo
  {pages} {092003} (\bibinfo {year} {2001})}\BibitemShut {NoStop}%
\bibitem [{\citenamefont {Asztalos}\ \emph {et~al.}(2010)\citenamefont
  {Asztalos} \emph {et~al.}}]{Asztalos:2009yp}%
  \BibitemOpen
  \bibfield  {author} {\bibinfo {author} {\bibfnamefont {S.~J.}\ \bibnamefont
  {Asztalos}} \emph {et~al.} (\bibinfo {collaboration} {ADMX}),\ }\href
  {\doibase 10.1103/PhysRevLett.104.041301} {\bibfield  {journal} {\bibinfo
  {journal} {Phys. Rev. Lett.}\ }\textbf {\bibinfo {volume} {104}},\ \bibinfo
  {pages} {041301} (\bibinfo {year} {2010})},\ \Eprint
  {http://arxiv.org/abs/0910.5914} {arXiv:0910.5914 [astro-ph.CO]} \BibitemShut
  {NoStop}%
\bibitem [{\citenamefont {Du}\ \emph {et~al.}(2018{\natexlab{a}})\citenamefont
  {Du} \emph {et~al.}}]{Du:2018uak}%
  \BibitemOpen
  \bibfield  {author} {\bibinfo {author} {\bibfnamefont {N.}~\bibnamefont {Du}}
  \emph {et~al.} (\bibinfo {collaboration} {ADMX}),\ }\href {\doibase
  10.1103/PhysRevLett.120.151301} {\bibfield  {journal} {\bibinfo  {journal}
  {Phys. Rev. Lett.}\ }\textbf {\bibinfo {volume} {120}},\ \bibinfo {pages}
  {151301} (\bibinfo {year} {2018}{\natexlab{a}})},\ \Eprint
  {http://arxiv.org/abs/1804.05750} {arXiv:1804.05750 [hep-ex]} \BibitemShut
  {NoStop}%
\bibitem [{\citenamefont {Braine}\ \emph
  {et~al.}(2020{\natexlab{a}})\citenamefont {Braine} \emph
  {et~al.}}]{Braine2020}%
  \BibitemOpen
  \bibfield  {author} {\bibinfo {author} {\bibfnamefont {T.}~\bibnamefont
  {Braine}} \emph {et~al.} (\bibinfo {collaboration} {ADMX}),\ }\href {\doibase
  10.1103/physrevlett.124.101303} {\bibfield  {journal} {\bibinfo  {journal}
  {Physical Review Letters}\ }\textbf {\bibinfo {volume} {124}} (\bibinfo
  {year} {2020}{\natexlab{a}}),\ 10.1103/physrevlett.124.101303}\BibitemShut
  {NoStop}%
\bibitem [{\citenamefont {Bradley}\ \emph {et~al.}(2003)\citenamefont {Bradley}
  \emph {et~al.}}]{Bradley2003}%
  \BibitemOpen
  \bibfield  {author} {\bibinfo {author} {\bibfnamefont {R.}~\bibnamefont
  {Bradley}} \emph {et~al.},\ }\href {\doibase 10.1103/RevModPhys.75.777}
  {\bibfield  {journal} {\bibinfo  {journal} {Rev. Mod. Phys.}\ }\textbf
  {\bibinfo {volume} {75}},\ \bibinfo {pages} {777} (\bibinfo {year}
  {2003})}\BibitemShut {NoStop}%
\bibitem [{\citenamefont {Asztalos}\ \emph {et~al.}(2004)\citenamefont
  {Asztalos} \emph {et~al.}}]{Bradley2004}%
  \BibitemOpen
  \bibfield  {author} {\bibinfo {author} {\bibfnamefont {S.~J.}\ \bibnamefont
  {Asztalos}} \emph {et~al.},\ }\href {\doibase 10.1103/PhysRevD.69.011101}
  {\bibfield  {journal} {\bibinfo  {journal} {Phys. Rev. D}\ }\textbf {\bibinfo
  {volume} {69}},\ \bibinfo {pages} {011101} (\bibinfo {year}
  {2004})}\BibitemShut {NoStop}%
\bibitem [{\citenamefont {Shokair}\ \emph {et~al.}(2014)\citenamefont {Shokair}
  \emph {et~al.}}]{Shokair2014}%
  \BibitemOpen
  \bibfield  {author} {\bibinfo {author} {\bibfnamefont {T.~M.}\ \bibnamefont
  {Shokair}} \emph {et~al.},\ }\href {\doibase 10.1142/s0217751x14430040}
  {\bibfield  {journal} {\bibinfo  {journal} {International Journal of Modern
  Physics A}\ }\textbf {\bibinfo {volume} {29}},\ \bibinfo {pages} {1443004}
  (\bibinfo {year} {2014})}\BibitemShut {NoStop}%
\bibitem [{\citenamefont {Brubaker}\ \emph {et~al.}(2017)\citenamefont
  {Brubaker} \emph {et~al.}}]{HAYSTAC}%
  \BibitemOpen
  \bibfield  {author} {\bibinfo {author} {\bibfnamefont {B.~M.}\ \bibnamefont
  {Brubaker}} \emph {et~al.} (\bibinfo {collaboration} {HAYSTAC}),\ }\href
  {\doibase 10.1103/PhysRevLett.118.061302} {\bibfield  {journal} {\bibinfo
  {journal} {Phys. Rev. Lett.}\ }\textbf {\bibinfo {volume} {118}},\ \bibinfo
  {pages} {061302} (\bibinfo {year} {2017})}\BibitemShut {NoStop}%
\bibitem [{\citenamefont {Zhong}\ \emph {et~al.}(2018)\citenamefont {Zhong}
  \emph {et~al.}}]{Zhong2018}%
  \BibitemOpen
  \bibfield  {author} {\bibinfo {author} {\bibfnamefont {L.}~\bibnamefont
  {Zhong}} \emph {et~al.} (\bibinfo {collaboration} {HAYSTAC}),\ }\href
  {\doibase 10.1103/physrevd.97.092001} {\bibfield  {journal} {\bibinfo
  {journal} {Physical Review D}\ }\textbf {\bibinfo {volume} {97}} (\bibinfo
  {year} {2018}),\ 10.1103/physrevd.97.092001}\BibitemShut {NoStop}%
\bibitem [{\citenamefont {Backes}\ \emph {et~al.}(2021)\citenamefont {Backes}
  \emph {et~al.}}]{Backes_2021}%
  \BibitemOpen
  \bibfield  {author} {\bibinfo {author} {\bibfnamefont {K.~M.}\ \bibnamefont
  {Backes}} \emph {et~al.} (\bibinfo {collaboration} {HAYSTAC}),\ }\href
  {\doibase 10.1038/s41586-021-03226-7} {\bibfield  {journal} {\bibinfo
  {journal} {Nature}\ }\textbf {\bibinfo {volume} {590}},\ \bibinfo {pages}
  {238} (\bibinfo {year} {2021})}\BibitemShut {NoStop}%
\bibitem [{\citenamefont {Choi}\ \emph {et~al.}(2021)\citenamefont {Choi},
  \citenamefont {Ahn}, \citenamefont {Ko}, \citenamefont {Lee},\ and\
  \citenamefont {Semertzidis}}]{Choi_2021}%
  \BibitemOpen
  \bibfield  {author} {\bibinfo {author} {\bibfnamefont {J.}~\bibnamefont
  {Choi}}, \bibinfo {author} {\bibfnamefont {S.}~\bibnamefont {Ahn}}, \bibinfo
  {author} {\bibfnamefont {B.}~\bibnamefont {Ko}}, \bibinfo {author}
  {\bibfnamefont {S.}~\bibnamefont {Lee}}, \ and\ \bibinfo {author}
  {\bibfnamefont {Y.}~\bibnamefont {Semertzidis}},\ }\href {\doibase
  10.1016/j.nima.2021.165667} {\bibfield  {journal} {\bibinfo  {journal}
  {NIM-A}\ }\textbf {\bibinfo {volume} {1013}},\ \bibinfo {pages} {165667}
  (\bibinfo {year} {2021})}\BibitemShut {NoStop}%
\bibitem [{\citenamefont {Crescini}\ \emph {et~al.}(2020)\citenamefont
  {Crescini} \emph {et~al.}}]{QUAX:2020adt}%
  \BibitemOpen
  \bibfield  {author} {\bibinfo {author} {\bibfnamefont {N.}~\bibnamefont
  {Crescini}} \emph {et~al.} (\bibinfo {collaboration} {QUAX}),\ }\href
  {\doibase 10.1103/PhysRevLett.124.171801} {\bibfield  {journal} {\bibinfo
  {journal} {Phys. Rev. Lett.}\ }\textbf {\bibinfo {volume} {124}},\ \bibinfo
  {pages} {171801} (\bibinfo {year} {2020})},\ \Eprint
  {http://arxiv.org/abs/2001.08940} {arXiv:2001.08940 [hep-ex]} \BibitemShut
  {NoStop}%
\bibitem [{\citenamefont {McAllister}\ \emph {et~al.}(2017)\citenamefont
  {McAllister} \emph {et~al.}}]{mcallister2017organ}%
  \BibitemOpen
  \bibfield  {author} {\bibinfo {author} {\bibfnamefont {B.~T.}\ \bibnamefont
  {McAllister}} \emph {et~al.},\ }\href@noop {} {\enquote {\bibinfo {title}
  {The organ experiment: An axion haloscope above 15 ghz},}\ } (\bibinfo {year}
  {2017}),\ \Eprint {http://arxiv.org/abs/1706.00209} {arXiv:1706.00209
  [physics.ins-det]} \BibitemShut {NoStop}%
\bibitem [{\citenamefont {Sikivie}(1983)}]{Sikivie:1983ip}%
  \BibitemOpen
  \bibfield  {author} {\bibinfo {author} {\bibfnamefont {P.}~\bibnamefont
  {Sikivie}},\ }\href {\doibase 10.1103/PhysRevLett.51.1415} {\bibfield
  {journal} {\bibinfo  {journal} {Phys. Rev. Lett.}\ }\textbf {\bibinfo
  {volume} {51}},\ \bibinfo {pages} {1415} (\bibinfo {year} {1983})},\ \bibinfo
  {note} {[Erratum: Phys.Rev.Lett. 52, 695 (1984)]}\BibitemShut {NoStop}%
\bibitem [{\citenamefont {Anastassopoulos}\ \emph {et~al.}(2017)\citenamefont
  {Anastassopoulos} \emph {et~al.}}]{CAST:2017uph}%
  \BibitemOpen
  \bibfield  {author} {\bibinfo {author} {\bibfnamefont {V.}~\bibnamefont
  {Anastassopoulos}} \emph {et~al.} (\bibinfo {collaboration} {CAST}),\ }\href
  {\doibase 10.1038/nphys4109} {\bibfield  {journal} {\bibinfo  {journal}
  {Nature Phys.}\ }\textbf {\bibinfo {volume} {13}},\ \bibinfo {pages} {584}
  (\bibinfo {year} {2017})},\ \Eprint {http://arxiv.org/abs/1705.02290}
  {arXiv:1705.02290 [hep-ex]} \BibitemShut {NoStop}%
\bibitem [{\citenamefont {Ehret}\ \emph {et~al.}(2010)\citenamefont {Ehret}
  \emph {et~al.}}]{Ehret:2010mh}%
  \BibitemOpen
  \bibfield  {author} {\bibinfo {author} {\bibfnamefont {K.}~\bibnamefont
  {Ehret}} \emph {et~al.} (\bibinfo {collaboration} {ALPS}),\ }\href {\doibase
  10.1016/j.physletb.2010.04.066} {\bibfield  {journal} {\bibinfo  {journal}
  {Phys. Lett.}\ }\textbf {\bibinfo {volume} {B689}},\ \bibinfo {pages} {149}
  (\bibinfo {year} {2010})},\ \Eprint {http://arxiv.org/abs/1004.1313}
  {arXiv:1004.1313 [hep-ex]} \BibitemShut {NoStop}%
\bibitem [{\citenamefont {{Diaz Ortiz}}\ \emph {et~al.}(2022)\citenamefont
  {{Diaz Ortiz}}, \citenamefont {Gleason}, \citenamefont {Grote}, \citenamefont
  {Hallal}, \citenamefont {Hartman}, \citenamefont {Hollis}, \citenamefont
  {Isleif}, \citenamefont {James}, \citenamefont {Karan}, \citenamefont
  {Kozlowski}, \citenamefont {Lindner}, \citenamefont {Messineo}, \citenamefont
  {Mueller}, \citenamefont {Põld}, \citenamefont {Smith}, \citenamefont
  {Spector}, \citenamefont {Tanner}, \citenamefont {Wei},\ and\ \citenamefont
  {Willke}}]{DIAZORTIZ2022100968}%
  \BibitemOpen
  \bibfield  {author} {\bibinfo {author} {\bibfnamefont {M.}~\bibnamefont
  {{Diaz Ortiz}}}, \bibinfo {author} {\bibfnamefont {J.}~\bibnamefont
  {Gleason}}, \bibinfo {author} {\bibfnamefont {H.}~\bibnamefont {Grote}},
  \bibinfo {author} {\bibfnamefont {A.}~\bibnamefont {Hallal}}, \bibinfo
  {author} {\bibfnamefont {M.}~\bibnamefont {Hartman}}, \bibinfo {author}
  {\bibfnamefont {H.}~\bibnamefont {Hollis}}, \bibinfo {author} {\bibfnamefont
  {K.-S.}\ \bibnamefont {Isleif}}, \bibinfo {author} {\bibfnamefont
  {A.}~\bibnamefont {James}}, \bibinfo {author} {\bibfnamefont
  {K.}~\bibnamefont {Karan}}, \bibinfo {author} {\bibfnamefont
  {T.}~\bibnamefont {Kozlowski}}, \bibinfo {author} {\bibfnamefont
  {A.}~\bibnamefont {Lindner}}, \bibinfo {author} {\bibfnamefont
  {G.}~\bibnamefont {Messineo}}, \bibinfo {author} {\bibfnamefont
  {G.}~\bibnamefont {Mueller}}, \bibinfo {author} {\bibfnamefont
  {J.}~\bibnamefont {Põld}}, \bibinfo {author} {\bibfnamefont
  {R.}~\bibnamefont {Smith}}, \bibinfo {author} {\bibfnamefont
  {A.}~\bibnamefont {Spector}}, \bibinfo {author} {\bibfnamefont
  {D.}~\bibnamefont {Tanner}}, \bibinfo {author} {\bibfnamefont {L.-W.}\
  \bibnamefont {Wei}}, \ and\ \bibinfo {author} {\bibfnamefont
  {B.}~\bibnamefont {Willke}},\ }\href {\doibase
  https://doi.org/10.1016/j.dark.2022.100968} {\bibfield  {journal} {\bibinfo
  {journal} {Physics of the Dark Universe}\ }\textbf {\bibinfo {volume} {35}},\
  \bibinfo {pages} {100968} (\bibinfo {year} {2022})}\BibitemShut {NoStop}%
\bibitem [{\citenamefont {Du}\ \emph {et~al.}(2018{\natexlab{b}})\citenamefont
  {Du} \emph {et~al.}}]{ADMX:2018gho}%
  \BibitemOpen
  \bibfield  {author} {\bibinfo {author} {\bibfnamefont {N.}~\bibnamefont {Du}}
  \emph {et~al.} (\bibinfo {collaboration} {ADMX}),\ }\href {\doibase
  10.1103/PhysRevLett.120.151301} {\bibfield  {journal} {\bibinfo  {journal}
  {Phys. Rev. Lett.}\ }\textbf {\bibinfo {volume} {120}},\ \bibinfo {pages}
  {151301} (\bibinfo {year} {2018}{\natexlab{b}})},\ \Eprint
  {http://arxiv.org/abs/1804.05750} {arXiv:1804.05750 [hep-ex]} \BibitemShut
  {NoStop}%
\bibitem [{\citenamefont {Braine}\ \emph
  {et~al.}(2020{\natexlab{b}})\citenamefont {Braine} \emph
  {et~al.}}]{ADMX:2019uok}%
  \BibitemOpen
  \bibfield  {author} {\bibinfo {author} {\bibfnamefont {T.}~\bibnamefont
  {Braine}} \emph {et~al.} (\bibinfo {collaboration} {ADMX}),\ }\href {\doibase
  10.1103/PhysRevLett.124.101303} {\bibfield  {journal} {\bibinfo  {journal}
  {Phys. Rev. Lett.}\ }\textbf {\bibinfo {volume} {124}},\ \bibinfo {pages}
  {101303} (\bibinfo {year} {2020}{\natexlab{b}})},\ \Eprint
  {http://arxiv.org/abs/1910.08638} {arXiv:1910.08638 [hep-ex]} \BibitemShut
  {NoStop}%
\bibitem [{\citenamefont {Bartram}\ \emph
  {et~al.}(2021{\natexlab{a}})\citenamefont {Bartram} \emph
  {et~al.}}]{ADMX:2020hay}%
  \BibitemOpen
  \bibfield  {author} {\bibinfo {author} {\bibfnamefont {C.}~\bibnamefont
  {Bartram}} \emph {et~al.} (\bibinfo {collaboration} {ADMX}),\ }\href
  {\doibase 10.1103/PhysRevD.103.032002} {\bibfield  {journal} {\bibinfo
  {journal} {Phys. Rev. D}\ }\textbf {\bibinfo {volume} {103}},\ \bibinfo
  {pages} {032002} (\bibinfo {year} {2021}{\natexlab{a}})},\ \Eprint
  {http://arxiv.org/abs/2010.06183} {arXiv:2010.06183 [astro-ph.CO]}
  \BibitemShut {NoStop}%
\bibitem [{\citenamefont {Bartram}\ \emph
  {et~al.}(2021{\natexlab{b}})\citenamefont {Bartram}, \citenamefont {Braine},
  \citenamefont {Burns}, \citenamefont {Cervantes}, \citenamefont {Crisosto},
  \citenamefont {Du}, \citenamefont {Korandla}, \citenamefont {Leum},
  \citenamefont {Mohapatra}, \citenamefont {Nitta}, \citenamefont {Rosenberg},
  \citenamefont {Rybka}, \citenamefont {Yang}, \citenamefont {Clarke},
  \citenamefont {Siddiqi}, \citenamefont {Agrawal}, \citenamefont {Dixit},
  \citenamefont {Awida}, \citenamefont {Chou}, \citenamefont {Hollister},
  \citenamefont {Knirck}, \citenamefont {Sonnenschein}, \citenamefont {Wester},
  \citenamefont {Gleason}, \citenamefont {Hipp}, \citenamefont {Jois},
  \citenamefont {Sikivie}, \citenamefont {Sullivan}, \citenamefont {Tanner},
  \citenamefont {Lentz}, \citenamefont {Khatiwada}, \citenamefont {Carosi},
  \citenamefont {Robertson}, \citenamefont {Woollett}, \citenamefont {Duffy},
  \citenamefont {Boutan}, \citenamefont {Jones}, \citenamefont {LaRoque},
  \citenamefont {Oblath}, \citenamefont {Taubman}, \citenamefont {Daw},
  \citenamefont {Perry}, \citenamefont {Buckley}, \citenamefont {Gaikwad},
  \citenamefont {Hoffman}, \citenamefont {Murch}, \citenamefont {Goryachev},
  \citenamefont {McAllister}, \citenamefont {Quiskamp}, \citenamefont
  {Thomson},\ and\ \citenamefont {Tobar}}]{PhysRevLett.127.261803}%
  \BibitemOpen
  \bibfield  {author} {\bibinfo {author} {\bibfnamefont {C.}~\bibnamefont
  {Bartram}}, \bibinfo {author} {\bibfnamefont {T.}~\bibnamefont {Braine}},
  \bibinfo {author} {\bibfnamefont {E.}~\bibnamefont {Burns}}, \bibinfo
  {author} {\bibfnamefont {R.}~\bibnamefont {Cervantes}}, \bibinfo {author}
  {\bibfnamefont {N.}~\bibnamefont {Crisosto}}, \bibinfo {author}
  {\bibfnamefont {N.}~\bibnamefont {Du}}, \bibinfo {author} {\bibfnamefont
  {H.}~\bibnamefont {Korandla}}, \bibinfo {author} {\bibfnamefont
  {G.}~\bibnamefont {Leum}}, \bibinfo {author} {\bibfnamefont {P.}~\bibnamefont
  {Mohapatra}}, \bibinfo {author} {\bibfnamefont {T.}~\bibnamefont {Nitta}},
  \bibinfo {author} {\bibfnamefont {L.~J.}\ \bibnamefont {Rosenberg}}, \bibinfo
  {author} {\bibfnamefont {G.}~\bibnamefont {Rybka}}, \bibinfo {author}
  {\bibfnamefont {J.}~\bibnamefont {Yang}}, \bibinfo {author} {\bibfnamefont
  {J.}~\bibnamefont {Clarke}}, \bibinfo {author} {\bibfnamefont
  {I.}~\bibnamefont {Siddiqi}}, \bibinfo {author} {\bibfnamefont
  {A.}~\bibnamefont {Agrawal}}, \bibinfo {author} {\bibfnamefont {A.~V.}\
  \bibnamefont {Dixit}}, \bibinfo {author} {\bibfnamefont {M.~H.}\ \bibnamefont
  {Awida}}, \bibinfo {author} {\bibfnamefont {A.~S.}\ \bibnamefont {Chou}},
  \bibinfo {author} {\bibfnamefont {M.}~\bibnamefont {Hollister}}, \bibinfo
  {author} {\bibfnamefont {S.}~\bibnamefont {Knirck}}, \bibinfo {author}
  {\bibfnamefont {A.}~\bibnamefont {Sonnenschein}}, \bibinfo {author}
  {\bibfnamefont {W.}~\bibnamefont {Wester}}, \bibinfo {author} {\bibfnamefont
  {J.~R.}\ \bibnamefont {Gleason}}, \bibinfo {author} {\bibfnamefont {A.~T.}\
  \bibnamefont {Hipp}}, \bibinfo {author} {\bibfnamefont {S.}~\bibnamefont
  {Jois}}, \bibinfo {author} {\bibfnamefont {P.}~\bibnamefont {Sikivie}},
  \bibinfo {author} {\bibfnamefont {N.~S.}\ \bibnamefont {Sullivan}}, \bibinfo
  {author} {\bibfnamefont {D.~B.}\ \bibnamefont {Tanner}}, \bibinfo {author}
  {\bibfnamefont {E.}~\bibnamefont {Lentz}}, \bibinfo {author} {\bibfnamefont
  {R.}~\bibnamefont {Khatiwada}}, \bibinfo {author} {\bibfnamefont
  {G.}~\bibnamefont {Carosi}}, \bibinfo {author} {\bibfnamefont
  {N.}~\bibnamefont {Robertson}}, \bibinfo {author} {\bibfnamefont
  {N.}~\bibnamefont {Woollett}}, \bibinfo {author} {\bibfnamefont {L.~D.}\
  \bibnamefont {Duffy}}, \bibinfo {author} {\bibfnamefont {C.}~\bibnamefont
  {Boutan}}, \bibinfo {author} {\bibfnamefont {M.}~\bibnamefont {Jones}},
  \bibinfo {author} {\bibfnamefont {B.~H.}\ \bibnamefont {LaRoque}}, \bibinfo
  {author} {\bibfnamefont {N.~S.}\ \bibnamefont {Oblath}}, \bibinfo {author}
  {\bibfnamefont {M.~S.}\ \bibnamefont {Taubman}}, \bibinfo {author}
  {\bibfnamefont {E.~J.}\ \bibnamefont {Daw}}, \bibinfo {author} {\bibfnamefont
  {M.~G.}\ \bibnamefont {Perry}}, \bibinfo {author} {\bibfnamefont {J.~H.}\
  \bibnamefont {Buckley}}, \bibinfo {author} {\bibfnamefont {C.}~\bibnamefont
  {Gaikwad}}, \bibinfo {author} {\bibfnamefont {J.}~\bibnamefont {Hoffman}},
  \bibinfo {author} {\bibfnamefont {K.~W.}\ \bibnamefont {Murch}}, \bibinfo
  {author} {\bibfnamefont {M.}~\bibnamefont {Goryachev}}, \bibinfo {author}
  {\bibfnamefont {B.~T.}\ \bibnamefont {McAllister}}, \bibinfo {author}
  {\bibfnamefont {A.}~\bibnamefont {Quiskamp}}, \bibinfo {author}
  {\bibfnamefont {C.}~\bibnamefont {Thomson}}, \ and\ \bibinfo {author}
  {\bibfnamefont {M.~E.}\ \bibnamefont {Tobar}} (\bibinfo {collaboration} {ADMX
  Collaboration}),\ }\href {\doibase 10.1103/PhysRevLett.127.261803} {\bibfield
   {journal} {\bibinfo  {journal} {Phys. Rev. Lett.}\ }\textbf {\bibinfo
  {volume} {127}},\ \bibinfo {pages} {261803} (\bibinfo {year}
  {2021}{\natexlab{b}})}\BibitemShut {NoStop}%
\bibitem [{\citenamefont {Ouellet}\ \emph {et~al.}(2019)\citenamefont {Ouellet}
  \emph {et~al.}}]{Ouellet:2018beu}%
  \BibitemOpen
  \bibfield  {author} {\bibinfo {author} {\bibfnamefont {J.~L.}\ \bibnamefont
  {Ouellet}} \emph {et~al.} (\bibinfo {collaboration} {ABRACADABRA}),\ }\href
  {\doibase 10.1103/PhysRevLett.122.121802} {\bibfield  {journal} {\bibinfo
  {journal} {Phys. Rev. Lett.}\ }\textbf {\bibinfo {volume} {122}},\ \bibinfo
  {pages} {121802} (\bibinfo {year} {2019})},\ \Eprint
  {http://arxiv.org/abs/1810.12257} {arXiv:1810.12257 [hep-ex]} \BibitemShut
  {NoStop}%
\bibitem [{\citenamefont {Salemi}\ \emph {et~al.}(2021)\citenamefont {Salemi}
  \emph {et~al.}}]{Salemi:2021gck}%
  \BibitemOpen
  \bibfield  {author} {\bibinfo {author} {\bibfnamefont {C.~P.}\ \bibnamefont
  {Salemi}} \emph {et~al.},\ }\href {\doibase 10.1103/PhysRevLett.127.081801}
  {\bibfield  {journal} {\bibinfo  {journal} {Phys. Rev. Lett.}\ }\textbf
  {\bibinfo {volume} {127}},\ \bibinfo {pages} {081801} (\bibinfo {year}
  {2021})},\ \Eprint {http://arxiv.org/abs/2102.06722} {arXiv:2102.06722
  [hep-ex]} \BibitemShut {NoStop}%
\bibitem [{\citenamefont {Graham}\ and\ \citenamefont
  {Rajendran}(2013)}]{Graham:2013gfa}%
  \BibitemOpen
  \bibfield  {author} {\bibinfo {author} {\bibfnamefont {P.~W.}\ \bibnamefont
  {Graham}}\ and\ \bibinfo {author} {\bibfnamefont {S.}~\bibnamefont
  {Rajendran}},\ }\href {\doibase 10.1103/PhysRevD.88.035023} {\bibfield
  {journal} {\bibinfo  {journal} {Phys. Rev. D}\ }\textbf {\bibinfo {volume}
  {88}},\ \bibinfo {pages} {035023} (\bibinfo {year} {2013})},\ \Eprint
  {http://arxiv.org/abs/1306.6088} {arXiv:1306.6088 [hep-ph]} \BibitemShut
  {NoStop}%
\bibitem [{\citenamefont {Budker}\ \emph {et~al.}(2014)\citenamefont {Budker},
  \citenamefont {Graham}, \citenamefont {Ledbetter}, \citenamefont
  {Rajendran},\ and\ \citenamefont {Sushkov}}]{Casper2013}%
  \BibitemOpen
  \bibfield  {author} {\bibinfo {author} {\bibfnamefont {D.}~\bibnamefont
  {Budker}}, \bibinfo {author} {\bibfnamefont {P.~W.}\ \bibnamefont {Graham}},
  \bibinfo {author} {\bibfnamefont {M.}~\bibnamefont {Ledbetter}}, \bibinfo
  {author} {\bibfnamefont {S.}~\bibnamefont {Rajendran}}, \ and\ \bibinfo
  {author} {\bibfnamefont {A.~O.}\ \bibnamefont {Sushkov}},\ }\href {\doibase
  10.1103/physrevx.4.021030} {\bibfield  {journal} {\bibinfo  {journal}
  {Physical Review X}\ }\textbf {\bibinfo {volume} {4}} (\bibinfo {year}
  {2014}),\ 10.1103/physrevx.4.021030}\BibitemShut {NoStop}%
\bibitem [{\citenamefont {Jackson~Kimball}\ \emph {et~al.}(2020)\citenamefont
  {Jackson~Kimball} \emph {et~al.}}]{JacksonKimball:2017elr}%
  \BibitemOpen
  \bibfield  {author} {\bibinfo {author} {\bibfnamefont {D.~F.}\ \bibnamefont
  {Jackson~Kimball}} \emph {et~al.},\ }\href {\doibase
  10.1007/978-3-030-43761-9_13} {\bibfield  {journal} {\bibinfo  {journal}
  {Springer Proc. Phys.}\ }\textbf {\bibinfo {volume} {245}},\ \bibinfo {pages}
  {105} (\bibinfo {year} {2020})},\ \Eprint {http://arxiv.org/abs/1711.08999}
  {arXiv:1711.08999 [physics.ins-det]} \BibitemShut {NoStop}%
\bibitem [{\citenamefont {Garcon}\ \emph {et~al.}(2019)\citenamefont {Garcon},
  \citenamefont {Blanchard}, \citenamefont {Centers}, \citenamefont {Figueroa},
  \citenamefont {Graham}, \citenamefont {Kimball}, \citenamefont {Rajendran},
  \citenamefont {Sushkov}, \citenamefont {Stadnik}, \citenamefont
  {Wickenbrock}, \citenamefont {Wu},\ and\ \citenamefont
  {Budker}}]{Garcon2019}%
  \BibitemOpen
  \bibfield  {author} {\bibinfo {author} {\bibfnamefont {A.}~\bibnamefont
  {Garcon}}, \bibinfo {author} {\bibfnamefont {J.~W.}\ \bibnamefont
  {Blanchard}}, \bibinfo {author} {\bibfnamefont {G.~P.}\ \bibnamefont
  {Centers}}, \bibinfo {author} {\bibfnamefont {N.~L.}\ \bibnamefont
  {Figueroa}}, \bibinfo {author} {\bibfnamefont {P.~W.}\ \bibnamefont
  {Graham}}, \bibinfo {author} {\bibfnamefont {D.~F.~J.}\ \bibnamefont
  {Kimball}}, \bibinfo {author} {\bibfnamefont {S.}~\bibnamefont {Rajendran}},
  \bibinfo {author} {\bibfnamefont {A.~O.}\ \bibnamefont {Sushkov}}, \bibinfo
  {author} {\bibfnamefont {Y.~V.}\ \bibnamefont {Stadnik}}, \bibinfo {author}
  {\bibfnamefont {A.}~\bibnamefont {Wickenbrock}}, \bibinfo {author}
  {\bibfnamefont {T.}~\bibnamefont {Wu}}, \ and\ \bibinfo {author}
  {\bibfnamefont {D.}~\bibnamefont {Budker}},\ }\href {\doibase
  10.1126/sciadv.aax4539} {\bibfield  {journal} {\bibinfo  {journal} {Science
  Advances}\ }\textbf {\bibinfo {volume} {5}} (\bibinfo {year} {2019}),\
  10.1126/sciadv.aax4539}\BibitemShut {NoStop}%
\bibitem [{\citenamefont {Wu}\ \emph {et~al.}(2019)\citenamefont {Wu},
  \citenamefont {Blanchard}, \citenamefont {Centers}, \citenamefont {Figueroa},
  \citenamefont {Garcon}, \citenamefont {Graham}, \citenamefont {Kimball},
  \citenamefont {Rajendran}, \citenamefont {Stadnik}, \citenamefont {Sushkov},
  \citenamefont {Wickenbrock},\ and\ \citenamefont {Budker}}]{Wu2019}%
  \BibitemOpen
  \bibfield  {author} {\bibinfo {author} {\bibfnamefont {T.}~\bibnamefont
  {Wu}}, \bibinfo {author} {\bibfnamefont {J.~W.}\ \bibnamefont {Blanchard}},
  \bibinfo {author} {\bibfnamefont {G.~P.}\ \bibnamefont {Centers}}, \bibinfo
  {author} {\bibfnamefont {N.~L.}\ \bibnamefont {Figueroa}}, \bibinfo {author}
  {\bibfnamefont {A.}~\bibnamefont {Garcon}}, \bibinfo {author} {\bibfnamefont
  {P.~W.}\ \bibnamefont {Graham}}, \bibinfo {author} {\bibfnamefont {D.~F.~J.}\
  \bibnamefont {Kimball}}, \bibinfo {author} {\bibfnamefont {S.}~\bibnamefont
  {Rajendran}}, \bibinfo {author} {\bibfnamefont {Y.~V.}\ \bibnamefont
  {Stadnik}}, \bibinfo {author} {\bibfnamefont {A.~O.}\ \bibnamefont
  {Sushkov}}, \bibinfo {author} {\bibfnamefont {A.}~\bibnamefont
  {Wickenbrock}}, \ and\ \bibinfo {author} {\bibfnamefont {D.}~\bibnamefont
  {Budker}},\ }\href {\doibase 10.1103/PhysRevLett.122.191302} {\bibfield
  {journal} {\bibinfo  {journal} {Phys. Rev. Lett.}\ }\textbf {\bibinfo
  {volume} {122}},\ \bibinfo {pages} {191302} (\bibinfo {year}
  {2019})}\BibitemShut {NoStop}%
\bibitem [{\citenamefont {Gao}\ \emph {et~al.}(2022)\citenamefont {Gao},
  \citenamefont {Halperin}, \citenamefont {Kahn}, \citenamefont {Nguyen},
  \citenamefont {Sch\"utte-Engel},\ and\ \citenamefont {Scott}}]{Gao:2022nuq}%
  \BibitemOpen
  \bibfield  {author} {\bibinfo {author} {\bibfnamefont {C.}~\bibnamefont
  {Gao}}, \bibinfo {author} {\bibfnamefont {W.}~\bibnamefont {Halperin}},
  \bibinfo {author} {\bibfnamefont {Y.}~\bibnamefont {Kahn}}, \bibinfo {author}
  {\bibfnamefont {M.}~\bibnamefont {Nguyen}}, \bibinfo {author} {\bibfnamefont
  {J.}~\bibnamefont {Sch\"utte-Engel}}, \ and\ \bibinfo {author} {\bibfnamefont
  {J.~W.}\ \bibnamefont {Scott}},\ }\href {\doibase
  10.1103/PhysRevLett.129.211801} {\bibfield  {journal} {\bibinfo  {journal}
  {Phys. Rev. Lett.}\ }\textbf {\bibinfo {volume} {129}},\ \bibinfo {pages}
  {211801} (\bibinfo {year} {2022})},\ \Eprint
  {http://arxiv.org/abs/2208.14454} {arXiv:2208.14454 [hep-ph]} \BibitemShut
  {NoStop}%
\bibitem [{\citenamefont {Foster}\ \emph {et~al.}(2023)\citenamefont {Foster},
  \citenamefont {Gao}, \citenamefont {Halperin}, \citenamefont {Kahn},
  \citenamefont {Mande}, \citenamefont {Nguyen}, \citenamefont
  {Sch\"utte-Engel},\ and\ \citenamefont {Scott}}]{Foster:2023bxl}%
  \BibitemOpen
  \bibfield  {author} {\bibinfo {author} {\bibfnamefont {J.~W.}\ \bibnamefont
  {Foster}}, \bibinfo {author} {\bibfnamefont {C.}~\bibnamefont {Gao}},
  \bibinfo {author} {\bibfnamefont {W.}~\bibnamefont {Halperin}}, \bibinfo
  {author} {\bibfnamefont {Y.}~\bibnamefont {Kahn}}, \bibinfo {author}
  {\bibfnamefont {A.}~\bibnamefont {Mande}}, \bibinfo {author} {\bibfnamefont
  {M.}~\bibnamefont {Nguyen}}, \bibinfo {author} {\bibfnamefont
  {J.}~\bibnamefont {Sch\"utte-Engel}}, \ and\ \bibinfo {author} {\bibfnamefont
  {J.~W.}\ \bibnamefont {Scott}},\ }\href@noop {} {\  (\bibinfo {year}
  {2023})},\ \Eprint {http://arxiv.org/abs/2310.07791} {arXiv:2310.07791
  [hep-ph]} \BibitemShut {NoStop}%
\bibitem [{\citenamefont {Garbrecht}\ and\ \citenamefont
  {McDonald}(2018)}]{Garbrecht:2018akc}%
  \BibitemOpen
  \bibfield  {author} {\bibinfo {author} {\bibfnamefont {B.}~\bibnamefont
  {Garbrecht}}\ and\ \bibinfo {author} {\bibfnamefont {J.~I.}\ \bibnamefont
  {McDonald}},\ }\href {\doibase 10.1088/1475-7516/2018/07/044} {\bibfield
  {journal} {\bibinfo  {journal} {JCAP}\ }\textbf {\bibinfo {volume} {07}},\
  \bibinfo {pages} {044} (\bibinfo {year} {2018})},\ \Eprint
  {http://arxiv.org/abs/1804.04224} {arXiv:1804.04224 [astro-ph.CO]}
  \BibitemShut {NoStop}%
\bibitem [{\citenamefont {Prabhu}(2021)}]{Prabhu2021}%
  \BibitemOpen
  \bibfield  {author} {\bibinfo {author} {\bibfnamefont {A.}~\bibnamefont
  {Prabhu}},\ }\href {\doibase 10.1103/physrevd.104.055038} {\bibfield
  {journal} {\bibinfo  {journal} {Physical Review D}\ }\textbf {\bibinfo
  {volume} {104}} (\bibinfo {year} {2021}),\
  10.1103/physrevd.104.055038}\BibitemShut {NoStop}%
\bibitem [{\citenamefont {Noordhuis}\ \emph {et~al.}(2023)\citenamefont
  {Noordhuis}, \citenamefont {Prabhu}, \citenamefont {Witte}, \citenamefont
  {Chen}, \citenamefont {Cruz},\ and\ \citenamefont
  {Weniger}}]{noordhuis2023novel}%
  \BibitemOpen
  \bibfield  {author} {\bibinfo {author} {\bibfnamefont {D.}~\bibnamefont
  {Noordhuis}}, \bibinfo {author} {\bibfnamefont {A.}~\bibnamefont {Prabhu}},
  \bibinfo {author} {\bibfnamefont {S.~J.}\ \bibnamefont {Witte}}, \bibinfo
  {author} {\bibfnamefont {A.~Y.}\ \bibnamefont {Chen}}, \bibinfo {author}
  {\bibfnamefont {F.}~\bibnamefont {Cruz}}, \ and\ \bibinfo {author}
  {\bibfnamefont {C.}~\bibnamefont {Weniger}},\ }\href {\doibase
  10.1103/PhysRevLett.131.111004} {\bibfield  {journal} {\bibinfo  {journal}
  {Phys. Rev. Lett.}\ }\textbf {\bibinfo {volume} {131}},\ \bibinfo {pages}
  {111004} (\bibinfo {year} {2023})},\ \Eprint
  {http://arxiv.org/abs/2209.09917} {arXiv:2209.09917 [hep-ph]} \BibitemShut
  {NoStop}%
\bibitem [{\citenamefont {Berlin}\ \emph {et~al.}(2020)\citenamefont {Berlin},
  \citenamefont {D'Agnolo}, \citenamefont {Ellis}, \citenamefont {Nantista},
  \citenamefont {Neilson}, \citenamefont {Schuster}, \citenamefont {Tantawi},
  \citenamefont {Toro},\ and\ \citenamefont {Zhou}}]{Berlin2020}%
  \BibitemOpen
  \bibfield  {author} {\bibinfo {author} {\bibfnamefont {A.}~\bibnamefont
  {Berlin}}, \bibinfo {author} {\bibfnamefont {R.~T.}\ \bibnamefont
  {D'Agnolo}}, \bibinfo {author} {\bibfnamefont {S.~A.~R.}\ \bibnamefont
  {Ellis}}, \bibinfo {author} {\bibfnamefont {C.}~\bibnamefont {Nantista}},
  \bibinfo {author} {\bibfnamefont {J.}~\bibnamefont {Neilson}}, \bibinfo
  {author} {\bibfnamefont {P.}~\bibnamefont {Schuster}}, \bibinfo {author}
  {\bibfnamefont {S.}~\bibnamefont {Tantawi}}, \bibinfo {author} {\bibfnamefont
  {N.}~\bibnamefont {Toro}}, \ and\ \bibinfo {author} {\bibfnamefont
  {K.}~\bibnamefont {Zhou}},\ }\href {\doibase 10.1007/jhep07(2020)088}
  {\bibfield  {journal} {\bibinfo  {journal} {Journal of High Energy Physics}\
  }\textbf {\bibinfo {volume} {2020}} (\bibinfo {year} {2020}),\
  10.1007/jhep07(2020)088}\BibitemShut {NoStop}%
\bibitem [{\citenamefont {Berlin}\ \emph {et~al.}(2021)\citenamefont {Berlin},
  \citenamefont {D'Agnolo}, \citenamefont {Ellis},\ and\ \citenamefont
  {Zhou}}]{Berlin2021}%
  \BibitemOpen
  \bibfield  {author} {\bibinfo {author} {\bibfnamefont {A.}~\bibnamefont
  {Berlin}}, \bibinfo {author} {\bibfnamefont {R.~T.}\ \bibnamefont
  {D'Agnolo}}, \bibinfo {author} {\bibfnamefont {S.~A.~R.}\ \bibnamefont
  {Ellis}}, \ and\ \bibinfo {author} {\bibfnamefont {K.}~\bibnamefont {Zhou}},\
  }\href {\doibase 10.1103/PhysRevD.104.L111701} {\bibfield  {journal}
  {\bibinfo  {journal} {Phys. Rev. D}\ }\textbf {\bibinfo {volume} {104}},\
  \bibinfo {pages} {L111701} (\bibinfo {year} {2021})}\BibitemShut {NoStop}%
\bibitem [{\citenamefont {Brouwer}\ \emph
  {et~al.}(2022{\natexlab{a}})\citenamefont {Brouwer} \emph
  {et~al.}}]{DMRadio:2022pkf}%
  \BibitemOpen
  \bibfield  {author} {\bibinfo {author} {\bibfnamefont {L.}~\bibnamefont
  {Brouwer}} \emph {et~al.} (\bibinfo {collaboration} {DMRadio}),\ }\href
  {\doibase 10.1103/PhysRevD.106.103008} {\bibfield  {journal} {\bibinfo
  {journal} {Phys. Rev. D}\ }\textbf {\bibinfo {volume} {106}},\ \bibinfo
  {pages} {103008} (\bibinfo {year} {2022}{\natexlab{a}})},\ \Eprint
  {http://arxiv.org/abs/2204.13781} {arXiv:2204.13781 [hep-ex]} \BibitemShut
  {NoStop}%
\bibitem [{\citenamefont {Brouwer}\ \emph
  {et~al.}(2022{\natexlab{b}})\citenamefont {Brouwer} \emph
  {et~al.}}]{DMRadio:2022jfv}%
  \BibitemOpen
  \bibfield  {author} {\bibinfo {author} {\bibfnamefont {L.}~\bibnamefont
  {Brouwer}} \emph {et~al.} (\bibinfo {collaboration} {DMRadio}),\ }\href
  {\doibase 10.1103/PhysRevD.106.112003} {\bibfield  {journal} {\bibinfo
  {journal} {Phys. Rev. D}\ }\textbf {\bibinfo {volume} {106}},\ \bibinfo
  {pages} {112003} (\bibinfo {year} {2022}{\natexlab{b}})},\ \Eprint
  {http://arxiv.org/abs/2203.11246} {arXiv:2203.11246 [hep-ex]} \BibitemShut
  {NoStop}%
\bibitem [{\citenamefont {{Deutsch}}(1955)}]{Deutsch1955}%
  \BibitemOpen
  \bibfield  {author} {\bibinfo {author} {\bibfnamefont {A.~J.}\ \bibnamefont
  {{Deutsch}}},\ }\href@noop {} {\bibfield  {journal} {\bibinfo  {journal}
  {Annales d'Astrophysique}\ }\textbf {\bibinfo {volume} {18}},\ \bibinfo
  {pages} {1} (\bibinfo {year} {1955})}\BibitemShut {NoStop}%
\bibitem [{\citenamefont {Hoyle}\ \emph {et~al.}(1964)\citenamefont {Hoyle},
  \citenamefont {Narlikar},\ and\ \citenamefont {Wheeler}}]{Hoyle1964}%
  \BibitemOpen
  \bibfield  {author} {\bibinfo {author} {\bibfnamefont {F.}~\bibnamefont
  {Hoyle}}, \bibinfo {author} {\bibfnamefont {J.~V.}\ \bibnamefont {Narlikar}},
  \ and\ \bibinfo {author} {\bibfnamefont {J.~A.}\ \bibnamefont {Wheeler}},\
  }\href {\doibase 10.1038/203914a0} {\bibfield  {journal} {\bibinfo  {journal}
  {Nature}\ }\textbf {\bibinfo {volume} {203}},\ \bibinfo {pages} {914}
  (\bibinfo {year} {1964})}\BibitemShut {NoStop}%
\bibitem [{\citenamefont {Pacini}(1967)}]{Pacini1967}%
  \BibitemOpen
  \bibfield  {author} {\bibinfo {author} {\bibfnamefont {F.}~\bibnamefont
  {Pacini}},\ }\href {\doibase 10.1038/216567a0} {\bibfield  {journal}
  {\bibinfo  {journal} {Nature}\ }\textbf {\bibinfo {volume} {216}},\ \bibinfo
  {pages} {567} (\bibinfo {year} {1967})}\BibitemShut {NoStop}%
\bibitem [{\citenamefont {Pacini}(1968)}]{Pacini1968}%
  \BibitemOpen
  \bibfield  {author} {\bibinfo {author} {\bibfnamefont {F.}~\bibnamefont
  {Pacini}},\ }\href {\doibase 10.1038/219145a0} {\bibfield  {journal}
  {\bibinfo  {journal} {Nature}\ }\textbf {\bibinfo {volume} {219}},\ \bibinfo
  {pages} {145} (\bibinfo {year} {1968})}\BibitemShut {NoStop}%
\bibitem [{\citenamefont {{Ostriker}}\ and\ \citenamefont
  {{Gunn}}(1969)}]{OstrikerGunn1969}%
  \BibitemOpen
  \bibfield  {author} {\bibinfo {author} {\bibfnamefont {J.~P.}\ \bibnamefont
  {{Ostriker}}}\ and\ \bibinfo {author} {\bibfnamefont {J.~E.}\ \bibnamefont
  {{Gunn}}},\ }\href {\doibase 10.1086/150160} {\bibfield  {journal} {\bibinfo
  {journal} {\apj}\ }\textbf {\bibinfo {volume} {157}},\ \bibinfo {pages}
  {1395} (\bibinfo {year} {1969})}\BibitemShut {NoStop}%
\bibitem [{\citenamefont {Melrose}\ and\ \citenamefont
  {Yuen}(2012)}]{Melrose2012}%
  \BibitemOpen
  \bibfield  {author} {\bibinfo {author} {\bibfnamefont {D.~B.}\ \bibnamefont
  {Melrose}}\ and\ \bibinfo {author} {\bibfnamefont {R.}~\bibnamefont {Yuen}},\
  }\href {\doibase 10.1088/0004-637x/745/2/169} {\bibfield  {journal} {\bibinfo
   {journal} {The Astrophysical Journal}\ }\textbf {\bibinfo {volume} {745}},\
  \bibinfo {pages} {169} (\bibinfo {year} {2012})}\BibitemShut {NoStop}%
\bibitem [{\citenamefont {{P{\'e}tri}}\ \emph {et~al.}(2002)\citenamefont
  {{P{\'e}tri}}, \citenamefont {{Heyvaerts}},\ and\ \citenamefont
  {{Bonazzola}}}]{2002A&A...384..414P}%
  \BibitemOpen
  \bibfield  {author} {\bibinfo {author} {\bibfnamefont {J.}~\bibnamefont
  {{P{\'e}tri}}}, \bibinfo {author} {\bibfnamefont {J.}~\bibnamefont
  {{Heyvaerts}}}, \ and\ \bibinfo {author} {\bibfnamefont {S.}~\bibnamefont
  {{Bonazzola}}},\ }\href {\doibase 10.1051/0004-6361:20020044} {\bibfield
  {journal} {\bibinfo  {journal} {Astronomy and Astrophysics}\ }\textbf
  {\bibinfo {volume} {384}},\ \bibinfo {pages} {414} (\bibinfo {year}
  {2002})}\BibitemShut {NoStop}%
\bibitem [{\citenamefont {{Philippov}}\ and\ \citenamefont
  {{Kramer}}(2022)}]{Philippov2022Review}%
  \BibitemOpen
  \bibfield  {author} {\bibinfo {author} {\bibfnamefont {A.}~\bibnamefont
  {{Philippov}}}\ and\ \bibinfo {author} {\bibfnamefont {M.}~\bibnamefont
  {{Kramer}}},\ }\href {\doibase 10.1146/annurev-astro-052920-112338}
  {\bibfield  {journal} {\bibinfo  {journal} {Annual Review of Astron and
  Astrophys}\ }\textbf {\bibinfo {volume} {60}},\ \bibinfo {pages} {495}
  (\bibinfo {year} {2022})}\BibitemShut {NoStop}%
\bibitem [{\citenamefont {Cruz}\ \emph {et~al.}(2023)\citenamefont {Cruz},
  \citenamefont {Grismayer}, \citenamefont {Chen}, \citenamefont {Spitkovsky},
  \citenamefont {Fonseca},\ and\ \citenamefont {Silva}}]{Cruz:2023vne}%
  \BibitemOpen
  \bibfield  {author} {\bibinfo {author} {\bibfnamefont {F.}~\bibnamefont
  {Cruz}}, \bibinfo {author} {\bibfnamefont {T.}~\bibnamefont {Grismayer}},
  \bibinfo {author} {\bibfnamefont {A.~Y.}\ \bibnamefont {Chen}}, \bibinfo
  {author} {\bibfnamefont {A.}~\bibnamefont {Spitkovsky}}, \bibinfo {author}
  {\bibfnamefont {R.~A.}\ \bibnamefont {Fonseca}}, \ and\ \bibinfo {author}
  {\bibfnamefont {L.~O.}\ \bibnamefont {Silva}},\ }\href@noop {} {\  (\bibinfo
  {year} {2023})},\ \Eprint {http://arxiv.org/abs/2309.04834} {arXiv:2309.04834
  [astro-ph.HE]} \BibitemShut {NoStop}%
\bibitem [{\citenamefont {Tolman}\ \emph {et~al.}(2022)\citenamefont {Tolman},
  \citenamefont {Philippov},\ and\ \citenamefont {Timokhin}}]{Tolman:2022unu}%
  \BibitemOpen
  \bibfield  {author} {\bibinfo {author} {\bibfnamefont {E.~A.}\ \bibnamefont
  {Tolman}}, \bibinfo {author} {\bibfnamefont {A.~A.}\ \bibnamefont
  {Philippov}}, \ and\ \bibinfo {author} {\bibfnamefont {A.~N.}\ \bibnamefont
  {Timokhin}},\ }\href@noop {} {\  (\bibinfo {year} {2022})},\ \Eprint
  {http://arxiv.org/abs/2202.01303} {arXiv:2202.01303 [astro-ph.HE]}
  \BibitemShut {NoStop}%
\bibitem [{\citenamefont {{Cruz}}\ \emph {et~al.}(2021)\citenamefont {{Cruz}},
  \citenamefont {{Grismayer}}, \citenamefont {{Chen}}, \citenamefont
  {{Spitkovsky}},\ and\ \citenamefont {{Silva}}}]{Cruz2021}%
  \BibitemOpen
  \bibfield  {author} {\bibinfo {author} {\bibfnamefont {F.}~\bibnamefont
  {{Cruz}}}, \bibinfo {author} {\bibfnamefont {T.}~\bibnamefont {{Grismayer}}},
  \bibinfo {author} {\bibfnamefont {A.~Y.}\ \bibnamefont {{Chen}}}, \bibinfo
  {author} {\bibfnamefont {A.}~\bibnamefont {{Spitkovsky}}}, \ and\ \bibinfo
  {author} {\bibfnamefont {L.~O.}\ \bibnamefont {{Silva}}},\ }\href {\doibase
  10.3847/2041-8213/ac2157} {\bibfield  {journal} {\bibinfo  {journal}
  {Astrophysical Journal Letters}\ }\textbf {\bibinfo {volume} {919}},\
  \bibinfo {eid} {L4} (\bibinfo {year} {2021})},\ \Eprint
  {http://arxiv.org/abs/2108.11702} {arXiv:2108.11702 [astro-ph.HE]}
  \BibitemShut {NoStop}%
\bibitem [{\citenamefont {Cruz}\ \emph {et~al.}(2022)\citenamefont {Cruz},
  \citenamefont {Grismayer}, \citenamefont {Iteanu}, \citenamefont {Tortone},\
  and\ \citenamefont {Silva}}]{Cruz:2022zyp}%
  \BibitemOpen
  \bibfield  {author} {\bibinfo {author} {\bibfnamefont {F.}~\bibnamefont
  {Cruz}}, \bibinfo {author} {\bibfnamefont {T.}~\bibnamefont {Grismayer}},
  \bibinfo {author} {\bibfnamefont {S.}~\bibnamefont {Iteanu}}, \bibinfo
  {author} {\bibfnamefont {P.}~\bibnamefont {Tortone}}, \ and\ \bibinfo
  {author} {\bibfnamefont {L.~O.}\ \bibnamefont {Silva}},\ }\href {\doibase
  10.1063/5.0085847} {\bibfield  {journal} {\bibinfo  {journal} {Phys.
  Plasmas}\ }\textbf {\bibinfo {volume} {29}},\ \bibinfo {pages} {052902}
  (\bibinfo {year} {2022})},\ \Eprint {http://arxiv.org/abs/2204.03766}
  {arXiv:2204.03766 [astro-ph.HE]} \BibitemShut {NoStop}%
\bibitem [{\citenamefont {{Goldreich}}\ and\ \citenamefont
  {{Julian}}(1969)}]{GoldreichJulian1969}%
  \BibitemOpen
  \bibfield  {author} {\bibinfo {author} {\bibfnamefont {P.}~\bibnamefont
  {{Goldreich}}}\ and\ \bibinfo {author} {\bibfnamefont {W.~H.}\ \bibnamefont
  {{Julian}}},\ }\href {\doibase 10.1086/150119} {\bibfield  {journal}
  {\bibinfo  {journal} {\apj}\ }\textbf {\bibinfo {volume} {157}},\ \bibinfo
  {pages} {869} (\bibinfo {year} {1969})}\BibitemShut {NoStop}%
\bibitem [{\citenamefont {{Ruderman}}\ and\ \citenamefont
  {{Sutherland}}(1975)}]{RudermanSutherland1975}%
  \BibitemOpen
  \bibfield  {author} {\bibinfo {author} {\bibfnamefont {M.~A.}\ \bibnamefont
  {{Ruderman}}}\ and\ \bibinfo {author} {\bibfnamefont {P.~G.}\ \bibnamefont
  {{Sutherland}}},\ }\href {\doibase 10.1086/153393} {\bibfield  {journal}
  {\bibinfo  {journal} {\apj}\ }\textbf {\bibinfo {volume} {196}},\ \bibinfo
  {pages} {51} (\bibinfo {year} {1975})}\BibitemShut {NoStop}%
\bibitem [{\citenamefont {Timokhin}\ and\ \citenamefont
  {Harding}(2015)}]{Timokhin:2015dua}%
  \BibitemOpen
  \bibfield  {author} {\bibinfo {author} {\bibfnamefont {A.~N.}\ \bibnamefont
  {Timokhin}}\ and\ \bibinfo {author} {\bibfnamefont {A.~K.}\ \bibnamefont
  {Harding}},\ }\href {\doibase 10.1088/0004-637X/810/2/144} {\bibfield
  {journal} {\bibinfo  {journal} {Astrophys. J.}\ }\textbf {\bibinfo {volume}
  {810}},\ \bibinfo {pages} {144} (\bibinfo {year} {2015})},\ \Eprint
  {http://arxiv.org/abs/1504.02194} {arXiv:1504.02194 [astro-ph.HE]}
  \BibitemShut {NoStop}%
\bibitem [{\citenamefont {Timokhin}\ and\ \citenamefont
  {Harding}(2019)}]{Timokhin:2018vdn}%
  \BibitemOpen
  \bibfield  {author} {\bibinfo {author} {\bibfnamefont {A.~N.}\ \bibnamefont
  {Timokhin}}\ and\ \bibinfo {author} {\bibfnamefont {A.~K.}\ \bibnamefont
  {Harding}},\ }\href {\doibase 10.3847/1538-4357/aaf050} {\bibfield  {journal}
  {\bibinfo  {journal} {Astrophys. J.}\ }\textbf {\bibinfo {volume} {871}},\
  \bibinfo {pages} {12} (\bibinfo {year} {2019})},\ \Eprint
  {http://arxiv.org/abs/1803.08924} {arXiv:1803.08924 [astro-ph.HE]}
  \BibitemShut {NoStop}%
\bibitem [{\citenamefont {Caputo}\ \emph {et~al.}(2023)\citenamefont {Caputo},
  \citenamefont {Witte}, \citenamefont {Philippov},\ and\ \citenamefont
  {Jacobson}}]{Caputo:2023cpv}%
  \BibitemOpen
  \bibfield  {author} {\bibinfo {author} {\bibfnamefont {A.}~\bibnamefont
  {Caputo}}, \bibinfo {author} {\bibfnamefont {S.~J.}\ \bibnamefont {Witte}},
  \bibinfo {author} {\bibfnamefont {A.~A.}\ \bibnamefont {Philippov}}, \ and\
  \bibinfo {author} {\bibfnamefont {T.}~\bibnamefont {Jacobson}},\ }\href@noop
  {} {\  (\bibinfo {year} {2023})},\ \Eprint {http://arxiv.org/abs/2311.14795}
  {arXiv:2311.14795 [hep-ph]} \BibitemShut {NoStop}%
\bibitem [{\citenamefont {{Manchester}}\ \emph {et~al.}(2005)\citenamefont
  {{Manchester}}, \citenamefont {{Hobbs}}, \citenamefont {{Teoh}},\ and\
  \citenamefont {{Hobbs}}}]{ATNF}%
  \BibitemOpen
  \bibfield  {author} {\bibinfo {author} {\bibfnamefont {R.~N.}\ \bibnamefont
  {{Manchester}}}, \bibinfo {author} {\bibfnamefont {G.~B.}\ \bibnamefont
  {{Hobbs}}}, \bibinfo {author} {\bibfnamefont {A.}~\bibnamefont {{Teoh}}}, \
  and\ \bibinfo {author} {\bibfnamefont {M.}~\bibnamefont {{Hobbs}}},\ }\href
  {\doibase 10.1086/428488} {\bibfield  {journal} {\bibinfo  {journal}
  {Astronomical Journal}\ }\textbf {\bibinfo {volume} {129}},\ \bibinfo {pages}
  {1993} (\bibinfo {year} {2005})},\ \Eprint
  {http://arxiv.org/abs/astro-ph/0412641} {arXiv:astro-ph/0412641 [astro-ph]}
  \BibitemShut {NoStop}%
\bibitem [{\citenamefont {Lyne}\ \emph {et~al.}(1993)\citenamefont {Lyne},
  \citenamefont {Pritchard},\ and\ \citenamefont {Graham~Smith}}]{Lyne1993}%
  \BibitemOpen
  \bibfield  {author} {\bibinfo {author} {\bibfnamefont {A.~G.}\ \bibnamefont
  {Lyne}}, \bibinfo {author} {\bibfnamefont {R.~S.}\ \bibnamefont {Pritchard}},
  \ and\ \bibinfo {author} {\bibfnamefont {F.}~\bibnamefont {Graham~Smith}},\
  }\href {\doibase 10.1093/mnras/265.4.1003} {\bibfield  {journal} {\bibinfo
  {journal} {Monthly Notices of the Royal Astronomical Society}\ }\textbf
  {\bibinfo {volume} {265}},\ \bibinfo {pages} {1003} (\bibinfo {year}
  {1993})},\ \Eprint
  {http://arxiv.org/abs/https://academic.oup.com/mnras/article-pdf/265/4/1003/3173877/mnras265-1003.pdf}
  {https://academic.oup.com/mnras/article-pdf/265/4/1003/3173877/mnras265-1003.pdf}
  \BibitemShut {NoStop}%
\bibitem [{\citenamefont {{Bejger}}\ and\ \citenamefont
  {{Haensel}}(2002)}]{Bejger2002}%
  \BibitemOpen
  \bibfield  {author} {\bibinfo {author} {\bibfnamefont {M.}~\bibnamefont
  {{Bejger}}}\ and\ \bibinfo {author} {\bibfnamefont {P.}~\bibnamefont
  {{Haensel}}},\ }\href {\doibase 10.1051/0004-6361:20021241} {\bibfield
  {journal} {\bibinfo  {journal} {Astronomy and Astrophysics}\ }\textbf
  {\bibinfo {volume} {396}},\ \bibinfo {pages} {917} (\bibinfo {year}
  {2002})},\ \Eprint {http://arxiv.org/abs/astro-ph/0209151}
  {arXiv:astro-ph/0209151 [astro-ph]} \BibitemShut {NoStop}%
\bibitem [{\citenamefont {Philippov}\ \emph {et~al.}(2014)\citenamefont
  {Philippov}, \citenamefont {Tchekhovskoy},\ and\ \citenamefont
  {Li}}]{Philippov2014}%
  \BibitemOpen
  \bibfield  {author} {\bibinfo {author} {\bibfnamefont {A.}~\bibnamefont
  {Philippov}}, \bibinfo {author} {\bibfnamefont {A.}~\bibnamefont
  {Tchekhovskoy}}, \ and\ \bibinfo {author} {\bibfnamefont {J.~G.}\
  \bibnamefont {Li}},\ }\href {\doibase 10.1093/mnras/stu591} {\bibfield
  {journal} {\bibinfo  {journal} {Monthly Notices of the Royal Astronomical
  Society}\ }\textbf {\bibinfo {volume} {441}},\ \bibinfo {pages} {1879–1887}
  (\bibinfo {year} {2014})}\BibitemShut {NoStop}%
\bibitem [{\citenamefont {Kou}\ and\ \citenamefont {Tong}(2015)}]{Kou2013}%
  \BibitemOpen
  \bibfield  {author} {\bibinfo {author} {\bibfnamefont {F.~F.}\ \bibnamefont
  {Kou}}\ and\ \bibinfo {author} {\bibfnamefont {H.}~\bibnamefont {Tong}},\
  }\href {\doibase 10.1093/mnras/stv734} {\bibfield  {journal} {\bibinfo
  {journal} {Monthly Notices of the Royal Astronomical Society}\ }\textbf
  {\bibinfo {volume} {450}},\ \bibinfo {pages} {1990} (\bibinfo {year}
  {2015})},\ \Eprint
  {http://arxiv.org/abs/https://academic.oup.com/mnras/article-pdf/450/2/1990/3076208/stv734.pdf}
  {https://academic.oup.com/mnras/article-pdf/450/2/1990/3076208/stv734.pdf}
  \BibitemShut {NoStop}%
\bibitem [{\citenamefont {Foster}\ \emph {et~al.}(2018)\citenamefont {Foster},
  \citenamefont {Rodd},\ and\ \citenamefont {Safdi}}]{Foster:2017hbq}%
  \BibitemOpen
  \bibfield  {author} {\bibinfo {author} {\bibfnamefont {J.~W.}\ \bibnamefont
  {Foster}}, \bibinfo {author} {\bibfnamefont {N.~L.}\ \bibnamefont {Rodd}}, \
  and\ \bibinfo {author} {\bibfnamefont {B.~R.}\ \bibnamefont {Safdi}},\ }\href
  {\doibase 10.1103/PhysRevD.97.123006} {\bibfield  {journal} {\bibinfo
  {journal} {Phys. Rev. D}\ }\textbf {\bibinfo {volume} {97}},\ \bibinfo
  {pages} {123006} (\bibinfo {year} {2018})},\ \Eprint
  {http://arxiv.org/abs/1711.10489} {arXiv:1711.10489 [astro-ph.CO]}
  \BibitemShut {NoStop}%
\bibitem [{\citenamefont {Dror}\ \emph {et~al.}(2021)\citenamefont {Dror},
  \citenamefont {Murayama},\ and\ \citenamefont {Rodd}}]{Dror2021}%
  \BibitemOpen
  \bibfield  {author} {\bibinfo {author} {\bibfnamefont {J.~A.}\ \bibnamefont
  {Dror}}, \bibinfo {author} {\bibfnamefont {H.}~\bibnamefont {Murayama}}, \
  and\ \bibinfo {author} {\bibfnamefont {N.~L.}\ \bibnamefont {Rodd}},\ }\href
  {\doibase 10.1103/physrevd.103.115004} {\bibfield  {journal} {\bibinfo
  {journal} {Physical Review D}\ }\textbf {\bibinfo {volume} {103}} (\bibinfo
  {year} {2021}),\ 10.1103/physrevd.103.115004}\BibitemShut {NoStop}%
\bibitem [{\citenamefont {Romanenko}\ \emph {et~al.}(2014)\citenamefont
  {Romanenko}, \citenamefont {Grassellino}, \citenamefont {Crawford},
  \citenamefont {Sergatskov},\ and\ \citenamefont {Melnychuk}}]{Romanenko2014}%
  \BibitemOpen
  \bibfield  {author} {\bibinfo {author} {\bibfnamefont {A.}~\bibnamefont
  {Romanenko}}, \bibinfo {author} {\bibfnamefont {A.}~\bibnamefont
  {Grassellino}}, \bibinfo {author} {\bibfnamefont {A.~C.}\ \bibnamefont
  {Crawford}}, \bibinfo {author} {\bibfnamefont {D.~A.}\ \bibnamefont
  {Sergatskov}}, \ and\ \bibinfo {author} {\bibfnamefont {O.}~\bibnamefont
  {Melnychuk}},\ }\href {\doibase 10.1063/1.4903808} {\bibfield  {journal}
  {\bibinfo  {journal} {Applied Physics Letters}\ }\textbf {\bibinfo {volume}
  {105}} (\bibinfo {year} {2014}),\ 10.1063/1.4903808}\BibitemShut {NoStop}%
\bibitem [{\citenamefont {Posen}\ \emph {et~al.}(2019)\citenamefont {Posen},
  \citenamefont {Wu}, \citenamefont {Grassellino}, \citenamefont {Harms},
  \citenamefont {Melnychuk}, \citenamefont {Sergatskov}, \citenamefont
  {Solyak}, \citenamefont {Romanenko}, \citenamefont {Palczewski},
  \citenamefont {Gonnella},\ and\ \citenamefont {Peterson}}]{Posen_2019}%
  \BibitemOpen
  \bibfield  {author} {\bibinfo {author} {\bibfnamefont {S.}~\bibnamefont
  {Posen}}, \bibinfo {author} {\bibfnamefont {G.}~\bibnamefont {Wu}}, \bibinfo
  {author} {\bibfnamefont {A.}~\bibnamefont {Grassellino}}, \bibinfo {author}
  {\bibfnamefont {E.}~\bibnamefont {Harms}}, \bibinfo {author} {\bibfnamefont
  {O.}~\bibnamefont {Melnychuk}}, \bibinfo {author} {\bibfnamefont
  {D.}~\bibnamefont {Sergatskov}}, \bibinfo {author} {\bibfnamefont
  {N.}~\bibnamefont {Solyak}}, \bibinfo {author} {\bibfnamefont
  {A.}~\bibnamefont {Romanenko}}, \bibinfo {author} {\bibfnamefont
  {A.}~\bibnamefont {Palczewski}}, \bibinfo {author} {\bibfnamefont
  {D.}~\bibnamefont {Gonnella}}, \ and\ \bibinfo {author} {\bibfnamefont
  {T.}~\bibnamefont {Peterson}},\ }\href {\doibase
  10.1103/physrevaccelbeams.22.032001} {\bibfield  {journal} {\bibinfo
  {journal} {Physical Review Accelerators and Beams}\ }\textbf {\bibinfo
  {volume} {22}} (\bibinfo {year} {2019}),\
  10.1103/physrevaccelbeams.22.032001}\BibitemShut {NoStop}%
\bibitem [{\citenamefont {Dror}\ \emph {et~al.}(2023)\citenamefont {Dror},
  \citenamefont {Gori}, \citenamefont {Leedom},\ and\ \citenamefont
  {Rodd}}]{Dror2022}%
  \BibitemOpen
  \bibfield  {author} {\bibinfo {author} {\bibfnamefont {J.~A.}\ \bibnamefont
  {Dror}}, \bibinfo {author} {\bibfnamefont {S.}~\bibnamefont {Gori}}, \bibinfo
  {author} {\bibfnamefont {J.~M.}\ \bibnamefont {Leedom}}, \ and\ \bibinfo
  {author} {\bibfnamefont {N.~L.}\ \bibnamefont {Rodd}},\ }\href {\doibase
  10.1103/PhysRevLett.130.181801} {\bibfield  {journal} {\bibinfo  {journal}
  {Phys. Rev. Lett.}\ }\textbf {\bibinfo {volume} {130}},\ \bibinfo {pages}
  {181801} (\bibinfo {year} {2023})}\BibitemShut {NoStop}%
\bibitem [{\citenamefont {Arik}\ \emph {et~al.}(2015)\citenamefont {Arik} \emph
  {et~al.}}]{CAST2015}%
  \BibitemOpen
  \bibfield  {author} {\bibinfo {author} {\bibfnamefont {M.}~\bibnamefont
  {Arik}} \emph {et~al.} (\bibinfo {collaboration} {CAST}),\ }\href {\doibase
  10.1103/PhysRevD.92.021101} {\bibfield  {journal} {\bibinfo  {journal} {Phys.
  Rev. D}\ }\textbf {\bibinfo {volume} {92}},\ \bibinfo {pages} {021101}
  (\bibinfo {year} {2015})}\BibitemShut {NoStop}%
\bibitem [{\citenamefont {Reyn\'es}\ \emph {et~al.}(2021)\citenamefont
  {Reyn\'es}, \citenamefont {Matthews}, \citenamefont {Reynolds}, \citenamefont
  {Russell}, \citenamefont {Smith},\ and\ \citenamefont
  {Marsh}}]{Reynes:2021bpe}%
  \BibitemOpen
  \bibfield  {author} {\bibinfo {author} {\bibfnamefont {J.~S.}\ \bibnamefont
  {Reyn\'es}}, \bibinfo {author} {\bibfnamefont {J.~H.}\ \bibnamefont
  {Matthews}}, \bibinfo {author} {\bibfnamefont {C.~S.}\ \bibnamefont
  {Reynolds}}, \bibinfo {author} {\bibfnamefont {H.~R.}\ \bibnamefont
  {Russell}}, \bibinfo {author} {\bibfnamefont {R.~N.}\ \bibnamefont {Smith}},
  \ and\ \bibinfo {author} {\bibfnamefont {M.~C.~D.}\ \bibnamefont {Marsh}},\
  }\href {\doibase 10.1093/mnras/stab3464} {\bibfield  {journal} {\bibinfo
  {journal} {Mon. Not. Roy. Astron. Soc.}\ }\textbf {\bibinfo {volume} {510}},\
  \bibinfo {pages} {1264} (\bibinfo {year} {2021})},\ \Eprint
  {http://arxiv.org/abs/2109.03261} {arXiv:2109.03261 [astro-ph.HE]}
  \BibitemShut {NoStop}%
\bibitem [{\citenamefont {Reynolds}\ \emph {et~al.}(2020)\citenamefont
  {Reynolds}, \citenamefont {Marsh}, \citenamefont {Russell}, \citenamefont
  {Fabian}, \citenamefont {Smith}, \citenamefont {Tombesi},\ and\ \citenamefont
  {Veilleux}}]{Reynolds:2019uqt}%
  \BibitemOpen
  \bibfield  {author} {\bibinfo {author} {\bibfnamefont {C.~S.}\ \bibnamefont
  {Reynolds}}, \bibinfo {author} {\bibfnamefont {M.~C.~D.}\ \bibnamefont
  {Marsh}}, \bibinfo {author} {\bibfnamefont {H.~R.}\ \bibnamefont {Russell}},
  \bibinfo {author} {\bibfnamefont {A.~C.}\ \bibnamefont {Fabian}}, \bibinfo
  {author} {\bibfnamefont {R.}~\bibnamefont {Smith}}, \bibinfo {author}
  {\bibfnamefont {F.}~\bibnamefont {Tombesi}}, \ and\ \bibinfo {author}
  {\bibfnamefont {S.}~\bibnamefont {Veilleux}},\ }\href {\doibase
  10.3847/1538-4357/ab6a0c} {\bibfield  {journal} {\bibinfo  {journal}
  {Astrophys. J.}\ }\textbf {\bibinfo {volume} {890}},\ \bibinfo {pages} {59}
  (\bibinfo {year} {2020})},\ \Eprint {http://arxiv.org/abs/1907.05475}
  {arXiv:1907.05475 [hep-ph]} \BibitemShut {NoStop}%
\bibitem [{\citenamefont {Marsh}\ \emph {et~al.}(2017)\citenamefont {Marsh},
  \citenamefont {Russell}, \citenamefont {Fabian}, \citenamefont {McNamara},
  \citenamefont {Nulsen},\ and\ \citenamefont {Reynolds}}]{Marsh:2017yvc}%
  \BibitemOpen
  \bibfield  {author} {\bibinfo {author} {\bibfnamefont {M.~C.~D.}\
  \bibnamefont {Marsh}}, \bibinfo {author} {\bibfnamefont {H.~R.}\ \bibnamefont
  {Russell}}, \bibinfo {author} {\bibfnamefont {A.~C.}\ \bibnamefont {Fabian}},
  \bibinfo {author} {\bibfnamefont {B.~P.}\ \bibnamefont {McNamara}}, \bibinfo
  {author} {\bibfnamefont {P.}~\bibnamefont {Nulsen}}, \ and\ \bibinfo {author}
  {\bibfnamefont {C.~S.}\ \bibnamefont {Reynolds}},\ }\href {\doibase
  10.1088/1475-7516/2017/12/036} {\bibfield  {journal} {\bibinfo  {journal}
  {JCAP}\ }\textbf {\bibinfo {volume} {12}},\ \bibinfo {pages} {036} (\bibinfo
  {year} {2017})},\ \Eprint {http://arxiv.org/abs/1703.07354} {arXiv:1703.07354
  [hep-ph]} \BibitemShut {NoStop}%
\bibitem [{\citenamefont {Wouters}\ and\ \citenamefont
  {Brun}(2013)}]{Wouters:2013hua}%
  \BibitemOpen
  \bibfield  {author} {\bibinfo {author} {\bibfnamefont {D.}~\bibnamefont
  {Wouters}}\ and\ \bibinfo {author} {\bibfnamefont {P.}~\bibnamefont {Brun}},\
  }\href {\doibase 10.1088/0004-637X/772/1/44} {\bibfield  {journal} {\bibinfo
  {journal} {Astrophys. J.}\ }\textbf {\bibinfo {volume} {772}},\ \bibinfo
  {pages} {44} (\bibinfo {year} {2013})},\ \Eprint
  {http://arxiv.org/abs/1304.0989} {arXiv:1304.0989 [astro-ph.HE]} \BibitemShut
  {NoStop}%
\bibitem [{\citenamefont {Dessert}\ \emph {et~al.}(2020)\citenamefont
  {Dessert}, \citenamefont {Foster},\ and\ \citenamefont
  {Safdi}}]{Dessert:2020lil}%
  \BibitemOpen
  \bibfield  {author} {\bibinfo {author} {\bibfnamefont {C.}~\bibnamefont
  {Dessert}}, \bibinfo {author} {\bibfnamefont {J.~W.}\ \bibnamefont {Foster}},
  \ and\ \bibinfo {author} {\bibfnamefont {B.~R.}\ \bibnamefont {Safdi}},\
  }\href {\doibase 10.1103/PhysRevLett.125.261102} {\bibfield  {journal}
  {\bibinfo  {journal} {Phys. Rev. Lett.}\ }\textbf {\bibinfo {volume} {125}},\
  \bibinfo {pages} {261102} (\bibinfo {year} {2020})},\ \Eprint
  {http://arxiv.org/abs/2008.03305} {arXiv:2008.03305 [hep-ph]} \BibitemShut
  {NoStop}%
\bibitem [{\citenamefont {Eby}\ and\ \citenamefont
  {Takhistov}(2024)}]{Eby:2024mhd}%
  \BibitemOpen
  \bibfield  {author} {\bibinfo {author} {\bibfnamefont {J.}~\bibnamefont
  {Eby}}\ and\ \bibinfo {author} {\bibfnamefont {V.}~\bibnamefont
  {Takhistov}},\ }\href@noop {} {\  (\bibinfo {year} {2024})},\ \Eprint
  {http://arxiv.org/abs/2402.00100} {arXiv:2402.00100 [hep-ph]} \BibitemShut
  {NoStop}%
\bibitem [{\citenamefont {Song}\ \emph {et~al.}(2024)\citenamefont {Song},
  \citenamefont {Su},\ and\ \citenamefont {Wu}}]{Song:2024rru}%
  \BibitemOpen
  \bibfield  {author} {\bibinfo {author} {\bibfnamefont {N.}~\bibnamefont
  {Song}}, \bibinfo {author} {\bibfnamefont {L.}~\bibnamefont {Su}}, \ and\
  \bibinfo {author} {\bibfnamefont {L.}~\bibnamefont {Wu}},\ }\href@noop {} {\
  (\bibinfo {year} {2024})},\ \Eprint {http://arxiv.org/abs/2402.15144}
  {arXiv:2402.15144 [hep-ph]} \BibitemShut {NoStop}%
\bibitem [{\citenamefont {Libanov}\ and\ \citenamefont
  {Troitsky}(2020)}]{Libanov:2019fzq}%
  \BibitemOpen
  \bibfield  {author} {\bibinfo {author} {\bibfnamefont {M.}~\bibnamefont
  {Libanov}}\ and\ \bibinfo {author} {\bibfnamefont {S.}~\bibnamefont
  {Troitsky}},\ }\href {\doibase 10.1016/j.physletb.2020.135252} {\bibfield
  {journal} {\bibinfo  {journal} {Phys. Lett. B}\ }\textbf {\bibinfo {volume}
  {802}},\ \bibinfo {pages} {135252} (\bibinfo {year} {2020})},\ \Eprint
  {http://arxiv.org/abs/1908.03084} {arXiv:1908.03084 [astro-ph.HE]}
  \BibitemShut {NoStop}%
\bibitem [{\citenamefont {Matthews}\ \emph {et~al.}(2022)\citenamefont
  {Matthews}, \citenamefont {Reynolds}, \citenamefont {Marsh}, \citenamefont
  {Sisk-Reyn\'es},\ and\ \citenamefont {Rodman}}]{Matthews:2022gqi}%
  \BibitemOpen
  \bibfield  {author} {\bibinfo {author} {\bibfnamefont {J.~H.}\ \bibnamefont
  {Matthews}}, \bibinfo {author} {\bibfnamefont {C.~S.}\ \bibnamefont
  {Reynolds}}, \bibinfo {author} {\bibfnamefont {M.~C.~D.}\ \bibnamefont
  {Marsh}}, \bibinfo {author} {\bibfnamefont {J.}~\bibnamefont
  {Sisk-Reyn\'es}}, \ and\ \bibinfo {author} {\bibfnamefont {P.~E.}\
  \bibnamefont {Rodman}},\ }\href {\doibase 10.3847/1538-4357/ac5625}
  {\bibfield  {journal} {\bibinfo  {journal} {Astrophys. J.}\ }\textbf
  {\bibinfo {volume} {930}},\ \bibinfo {pages} {90} (\bibinfo {year} {2022})},\
  \Eprint {http://arxiv.org/abs/2202.08875} {arXiv:2202.08875 [astro-ph.HE]}
  \BibitemShut {NoStop}%
\bibitem [{\citenamefont {Lee}\ \emph {et~al.}(2023)\citenamefont {Lee},
  \citenamefont {Lisanti}, \citenamefont {Terrano},\ and\ \citenamefont
  {Romalis}}]{Lee2023}%
  \BibitemOpen
  \bibfield  {author} {\bibinfo {author} {\bibfnamefont {J.}~\bibnamefont
  {Lee}}, \bibinfo {author} {\bibfnamefont {M.}~\bibnamefont {Lisanti}},
  \bibinfo {author} {\bibfnamefont {W.~A.}\ \bibnamefont {Terrano}}, \ and\
  \bibinfo {author} {\bibfnamefont {M.}~\bibnamefont {Romalis}},\ }\href
  {\doibase 10.1103/physrevx.13.011050} {\bibfield  {journal} {\bibinfo
  {journal} {Physical Review X}\ }\textbf {\bibinfo {volume} {13}} (\bibinfo
  {year} {2023}),\ 10.1103/physrevx.13.011050}\BibitemShut {NoStop}%
\bibitem [{\citenamefont {Bloch}\ \emph {et~al.}(2022)\citenamefont {Bloch},
  \citenamefont {Ronen}, \citenamefont {Shaham}, \citenamefont {Katz},
  \citenamefont {Volansky},\ and\ \citenamefont {Katz}}]{Bloch2022}%
  \BibitemOpen
  \bibfield  {author} {\bibinfo {author} {\bibfnamefont {I.~M.}\ \bibnamefont
  {Bloch}}, \bibinfo {author} {\bibfnamefont {G.}~\bibnamefont {Ronen}},
  \bibinfo {author} {\bibfnamefont {R.}~\bibnamefont {Shaham}}, \bibinfo
  {author} {\bibfnamefont {O.}~\bibnamefont {Katz}}, \bibinfo {author}
  {\bibfnamefont {T.}~\bibnamefont {Volansky}}, \ and\ \bibinfo {author}
  {\bibfnamefont {O.}~\bibnamefont {Katz}},\ }\href {\doibase
  10.1126/sciadv.abl8919} {\bibfield  {journal} {\bibinfo  {journal} {Science
  Advances}\ }\textbf {\bibinfo {volume} {8}} (\bibinfo {year} {2022}),\
  10.1126/sciadv.abl8919}\BibitemShut {NoStop}%
\bibitem [{\citenamefont {Buschmann}\ \emph {et~al.}(2022)\citenamefont
  {Buschmann}, \citenamefont {Dessert}, \citenamefont {Foster}, \citenamefont
  {Long},\ and\ \citenamefont {Safdi}}]{Buschmann:2021juv}%
  \BibitemOpen
  \bibfield  {author} {\bibinfo {author} {\bibfnamefont {M.}~\bibnamefont
  {Buschmann}}, \bibinfo {author} {\bibfnamefont {C.}~\bibnamefont {Dessert}},
  \bibinfo {author} {\bibfnamefont {J.~W.}\ \bibnamefont {Foster}}, \bibinfo
  {author} {\bibfnamefont {A.~J.}\ \bibnamefont {Long}}, \ and\ \bibinfo
  {author} {\bibfnamefont {B.~R.}\ \bibnamefont {Safdi}},\ }\href {\doibase
  10.1103/PhysRevLett.128.091102} {\bibfield  {journal} {\bibinfo  {journal}
  {Phys. Rev. Lett.}\ }\textbf {\bibinfo {volume} {128}},\ \bibinfo {pages}
  {091102} (\bibinfo {year} {2022})},\ \Eprint
  {http://arxiv.org/abs/2111.09892} {arXiv:2111.09892 [hep-ph]} \BibitemShut
  {NoStop}%
\bibitem [{\citenamefont {Payez}\ \emph {et~al.}(2015)\citenamefont {Payez},
  \citenamefont {Evoli}, \citenamefont {Fischer}, \citenamefont {Giannotti},
  \citenamefont {Mirizzi},\ and\ \citenamefont {Ringwald}}]{Payez:2014xsa}%
  \BibitemOpen
  \bibfield  {author} {\bibinfo {author} {\bibfnamefont {A.}~\bibnamefont
  {Payez}}, \bibinfo {author} {\bibfnamefont {C.}~\bibnamefont {Evoli}},
  \bibinfo {author} {\bibfnamefont {T.}~\bibnamefont {Fischer}}, \bibinfo
  {author} {\bibfnamefont {M.}~\bibnamefont {Giannotti}}, \bibinfo {author}
  {\bibfnamefont {A.}~\bibnamefont {Mirizzi}}, \ and\ \bibinfo {author}
  {\bibfnamefont {A.}~\bibnamefont {Ringwald}},\ }\href {\doibase
  10.1088/1475-7516/2015/02/006} {\bibfield  {journal} {\bibinfo  {journal}
  {JCAP}\ }\textbf {\bibinfo {volume} {02}},\ \bibinfo {pages} {006} (\bibinfo
  {year} {2015})},\ \Eprint {http://arxiv.org/abs/1410.3747} {arXiv:1410.3747
  [astro-ph.HE]} \BibitemShut {NoStop}%
\bibitem [{\citenamefont {Manzari}\ \emph {et~al.}(pear)\citenamefont
  {Manzari}, \citenamefont {Park}, \citenamefont {Savoray},\ and\ \citenamefont
  {Safdi}}]{safdi2024}%
  \BibitemOpen
  \bibfield  {author} {\bibinfo {author} {\bibfnamefont {C.~A.}\ \bibnamefont
  {Manzari}}, \bibinfo {author} {\bibfnamefont {Y.}~\bibnamefont {Park}},
  \bibinfo {author} {\bibfnamefont {I.}~\bibnamefont {Savoray}}, \ and\
  \bibinfo {author} {\bibfnamefont {B.~R.}\ \bibnamefont {Safdi}},\ }\href@noop
  {} {\  (\bibinfo {year} {2024 to-appear})}\BibitemShut {NoStop}%
\bibitem [{\citenamefont {Spitkovsky}(2006)}]{Spitkovsky2006}%
  \BibitemOpen
  \bibfield  {author} {\bibinfo {author} {\bibfnamefont {A.}~\bibnamefont
  {Spitkovsky}},\ }\href {\doibase 10.1086/507518} {\bibfield  {journal}
  {\bibinfo  {journal} {The Astrophysical Journal}\ }\textbf {\bibinfo {volume}
  {648}},\ \bibinfo {pages} {L51} (\bibinfo {year} {2006})}\BibitemShut
  {NoStop}%
\bibitem [{\citenamefont {{Kalapotharakos}}\ and\ \citenamefont
  {{Contopoulos}}(2009)}]{Kalapotharakos2009}%
  \BibitemOpen
  \bibfield  {author} {\bibinfo {author} {\bibfnamefont {C.}~\bibnamefont
  {{Kalapotharakos}}}\ and\ \bibinfo {author} {\bibfnamefont {I.}~\bibnamefont
  {{Contopoulos}}},\ }\href {\doibase 10.1051/0004-6361:200810281} {\bibfield
  {journal} {\bibinfo  {journal} {Astronomy and Astrophysics}\ }\textbf
  {\bibinfo {volume} {496}},\ \bibinfo {pages} {495} (\bibinfo {year}
  {2009})},\ \Eprint {http://arxiv.org/abs/0811.2863} {arXiv:0811.2863
  [astro-ph]} \BibitemShut {NoStop}%
\bibitem [{\citenamefont {P\'{e}tri}(2012)}]{Petri2012}%
  \BibitemOpen
  \bibfield  {author} {\bibinfo {author} {\bibfnamefont {J.}~\bibnamefont
  {P\'{e}tri}},\ }\href {\doibase 10.1111/j.1365-2966.2012.21238.x} {\bibfield
  {journal} {\bibinfo  {journal} {Monthly Notices of the Royal Astronomical
  Society}\ }\textbf {\bibinfo {volume} {424}},\ \bibinfo {pages} {605}
  (\bibinfo {year} {2012})}\BibitemShut {NoStop}%
\bibitem [{\citenamefont {P\'erez}\ and\ \citenamefont
  {Granger}(2007)}]{PER-GRA:2007}%
  \BibitemOpen
  \bibfield  {author} {\bibinfo {author} {\bibfnamefont {F.}~\bibnamefont
  {P\'erez}}\ and\ \bibinfo {author} {\bibfnamefont {B.~E.}\ \bibnamefont
  {Granger}},\ }\href {\doibase 10.1109/MCSE.2007.53} {\bibfield  {journal}
  {\bibinfo  {journal} {Computing in Science and Engineering}\ }\textbf
  {\bibinfo {volume} {9}},\ \bibinfo {pages} {21} (\bibinfo {year}
  {2007})}\BibitemShut {NoStop}%
\bibitem [{\citenamefont {Hunter}(2007)}]{Hunter:2007}%
  \BibitemOpen
  \bibfield  {author} {\bibinfo {author} {\bibfnamefont {J.~D.}\ \bibnamefont
  {Hunter}},\ }\href {\doibase 10.1109/MCSE.2007.55} {\bibfield  {journal}
  {\bibinfo  {journal} {Computing in Science \& Engineering}\ }\textbf
  {\bibinfo {volume} {9}},\ \bibinfo {pages} {90} (\bibinfo {year}
  {2007})}\BibitemShut {NoStop}%
\bibitem [{\citenamefont {Kluyver}\ \emph {et~al.}(2016)\citenamefont
  {Kluyver}, \citenamefont {Ragan-Kelley}, \citenamefont {P{\'e}rez},
  \citenamefont {Granger}, \citenamefont {Bussonnier}, \citenamefont
  {Frederic}, \citenamefont {Kelley}, \citenamefont {Hamrick}, \citenamefont
  {Grout}, \citenamefont {Corlay}, \citenamefont {Ivanov}, \citenamefont
  {Avila}, \citenamefont {Abdalla},\ and\ \citenamefont
  {Willing}}]{Kluyver2016jupyter}%
  \BibitemOpen
  \bibfield  {author} {\bibinfo {author} {\bibfnamefont {T.}~\bibnamefont
  {Kluyver}}, \bibinfo {author} {\bibfnamefont {B.}~\bibnamefont
  {Ragan-Kelley}}, \bibinfo {author} {\bibfnamefont {F.}~\bibnamefont
  {P{\'e}rez}}, \bibinfo {author} {\bibfnamefont {B.}~\bibnamefont {Granger}},
  \bibinfo {author} {\bibfnamefont {M.}~\bibnamefont {Bussonnier}}, \bibinfo
  {author} {\bibfnamefont {J.}~\bibnamefont {Frederic}}, \bibinfo {author}
  {\bibfnamefont {K.}~\bibnamefont {Kelley}}, \bibinfo {author} {\bibfnamefont
  {J.}~\bibnamefont {Hamrick}}, \bibinfo {author} {\bibfnamefont
  {J.}~\bibnamefont {Grout}}, \bibinfo {author} {\bibfnamefont
  {S.}~\bibnamefont {Corlay}}, \bibinfo {author} {\bibfnamefont
  {P.}~\bibnamefont {Ivanov}}, \bibinfo {author} {\bibfnamefont
  {D.}~\bibnamefont {Avila}}, \bibinfo {author} {\bibfnamefont
  {S.}~\bibnamefont {Abdalla}}, \ and\ \bibinfo {author} {\bibfnamefont
  {C.}~\bibnamefont {Willing}},\ }in\ \href@noop {} {\emph {\bibinfo
  {booktitle} {Positioning and Power in Academic Publishing: Players, Agents
  and Agendas}}},\ \bibinfo {editor} {edited by\ \bibinfo {editor}
  {\bibfnamefont {F.}~\bibnamefont {Loizides}}\ and\ \bibinfo {editor}
  {\bibfnamefont {B.}~\bibnamefont {Schmidt}}}\ (\bibinfo {organization} {IOS
  Press},\ \bibinfo {year} {2016})\ pp.\ \bibinfo {pages} {87 --
  90}\BibitemShut {NoStop}%
\bibitem [{\citenamefont {Harris}\ \emph {et~al.}(2020)\citenamefont {Harris},
  \citenamefont {Millman}, \citenamefont {van~der Walt}, \citenamefont
  {Gommers}, \citenamefont {Virtanen}, \citenamefont {Cournapeau},
  \citenamefont {Wieser}, \citenamefont {Taylor}, \citenamefont {Berg},
  \citenamefont {Smith}, \citenamefont {Kern}, \citenamefont {Picus},
  \citenamefont {Hoyer}, \citenamefont {van Kerkwijk}, \citenamefont {Brett},
  \citenamefont {Haldane}, \citenamefont {del R{\'{i}}o}, \citenamefont
  {Wiebe}, \citenamefont {Peterson}, \citenamefont {G{\'{e}}rard-Marchant},
  \citenamefont {Sheppard}, \citenamefont {Reddy}, \citenamefont {Weckesser},
  \citenamefont {Abbasi}, \citenamefont {Gohlke},\ and\ \citenamefont
  {Oliphant}}]{harris2020array}%
  \BibitemOpen
  \bibfield  {author} {\bibinfo {author} {\bibfnamefont {C.~R.}\ \bibnamefont
  {Harris}}, \bibinfo {author} {\bibfnamefont {K.~J.}\ \bibnamefont {Millman}},
  \bibinfo {author} {\bibfnamefont {S.~J.}\ \bibnamefont {van~der Walt}},
  \bibinfo {author} {\bibfnamefont {R.}~\bibnamefont {Gommers}}, \bibinfo
  {author} {\bibfnamefont {P.}~\bibnamefont {Virtanen}}, \bibinfo {author}
  {\bibfnamefont {D.}~\bibnamefont {Cournapeau}}, \bibinfo {author}
  {\bibfnamefont {E.}~\bibnamefont {Wieser}}, \bibinfo {author} {\bibfnamefont
  {J.}~\bibnamefont {Taylor}}, \bibinfo {author} {\bibfnamefont
  {S.}~\bibnamefont {Berg}}, \bibinfo {author} {\bibfnamefont {N.~J.}\
  \bibnamefont {Smith}}, \bibinfo {author} {\bibfnamefont {R.}~\bibnamefont
  {Kern}}, \bibinfo {author} {\bibfnamefont {M.}~\bibnamefont {Picus}},
  \bibinfo {author} {\bibfnamefont {S.}~\bibnamefont {Hoyer}}, \bibinfo
  {author} {\bibfnamefont {M.~H.}\ \bibnamefont {van Kerkwijk}}, \bibinfo
  {author} {\bibfnamefont {M.}~\bibnamefont {Brett}}, \bibinfo {author}
  {\bibfnamefont {A.}~\bibnamefont {Haldane}}, \bibinfo {author} {\bibfnamefont
  {J.~F.}\ \bibnamefont {del R{\'{i}}o}}, \bibinfo {author} {\bibfnamefont
  {M.}~\bibnamefont {Wiebe}}, \bibinfo {author} {\bibfnamefont
  {P.}~\bibnamefont {Peterson}}, \bibinfo {author} {\bibfnamefont
  {P.}~\bibnamefont {G{\'{e}}rard-Marchant}}, \bibinfo {author} {\bibfnamefont
  {K.}~\bibnamefont {Sheppard}}, \bibinfo {author} {\bibfnamefont
  {T.}~\bibnamefont {Reddy}}, \bibinfo {author} {\bibfnamefont
  {W.}~\bibnamefont {Weckesser}}, \bibinfo {author} {\bibfnamefont
  {H.}~\bibnamefont {Abbasi}}, \bibinfo {author} {\bibfnamefont
  {C.}~\bibnamefont {Gohlke}}, \ and\ \bibinfo {author} {\bibfnamefont {T.~E.}\
  \bibnamefont {Oliphant}},\ }\href {\doibase 10.1038/s41586-020-2649-2}
  {\bibfield  {journal} {\bibinfo  {journal} {Nature}\ }\textbf {\bibinfo
  {volume} {585}},\ \bibinfo {pages} {357} (\bibinfo {year}
  {2020})}\BibitemShut {NoStop}%
\bibitem [{\citenamefont {Virtanen}\ \emph {et~al.}(2020)\citenamefont
  {Virtanen}, \citenamefont {Gommers}, \citenamefont {Oliphant}, \citenamefont
  {Haberland}, \citenamefont {Reddy}, \citenamefont {Cournapeau}, \citenamefont
  {Burovski}, \citenamefont {Peterson}, \citenamefont {Weckesser},
  \citenamefont {Bright}, \citenamefont {{van der Walt}}, \citenamefont
  {Brett}, \citenamefont {Wilson}, \citenamefont {Millman}, \citenamefont
  {Mayorov}, \citenamefont {Nelson}, \citenamefont {Jones}, \citenamefont
  {Kern}, \citenamefont {Larson}, \citenamefont {Carey}, \citenamefont {Polat},
  \citenamefont {Feng}, \citenamefont {Moore}, \citenamefont {{VanderPlas}},
  \citenamefont {Laxalde}, \citenamefont {Perktold}, \citenamefont {Cimrman},
  \citenamefont {Henriksen}, \citenamefont {Quintero}, \citenamefont {Harris},
  \citenamefont {Archibald}, \citenamefont {Ribeiro}, \citenamefont
  {Pedregosa}, \citenamefont {{van Mulbregt}},\ and\ \citenamefont {{SciPy 1.0
  Contributors}}}]{2020SciPy-NMeth}%
  \BibitemOpen
  \bibfield  {author} {\bibinfo {author} {\bibfnamefont {P.}~\bibnamefont
  {Virtanen}}, \bibinfo {author} {\bibfnamefont {R.}~\bibnamefont {Gommers}},
  \bibinfo {author} {\bibfnamefont {T.~E.}\ \bibnamefont {Oliphant}}, \bibinfo
  {author} {\bibfnamefont {M.}~\bibnamefont {Haberland}}, \bibinfo {author}
  {\bibfnamefont {T.}~\bibnamefont {Reddy}}, \bibinfo {author} {\bibfnamefont
  {D.}~\bibnamefont {Cournapeau}}, \bibinfo {author} {\bibfnamefont
  {E.}~\bibnamefont {Burovski}}, \bibinfo {author} {\bibfnamefont
  {P.}~\bibnamefont {Peterson}}, \bibinfo {author} {\bibfnamefont
  {W.}~\bibnamefont {Weckesser}}, \bibinfo {author} {\bibfnamefont
  {J.}~\bibnamefont {Bright}}, \bibinfo {author} {\bibfnamefont {S.~J.}\
  \bibnamefont {{van der Walt}}}, \bibinfo {author} {\bibfnamefont
  {M.}~\bibnamefont {Brett}}, \bibinfo {author} {\bibfnamefont
  {J.}~\bibnamefont {Wilson}}, \bibinfo {author} {\bibfnamefont {K.~J.}\
  \bibnamefont {Millman}}, \bibinfo {author} {\bibfnamefont {N.}~\bibnamefont
  {Mayorov}}, \bibinfo {author} {\bibfnamefont {A.~R.~J.}\ \bibnamefont
  {Nelson}}, \bibinfo {author} {\bibfnamefont {E.}~\bibnamefont {Jones}},
  \bibinfo {author} {\bibfnamefont {R.}~\bibnamefont {Kern}}, \bibinfo {author}
  {\bibfnamefont {E.}~\bibnamefont {Larson}}, \bibinfo {author} {\bibfnamefont
  {C.~J.}\ \bibnamefont {Carey}}, \bibinfo {author} {\bibfnamefont
  {{\.I}.}~\bibnamefont {Polat}}, \bibinfo {author} {\bibfnamefont
  {Y.}~\bibnamefont {Feng}}, \bibinfo {author} {\bibfnamefont {E.~W.}\
  \bibnamefont {Moore}}, \bibinfo {author} {\bibfnamefont {J.}~\bibnamefont
  {{VanderPlas}}}, \bibinfo {author} {\bibfnamefont {D.}~\bibnamefont
  {Laxalde}}, \bibinfo {author} {\bibfnamefont {J.}~\bibnamefont {Perktold}},
  \bibinfo {author} {\bibfnamefont {R.}~\bibnamefont {Cimrman}}, \bibinfo
  {author} {\bibfnamefont {I.}~\bibnamefont {Henriksen}}, \bibinfo {author}
  {\bibfnamefont {E.~A.}\ \bibnamefont {Quintero}}, \bibinfo {author}
  {\bibfnamefont {C.~R.}\ \bibnamefont {Harris}}, \bibinfo {author}
  {\bibfnamefont {A.~M.}\ \bibnamefont {Archibald}}, \bibinfo {author}
  {\bibfnamefont {A.~H.}\ \bibnamefont {Ribeiro}}, \bibinfo {author}
  {\bibfnamefont {F.}~\bibnamefont {Pedregosa}}, \bibinfo {author}
  {\bibfnamefont {P.}~\bibnamefont {{van Mulbregt}}}, \ and\ \bibinfo {author}
  {\bibnamefont {{SciPy 1.0 Contributors}}},\ }\href {\doibase
  10.1038/s41592-019-0686-2} {\bibfield  {journal} {\bibinfo  {journal} {Nature
  Methods}\ }\textbf {\bibinfo {volume} {17}},\ \bibinfo {pages} {261}
  (\bibinfo {year} {2020})}\BibitemShut {NoStop}%
\bibitem [{\citenamefont {Bogolyubov}\ and\ \citenamefont
  {Shirkov}(1959)}]{Bogolyubov1959}%
  \BibitemOpen
  \bibfield  {author} {\bibinfo {author} {\bibfnamefont {N.~N.}\ \bibnamefont
  {Bogolyubov}}\ and\ \bibinfo {author} {\bibfnamefont {D.~V.}\ \bibnamefont
  {Shirkov}},\ }\href@noop {} {\emph {\bibinfo {title} {{Introduction To the
  Theory of Quantized Fields}}}},\ Vol.~\bibinfo {volume} {3}\ (\bibinfo {year}
  {1959})\BibitemShut {NoStop}%
\bibitem [{\citenamefont {Economou}(2006)}]{Economou2006}%
  \BibitemOpen
  \bibfield  {author} {\bibinfo {author} {\bibfnamefont {E.}~\bibnamefont
  {Economou}},\ }\href@noop {} {\emph {\bibinfo {title} {Green's Functions in
  Quantum Physics}}},\ Springer Series in Solid-State Sciences\ (\bibinfo
  {publisher} {Springer Berlin Heidelberg},\ \bibinfo {year}
  {2006})\BibitemShut {NoStop}%
\bibitem [{\citenamefont {Li}\ and\ \citenamefont {Ruffini}(1986)}]{Li1986}%
  \BibitemOpen
  \bibfield  {author} {\bibinfo {author} {\bibfnamefont {M.}~\bibnamefont
  {Li}}\ and\ \bibinfo {author} {\bibfnamefont {R.}~\bibnamefont {Ruffini}},\
  }\href {\doibase https://doi.org/10.1016/0375-9601(86)90349-X} {\bibfield
  {journal} {\bibinfo  {journal} {Physics Letters A}\ }\textbf {\bibinfo
  {volume} {116}},\ \bibinfo {pages} {20} (\bibinfo {year} {1986})}\BibitemShut
  {NoStop}%
\bibitem [{\citenamefont {Krause}\ \emph {et~al.}(1994)\citenamefont {Krause},
  \citenamefont {Kloor},\ and\ \citenamefont {Fischbach}}]{Krause1994}%
  \BibitemOpen
  \bibfield  {author} {\bibinfo {author} {\bibfnamefont {D.~E.}\ \bibnamefont
  {Krause}}, \bibinfo {author} {\bibfnamefont {H.~T.}\ \bibnamefont {Kloor}}, \
  and\ \bibinfo {author} {\bibfnamefont {E.}~\bibnamefont {Fischbach}},\ }\href
  {\doibase 10.1103/PhysRevD.49.6892} {\bibfield  {journal} {\bibinfo
  {journal} {Phys. Rev. D}\ }\textbf {\bibinfo {volume} {49}},\ \bibinfo
  {pages} {6892} (\bibinfo {year} {1994})}\BibitemShut {NoStop}%
\bibitem [{\citenamefont {Kahn}\ \emph {et~al.}(2016)\citenamefont {Kahn},
  \citenamefont {Safdi},\ and\ \citenamefont {Thaler}}]{ABRACADABRA}%
  \BibitemOpen
  \bibfield  {author} {\bibinfo {author} {\bibfnamefont {Y.}~\bibnamefont
  {Kahn}}, \bibinfo {author} {\bibfnamefont {B.~R.}\ \bibnamefont {Safdi}}, \
  and\ \bibinfo {author} {\bibfnamefont {J.}~\bibnamefont {Thaler}},\ }\href
  {\doibase 10.1103/PhysRevLett.117.141801} {\bibfield  {journal} {\bibinfo
  {journal} {Phys. Rev. Lett.}\ }\textbf {\bibinfo {volume} {117}},\ \bibinfo
  {pages} {141801} (\bibinfo {year} {2016})}\BibitemShut {NoStop}%
\bibitem [{\citenamefont {Benabou}\ \emph {et~al.}(2023)\citenamefont
  {Benabou}, \citenamefont {Foster}, \citenamefont {Kahn}, \citenamefont
  {Safdi},\ and\ \citenamefont {Salemi}}]{Benabou:2022qpv}%
  \BibitemOpen
  \bibfield  {author} {\bibinfo {author} {\bibfnamefont {J.~N.}\ \bibnamefont
  {Benabou}}, \bibinfo {author} {\bibfnamefont {J.~W.}\ \bibnamefont {Foster}},
  \bibinfo {author} {\bibfnamefont {Y.}~\bibnamefont {Kahn}}, \bibinfo {author}
  {\bibfnamefont {B.~R.}\ \bibnamefont {Safdi}}, \ and\ \bibinfo {author}
  {\bibfnamefont {C.~P.}\ \bibnamefont {Salemi}},\ }\href {\doibase
  10.1103/PhysRevD.108.035009} {\bibfield  {journal} {\bibinfo  {journal}
  {Phys. Rev. D}\ }\textbf {\bibinfo {volume} {108}},\ \bibinfo {pages}
  {035009} (\bibinfo {year} {2023})},\ \Eprint
  {http://arxiv.org/abs/2211.00008} {arXiv:2211.00008 [hep-ph]} \BibitemShut
  {NoStop}%
\bibitem [{\citenamefont {{Anton}}\ \emph {et~al.}(2013)\citenamefont
  {{Anton}}, \citenamefont {{Birenbaum}}, \citenamefont {{O'Kelley}},
  \citenamefont {{Bolkhovsky}}, \citenamefont {{Braje}}, \citenamefont
  {{Fitch}}, \citenamefont {{Neeley}}, \citenamefont {{Hilton}}, \citenamefont
  {{Cho}}, \citenamefont {{Irwin}}, \citenamefont {{Wellstood}}, \citenamefont
  {{Oliver}}, \citenamefont {{Shnirman}},\ and\ \citenamefont
  {{Clarke}}}]{2013PhRvL.110n7002A}%
  \BibitemOpen
  \bibfield  {author} {\bibinfo {author} {\bibfnamefont {S.~M.}\ \bibnamefont
  {{Anton}}}, \bibinfo {author} {\bibfnamefont {J.~S.}\ \bibnamefont
  {{Birenbaum}}}, \bibinfo {author} {\bibfnamefont {S.~R.}\ \bibnamefont
  {{O'Kelley}}}, \bibinfo {author} {\bibfnamefont {V.}~\bibnamefont
  {{Bolkhovsky}}}, \bibinfo {author} {\bibfnamefont {D.~A.}\ \bibnamefont
  {{Braje}}}, \bibinfo {author} {\bibfnamefont {G.}~\bibnamefont {{Fitch}}},
  \bibinfo {author} {\bibfnamefont {M.}~\bibnamefont {{Neeley}}}, \bibinfo
  {author} {\bibfnamefont {G.~C.}\ \bibnamefont {{Hilton}}}, \bibinfo {author}
  {\bibfnamefont {H.~M.}\ \bibnamefont {{Cho}}}, \bibinfo {author}
  {\bibfnamefont {K.~D.}\ \bibnamefont {{Irwin}}}, \bibinfo {author}
  {\bibfnamefont {F.~C.}\ \bibnamefont {{Wellstood}}}, \bibinfo {author}
  {\bibfnamefont {W.~D.}\ \bibnamefont {{Oliver}}}, \bibinfo {author}
  {\bibfnamefont {A.}~\bibnamefont {{Shnirman}}}, \ and\ \bibinfo {author}
  {\bibfnamefont {J.}~\bibnamefont {{Clarke}}},\ }\href {\doibase
  10.1103/PhysRevLett.110.147002} {\bibfield  {journal} {\bibinfo  {journal}
  {\prl}\ }\textbf {\bibinfo {volume} {110}},\ \bibinfo {eid} {147002}
  (\bibinfo {year} {2013})}\BibitemShut {NoStop}%
\bibitem [{\citenamefont {Brouwer}\ \emph
  {et~al.}(2022{\natexlab{c}})\citenamefont {Brouwer}, \citenamefont
  {Chaudhuri}, \citenamefont {Cho}, \citenamefont {Corbin}, \citenamefont
  {Craddock}, \citenamefont {Dawson}, \citenamefont {Droster}, \citenamefont
  {Foster}, \citenamefont {Fry}, \citenamefont {Graham}, \citenamefont
  {Henning}, \citenamefont {Irwin}, \citenamefont {Kadribasic}, \citenamefont
  {Kahn}, \citenamefont {Keller}, \citenamefont {Kolevatov}, \citenamefont
  {Kuenstner}, \citenamefont {Leder}, \citenamefont {Li}, \citenamefont
  {Ouellet}, \citenamefont {Pappas}, \citenamefont {Phipps}, \citenamefont
  {Rapidis}, \citenamefont {Safdi}, \citenamefont {Salemi}, \citenamefont
  {Simanovskaia}, \citenamefont {Singh}, \citenamefont {van Assendelft},
  \citenamefont {van Bibber}, \citenamefont {Wells}, \citenamefont {Winslow},
  \citenamefont {Wisniewski},\ and\ \citenamefont {and}}]{Brouwer2022}%
  \BibitemOpen
  \bibfield  {author} {\bibinfo {author} {\bibfnamefont {L.}~\bibnamefont
  {Brouwer}}, \bibinfo {author} {\bibfnamefont {S.}~\bibnamefont {Chaudhuri}},
  \bibinfo {author} {\bibfnamefont {H.-M.}\ \bibnamefont {Cho}}, \bibinfo
  {author} {\bibfnamefont {J.}~\bibnamefont {Corbin}}, \bibinfo {author}
  {\bibfnamefont {W.}~\bibnamefont {Craddock}}, \bibinfo {author}
  {\bibfnamefont {C.}~\bibnamefont {Dawson}}, \bibinfo {author} {\bibfnamefont
  {A.}~\bibnamefont {Droster}}, \bibinfo {author} {\bibfnamefont
  {J.}~\bibnamefont {Foster}}, \bibinfo {author} {\bibfnamefont
  {J.}~\bibnamefont {Fry}}, \bibinfo {author} {\bibfnamefont {P.}~\bibnamefont
  {Graham}}, \bibinfo {author} {\bibfnamefont {R.}~\bibnamefont {Henning}},
  \bibinfo {author} {\bibfnamefont {K.}~\bibnamefont {Irwin}}, \bibinfo
  {author} {\bibfnamefont {F.}~\bibnamefont {Kadribasic}}, \bibinfo {author}
  {\bibfnamefont {Y.}~\bibnamefont {Kahn}}, \bibinfo {author} {\bibfnamefont
  {A.}~\bibnamefont {Keller}}, \bibinfo {author} {\bibfnamefont
  {R.}~\bibnamefont {Kolevatov}}, \bibinfo {author} {\bibfnamefont
  {S.}~\bibnamefont {Kuenstner}}, \bibinfo {author} {\bibfnamefont
  {A.}~\bibnamefont {Leder}}, \bibinfo {author} {\bibfnamefont
  {D.}~\bibnamefont {Li}}, \bibinfo {author} {\bibfnamefont {J.}~\bibnamefont
  {Ouellet}}, \bibinfo {author} {\bibfnamefont {K.}~\bibnamefont {Pappas}},
  \bibinfo {author} {\bibfnamefont {A.}~\bibnamefont {Phipps}}, \bibinfo
  {author} {\bibfnamefont {N.}~\bibnamefont {Rapidis}}, \bibinfo {author}
  {\bibfnamefont {B.}~\bibnamefont {Safdi}}, \bibinfo {author} {\bibfnamefont
  {C.}~\bibnamefont {Salemi}}, \bibinfo {author} {\bibfnamefont
  {M.}~\bibnamefont {Simanovskaia}}, \bibinfo {author} {\bibfnamefont
  {J.}~\bibnamefont {Singh}}, \bibinfo {author} {\bibfnamefont
  {E.}~\bibnamefont {van Assendelft}}, \bibinfo {author} {\bibfnamefont
  {K.}~\bibnamefont {van Bibber}}, \bibinfo {author} {\bibfnamefont
  {K.}~\bibnamefont {Wells}}, \bibinfo {author} {\bibfnamefont
  {L.}~\bibnamefont {Winslow}}, \bibinfo {author} {\bibfnamefont
  {W.}~\bibnamefont {Wisniewski}}, \ and\ \bibinfo {author} {\bibfnamefont
  {B.~Y.}\ \bibnamefont {and}},\ }\href {\doibase 10.1103/physrevd.106.103008}
  {\bibfield  {journal} {\bibinfo  {journal} {Physical Review D}\ }\textbf
  {\bibinfo {volume} {106}} (\bibinfo {year} {2022}{\natexlab{c}}),\
  10.1103/physrevd.106.103008}\BibitemShut {NoStop}%
\bibitem [{\citenamefont {{Barnard}}\ and\ \citenamefont
  {{Arons}}(1982)}]{1982ApJ...254..713B}%
  \BibitemOpen
  \bibfield  {author} {\bibinfo {author} {\bibfnamefont {J.~J.}\ \bibnamefont
  {{Barnard}}}\ and\ \bibinfo {author} {\bibfnamefont {J.}~\bibnamefont
  {{Arons}}},\ }\href {\doibase 10.1086/159784} {\bibfield  {journal} {\bibinfo
   {journal} {\apj}\ }\textbf {\bibinfo {volume} {254}},\ \bibinfo {pages}
  {713} (\bibinfo {year} {1982})}\BibitemShut {NoStop}%
\bibitem [{\citenamefont {Gralla}\ \emph {et~al.}(2017)\citenamefont {Gralla},
  \citenamefont {Lupsasca},\ and\ \citenamefont {Philippov}}]{Gralla_2017}%
  \BibitemOpen
  \bibfield  {author} {\bibinfo {author} {\bibfnamefont {S.~E.}\ \bibnamefont
  {Gralla}}, \bibinfo {author} {\bibfnamefont {A.}~\bibnamefont {Lupsasca}}, \
  and\ \bibinfo {author} {\bibfnamefont {A.}~\bibnamefont {Philippov}},\ }\href
  {\doibase 10.3847/1538-4357/aa978d} {\bibfield  {journal} {\bibinfo
  {journal} {The Astrophysical Journal}\ }\textbf {\bibinfo {volume} {851}},\
  \bibinfo {pages} {137} (\bibinfo {year} {2017})}\BibitemShut {NoStop}%
\bibitem [{\citenamefont {Kalapotharakos}\ \emph {et~al.}(2021)\citenamefont
  {Kalapotharakos}, \citenamefont {Wadiasingh}, \citenamefont {Harding},\ and\
  \citenamefont {Kazanas}}]{Kalapotharakos_2021}%
  \BibitemOpen
  \bibfield  {author} {\bibinfo {author} {\bibfnamefont {C.}~\bibnamefont
  {Kalapotharakos}}, \bibinfo {author} {\bibfnamefont {Z.}~\bibnamefont
  {Wadiasingh}}, \bibinfo {author} {\bibfnamefont {A.~K.}\ \bibnamefont
  {Harding}}, \ and\ \bibinfo {author} {\bibfnamefont {D.}~\bibnamefont
  {Kazanas}},\ }\href {\doibase 10.3847/1538-4357/abcec0} {\ \textbf {\bibinfo
  {volume} {907}},\ \bibinfo {pages} {63} (\bibinfo {year} {2021})}\BibitemShut
  {NoStop}%
\bibitem [{\citenamefont {{Arons}}\ and\ \citenamefont
  {{Scharlemann}}(1979)}]{1979ApJ...231..854A}%
  \BibitemOpen
  \bibfield  {author} {\bibinfo {author} {\bibfnamefont {J.}~\bibnamefont
  {{Arons}}}\ and\ \bibinfo {author} {\bibfnamefont {E.~T.}\ \bibnamefont
  {{Scharlemann}}},\ }\href {\doibase 10.1086/157250} {\bibfield  {journal}
  {\bibinfo  {journal} {\apj}\ }\textbf {\bibinfo {volume} {231}},\ \bibinfo
  {pages} {854} (\bibinfo {year} {1979})}\BibitemShut {NoStop}%
\bibitem [{\citenamefont {Usov}\ and\ \citenamefont
  {Melrose}(1995)}]{Usov1995}%
  \BibitemOpen
  \bibfield  {author} {\bibinfo {author} {\bibfnamefont {V.}~\bibnamefont
  {Usov}}\ and\ \bibinfo {author} {\bibfnamefont {D.}~\bibnamefont {Melrose}},\
  }\href {\doibase 10.1071/ph950571} {\bibfield  {journal} {\bibinfo  {journal}
  {Australian Journal of Physics}\ }\textbf {\bibinfo {volume} {48}},\ \bibinfo
  {pages} {571} (\bibinfo {year} {1995})}\BibitemShut {NoStop}%
\bibitem [{\citenamefont {{Harding}}\ and\ \citenamefont
  {{Muslimov}}(1998)}]{1998ApJ...508..328H}%
  \BibitemOpen
  \bibfield  {author} {\bibinfo {author} {\bibfnamefont {A.~K.}\ \bibnamefont
  {{Harding}}}\ and\ \bibinfo {author} {\bibfnamefont {A.~G.}\ \bibnamefont
  {{Muslimov}}},\ }\href {\doibase 10.1086/306394} {\bibfield  {journal}
  {\bibinfo  {journal} {\apj}\ }\textbf {\bibinfo {volume} {508}},\ \bibinfo
  {pages} {328} (\bibinfo {year} {1998})},\ \Eprint
  {http://arxiv.org/abs/astro-ph/9805132} {arXiv:astro-ph/9805132 [astro-ph]}
  \BibitemShut {NoStop}%
\bibitem [{\citenamefont
  {Harding}(2007)}]{harding2007pulsarhighenergyemissionpolar}%
  \BibitemOpen
  \bibfield  {author} {\bibinfo {author} {\bibfnamefont {A.~K.}\ \bibnamefont
  {Harding}},\ }\href {https://arxiv.org/abs/0710.3517} {\enquote {\bibinfo
  {title} {Pulsar high-energy emission from the polar cap and slot gap},}\ }
  (\bibinfo {year} {2007}),\ \Eprint {http://arxiv.org/abs/0710.3517}
  {arXiv:0710.3517 [astro-ph]} \BibitemShut {NoStop}%
\bibitem [{\citenamefont {{Philippov}}\ \emph {et~al.}(2020)\citenamefont
  {{Philippov}}, \citenamefont {{Timokhin}},\ and\ \citenamefont
  {{Spitkovsky}}}]{Philippov2020}%
  \BibitemOpen
  \bibfield  {author} {\bibinfo {author} {\bibfnamefont {A.}~\bibnamefont
  {{Philippov}}}, \bibinfo {author} {\bibfnamefont {A.}~\bibnamefont
  {{Timokhin}}}, \ and\ \bibinfo {author} {\bibfnamefont {A.}~\bibnamefont
  {{Spitkovsky}}},\ }\href {\doibase 10.1103/PhysRevLett.124.245101} {\bibfield
   {journal} {\bibinfo  {journal} {\prl}\ }\textbf {\bibinfo {volume} {124}},\
  \bibinfo {eid} {245101} (\bibinfo {year} {2020})},\ \Eprint
  {http://arxiv.org/abs/2001.02236} {arXiv:2001.02236 [astro-ph.HE]}
  \BibitemShut {NoStop}%
\bibitem [{\citenamefont {Philippov}\ \emph {et~al.}(2019)\citenamefont
  {Philippov}, \citenamefont {Uzdensky}, \citenamefont {Spitkovsky},\ and\
  \citenamefont {Cerutti}}]{Philippov2019}%
  \BibitemOpen
  \bibfield  {author} {\bibinfo {author} {\bibfnamefont {A.}~\bibnamefont
  {Philippov}}, \bibinfo {author} {\bibfnamefont {D.~A.}\ \bibnamefont
  {Uzdensky}}, \bibinfo {author} {\bibfnamefont {A.}~\bibnamefont
  {Spitkovsky}}, \ and\ \bibinfo {author} {\bibfnamefont {B.}~\bibnamefont
  {Cerutti}},\ }\href {\doibase 10.3847/2041-8213/ab1590} {\bibfield  {journal}
  {\bibinfo  {journal} {The Astrophysical Journal}\ }\textbf {\bibinfo {volume}
  {876}},\ \bibinfo {pages} {L6} (\bibinfo {year} {2019})}\BibitemShut
  {NoStop}%
\bibitem [{\citenamefont {Eilek}\ and\ \citenamefont
  {Hankins}(2016)}]{Eilek_2016}%
  \BibitemOpen
  \bibfield  {author} {\bibinfo {author} {\bibfnamefont {J.~A.}\ \bibnamefont
  {Eilek}}\ and\ \bibinfo {author} {\bibfnamefont {T.~H.}\ \bibnamefont
  {Hankins}},\ }\href {\doibase 10.1017/s002237781600043x} {\bibfield
  {journal} {\bibinfo  {journal} {Journal of Plasma Physics}\ }\textbf
  {\bibinfo {volume} {82}} (\bibinfo {year} {2016}),\
  10.1017/s002237781600043x}\BibitemShut {NoStop}%
\bibitem [{\citenamefont {Bransgrove}\ \emph {et~al.}(2023)\citenamefont
  {Bransgrove}, \citenamefont {Beloborodov},\ and\ \citenamefont
  {Levin}}]{Bransgrove:2022afn}%
  \BibitemOpen
  \bibfield  {author} {\bibinfo {author} {\bibfnamefont {A.}~\bibnamefont
  {Bransgrove}}, \bibinfo {author} {\bibfnamefont {A.~M.}\ \bibnamefont
  {Beloborodov}}, \ and\ \bibinfo {author} {\bibfnamefont {Y.}~\bibnamefont
  {Levin}},\ }\href {\doibase 10.3847/2041-8213/ad0556} {\bibfield  {journal}
  {\bibinfo  {journal} {Astrophys. J. Lett.}\ }\textbf {\bibinfo {volume}
  {958}},\ \bibinfo {pages} {L9} (\bibinfo {year} {2023})},\ \Eprint
  {http://arxiv.org/abs/2209.11362} {arXiv:2209.11362 [astro-ph.HE]}
  \BibitemShut {NoStop}%
\bibitem [{\citenamefont {Philippov}\ \emph {et~al.}(2020)\citenamefont
  {Philippov}, \citenamefont {Timokhin},\ and\ \citenamefont
  {Spitkovsky}}]{Philippov:2020jxu}%
  \BibitemOpen
  \bibfield  {author} {\bibinfo {author} {\bibfnamefont {A.}~\bibnamefont
  {Philippov}}, \bibinfo {author} {\bibfnamefont {A.}~\bibnamefont {Timokhin}},
  \ and\ \bibinfo {author} {\bibfnamefont {A.}~\bibnamefont {Spitkovsky}},\
  }\href {\doibase 10.1103/PhysRevLett.124.245101} {\bibfield  {journal}
  {\bibinfo  {journal} {Phys. Rev. Lett.}\ }\textbf {\bibinfo {volume} {124}},\
  \bibinfo {pages} {245101} (\bibinfo {year} {2020})},\ \Eprint
  {http://arxiv.org/abs/2001.02236} {arXiv:2001.02236 [astro-ph.HE]}
  \BibitemShut {NoStop}%
\bibitem [{\citenamefont {Okawa}\ and\ \citenamefont
  {Chen}(2024)}]{okawa2024modelpairproductionlimit}%
  \BibitemOpen
  \bibfield  {author} {\bibinfo {author} {\bibfnamefont {T.}~\bibnamefont
  {Okawa}}\ and\ \bibinfo {author} {\bibfnamefont {A.~Y.}\ \bibnamefont
  {Chen}},\ }\href {https://arxiv.org/abs/2402.19436} {\enquote {\bibinfo
  {title} {A model for pair production limit cycles in pulsar
  magnetospheres},}\ } (\bibinfo {year} {2024}),\ \Eprint
  {http://arxiv.org/abs/2402.19436} {arXiv:2402.19436 [astro-ph.HE]}
  \BibitemShut {NoStop}%
\end{thebibliography}%

\clearpage

\onecolumngrid

\begin{center}
  \textbf{\large Supplementary Material for An Axion Pulsarscope}\\[.2cm]
  \vspace{0.05in}
  {Mariia Khelashvili, Mariangela Lisanti, Anirudh Prabhu, and Benjamin R. Safdi}
\end{center}

\vspace{0.2in}
\setcounter{equation}{0}
\setcounter{figure}{0}
\setcounter{table}{0}
\setcounter{section}{0}
\setcounter{page}{1}
\thispagestyle{empty}
\makeatletter
\renewcommand{\theequation}{S\arabic{equation}}
\renewcommand{\thefigure}{S\arabic{figure}}
\renewcommand{\thetable}{S\arabic{table}}
\def\set@footnotewidth{\onecolumngrid}

\vspace{-0.4in}
\section{Axion Emission in a Pulsar's Electromagnetic Field}
\label{sec:dipole_radiation}

The derivation of the axion power in the vacuum dipole model~(VDM) and the polar cap~(PC) model in the main text assumes the axion is massless. This section generalizes the derivation to finite mass $m_a \le \Omega$, where $\Omega$ is the pulsar's rotational frequency. The evolution of the axion field exterior to the pulsar is governed by the Klein-Gordon equation, $(\Box + m_a^2) a(x) = \gagg \edb$, where $\bfE$ ($\bfB$) is the electric (magnetic) field in the pulsar magnetosphere. The Klein-Gordon equation with a source can be solved using the retarded Green function~\citep{Bogolyubov1959, Economou2006}:
\begin{equation}
     G(\bold{x},\tau) = \frac{\theta(\tau)}{2\pi} \left(\delta(s^2) - \frac{m_a}{2s}\theta(s^2)\, J_1 (m_a s) \right)\,, \qquad s^2 = \tau^2 - \bold{x}^2\, \, ,
\end{equation}
where $J_1(x)$ is the Bessel function, $\theta$ is the Heaviside function, and $\delta$ is the delta function. The solution can be simplified for the case of localized periodic sources using the method of~\citep{Li1986, Krause1994}. In this case, the radiation from the pulsar magnetosphere  
is sourced by the dipole moment of $\edb$ to leading-order (see, {\it e.g.}, (\ref{eq:p_dipole}) in the main text). We briefly summarize the calculations of axion dipole radiation and then apply the results to the two magnetosphere models considered in this work. 

Assuming that the source periodically rotates at the pulsar rotation frequency $\Omega$,  the axion-field solution takes the form~\cite{Krause1994}
\begin{equation}
    a(\br, t) = \sum\limits_{m = -\infty}^{\infty} e^{-i m\Omega t} a_{m}(\br)\,.
\end{equation}

In general, periodically rotating sources radiate at the fundamental frequency $\Omega$ and its integer multiples. For any realistic pulsar, however, the near-surface region rotates with velocity $v \sim \Omega \rns \ll 1$, where $\rns$ is the neutron star's radius, and radiated power at higher frequencies is highly suppressed compared to the fundamental frequency {by powers of $v$}. Thus, we only consider radiation at the fundamental frequency:
\begin{equation}
    a(\br, t) = \expom a_{\scriptscriptstyle{\Omega}}(\br) + \expomp a_{\scriptscriptstyle{-\Omega}}(\br)\,.
\end{equation}
The corresponding Fourier component of the charge density, $\rho_{\rm eff}(\br, t) \equiv \gagg {\edb} (\br, t)$, is 
\begin{equation}\label{charge-comp}
    \rho_{\scriptscriptstyle{\pm\Omega}}(\br) = \frac{1}{T} \int\limits_{0}^{T} dt\, \expompm \rho_{\rm eff}(\br, t)\,, 
\end{equation}
where $T = 2\pi/\Omega$ is the source rotation period.
The axion field obeys the equation of motion:
\begin{equation}
    (\nabla^2 + k^2) a_{\scriptscriptstyle{\pm\Omega}}(\br) = -\rho_{\scriptscriptstyle{\pm\Omega}}(\br)\,, \quad k = \sqrt{\Omega^2 - m_a^2}\,.
\end{equation}
We may solve this equation using the outgoing-wave Green function~\cite{Krause1994},
\begin{equation}\label{outgoing-greenf}
    G(\br,\br') = \frac{1}{4\pi}\frac{e^{\pm i k\moddist}}{\moddist}\,,
\end{equation}
leading to
\begin{equation}
    a_{\scriptscriptstyle{\pm\Omega}}(\br) = \frac{1}{4\pi} \int d^3r' \frac{e^{\pm i k\moddist}}{\moddist} \rho_{\scriptscriptstyle{\pm\Omega}}(\br')\,. 
\end{equation}
In the long-wavelength ($k |\br'| \ll 1$), far-field ($|\br| \gg |\br'|$) limit, the solution takes the form 
\begin{equation}
    a_{\scriptscriptstyle{\pm\Omega}}(\br) = \frac{e^{\pm ikr}}{4\pi r} \left( \int d^3r'\rho_{\scriptscriptstyle{\pm\Omega}}(\br') \pm ik\hat{n} \cdot \int d^3r'\rho_{\scriptscriptstyle{\pm\Omega}}(\br') \br' + \cdots\right) = \frac{e^{\pm ikr}}{4\pi r} \left[Q_{\scriptscriptstyle{\pm\Omega}} \pm ik\hat{n}\cdot\bp_{\scriptscriptstyle{\pm\Omega}} + \cdots \right] \,, 
\end{equation}
{where $\hat{n}$ is the unit vector in the direction of $\br$,} and $Q_{\scriptscriptstyle{\pm\Omega}}$ ($\bp_{\scriptscriptstyle{\pm\Omega}}$) is the monopole (dipole) moment of $\rho_{\scriptscriptstyle{\pm\Omega}}$, given by
\begin{equation}
    Q_{\scriptscriptstyle{\pm\Omega}} = \int d^3r'\rho_{\scriptscriptstyle{\pm\Omega}}(\br')\,, \qquad \bp_{\scriptscriptstyle{\pm\Omega}} = \int d^3r'\rho_{\scriptscriptstyle{\pm\Omega}}(\br') \br'\,.
\end{equation}

In the VDM and PC models, the monopole term vanishes and the leading-order term is that of the dipole. The time-dependent solution for the outgoing axion wave radiated by the dipole is
\begin{equation}\label{axion-field-dipole}
    a(\br, t) = \frac{ik}{4\pi r} \left(\hat{n}\cdot \bp_{\scriptscriptstyle{+\Omega}}\, e^{-i(\Omega t - kr)} - \hat{n}\cdot \bp_{\scriptscriptstyle{-\Omega}}\, e^{i(\Omega t - kr)} \right) \,.
\end{equation}
The radiated axion power follows:
\begin{equation}
    P_a = \int \bpi \cdot d\boldsymbol{\sigma}\,, \qquad \langle\bpi\rangle_T = \langle \dot{a}^* \boldnabla a \rangle_T = \langle \dot{a}_{\scriptscriptstyle{\Omega}}^*(\br, t)\boldnabla a_{\scriptscriptstyle{\Omega}}(\br, t) + \dot{a}_{\scriptscriptstyle{-\Omega}}^*(\br, t)\boldnabla a_{\scriptscriptstyle{-\Omega}}(\br, t) \rangle_T\,,
\end{equation}
where $\left\langle .. \right\rangle_T$ represents the time-average over the rotational period. For a real source, $\operatorname{Im} \rho(\br, t) = 0$ and the Fourier components are $\rho_{{\scriptscriptstyle{-\Omega}}}^* = \rho_{{\scriptscriptstyle{\Omega}}}$, therefore $\bp_{-\scriptscriptstyle{\Omega}}^* = \bp_{\scriptscriptstyle{\Omega}}$ and the total axion dipole radiation becomes:
\begin{equation}\label{axion-dipole-general-eq}
    \langle P_a \rangle_T = \frac{1}{8\pi^2} \Omega k^3 \iint d\omega_n  |\bp_{\scriptscriptstyle{\Omega}}\cdot \hat{n}|^2\,.
\end{equation}

The integration is performed over the solid angle $\omega_n$, and the final expression includes the Fourier component of the dipole moment:
\begin{equation}\label{d-momentum-fourier}
    \bp_{\scriptscriptstyle{\Omega}} = \frac{1}{T}  \int\limits_0^{T} dt\, e^{i\Omega t} \int d^3r \rho_{\rm eff}(\br, t)\br\,.
\end{equation}

\subsection{Vacuum {Dipole Model}}

For a fully unscreened magnetosphere, {the distribution of effective charge density, $\gagg\edb$, in the pulsar's magnetosphere is given by~\eqref{eqn:edbvac}.} The radiation of axions in this scenario can be modeled by a rotating dipole with effective dipole moment~\eqref{eq:p_dipole}. From (\ref{d-momentum-fourier}), 
\begin{equation}
    \hat{n}\cdot\bp_{\scriptscriptstyle{\Omega}} \approx \frac{i\pi}{12} \gagg \frac{|\bfmu|^2\Omega}{\rns} \sin(2\theta_m) \sin\theta_n e^{i\phi_n} \,,
\end{equation}
where $|\bfmu| = B_0 \rns^3$, $B_0$ is the surface magnetic field of the pulsar, $\theta_m$ is the angle between the pulsar's magnetic dipole axis and its rotation axis, and  $(\theta_n, \phi_n)$ are the angles in spherical coordinates that define the direction of $\hat{n}$. From the dipole radiation power~\eqref{axion-dipole-general-eq}, it follows that
\begin{equation}
    \langle P_a \rangle_T = \frac{\pi}{432} g_{a\gamma\gamma}^2 B_0^4 \Omega^6 \rns^{10} \sin^2 (2\theta_m) \left(1 - \frac{m_a^2}{\Omega^2} \right)^{3/2} \,,
\end{equation}
which reduces to~\eqref{eqn:vacuumdipolepower} in the limit of a massless axion.

\subsection{Polar Cap Model}

We now discuss the model of a mostly-screened magnetosphere, where the axion is sourced only in the small polar cap (PC) regions above the magnetic poles. 
The size of the PC, $r_\text{pc} \ll \rns \ll 1 / \Omega$, can be neglected, so we model the PCs as two rotating point-like sources with opposite charges ($Q_a^N = - Q_a^S$) and opposite velocity and position  ($\bold{v}_{\rm pc}^N = - \bold{v}_{\rm pc}^S$ and $\bold{r}_{\rm pc}^N = - \bold{r}_{\rm pc}^S = \bq(t)$), where $N$ ($S$) represents the PC at the northern (southern) pole. 
The effective charge density for the point-like PC model is 
\begin{equation}
    \rho(\br, t) = Q_a\left[ \delta(\br - \bq(t)) - \delta(\br + \bq(t))\right]\,,
\end{equation}
where 
\begin{equation}
    \bq(t) = \rns \left(\sin \theta_m \left(\cos(\Omega t) \hat{\bf x} + \sin (\Omega t)\hat{\bf y} \right) +  \cos \theta_m \hat{\bf z}\right)\,.
\end{equation}
We can find the Fourier component of the dipole moment~\eqref{d-momentum-fourier}:
\begin{equation}
    \hat{n} \cdot \bp_{\scriptscriptstyle{\Omega}} =  Q_a \rns \sin\theta_m \sin \theta_n e^{-i\phi_n}\,,
\end{equation}
where $\theta_n, \phi_n$  are spherical coordinate angles of the observer. Finally, the resulting axion radiation power is:
\begin{equation}
    \langle P_a \rangle_T = \frac{1}{3\pi} \rns^2 \Omega^4 Q_a^2 \sin^2 \theta_m \left(1 - \frac{m_a^2}{\Omega^2} \right)^{3/2}\,,
\end{equation}
which is the same as~\eqref{eq:P_a_PC} in the ultra-relativistic limit ($m_a \ll \Omega$).

\vspace{0.2in}

\section{Dark SRF Projection}

This section provides details for computing  the sensitivity of superconducting radio frequency~(SRF) cavities to relativistic monochromatic axion signals, using the conventions described in~\cite{Berlin2020, Berlin2021}. The search for relativistic axions does not exactly map onto the search for non-relativistic axion DM, since spatial gradients cannot be ignored in the former case. In particular, the wave equation takes the form 
\begin{align} \label{eqn:waveqn}
    \left({\partial^2 \over \partial t^2} - \nablabold^2 \right) \bfE = \gagg \left((\nablabold a \cdot \nablabold) \bfB_p - \dot{a} {\partial \bfB_p \over \partial t} \right)\, , 
\end{align}
where $a$ is the axion field and $\bfB_p$ is the magnetic field associated with the pump mode. {The first term on the right-hand side in (\ref{eqn:waveqn}) can be neglected as it is incapable of exciting resonant modes of a cavity~\cite{Dror2021}.}
The axion field drives power into the signal mode, with the following {power spectral density}~(PSD)~(see {\it e.g.},~\cite{Berlin2020}): 
\begin{align}
    S_{\rm sig}(\omega) &= {\omega_1 \over 4 Q_1} (\gagg \eta_{10} B_p)^2 V_{\rm cav} \ {\omega^2 (\omega - \omega_0)^2 S_a(\omega - \omega_0) \over (\omega^2 - \omega_1^2)^2 + (\omega \omega_1 / Q_1)^2} \, ,
\end{align}
where $S_a$ is the PSD of the axion signal, $\omega_0$ and $\omega_1$ are the frequencies of the pump and signal modes, respectively, $Q_1$ is the quality factor of the signal mode, $\eta_{10}$ is the mode overlap factor, and $V_{\rm cav}$ is the volume of the cavity. The PSD is defined using the following normalization:
\begin{align}\label{eqn:psdnormalization}
     \displaystyle\int S_a(\omega) d\omega = (2\pi)^2 {\rhops \over \Omega^2} \, ,
\end{align}
where $\rhops$ is the local axion energy density and $\Omega$ is the frequency of the axion signal. Note the distinction with the DM search in which the frequency is equal to the axion mass.

Since the axion coherence time is much greater than the integration time of the experiment, the axion signal PSD falls in a single frequency bin. The signal and noise powers measured in a frequency bin are determined by integrating their respective PSDs over the frequencies in the bin. On resonance, the splitting between the two modes is equal to the frequency of the axion signal: $\omega_1 = \omega_0 + \Omega$. The signal PSD may be approximated as a delta function, giving a total power of
\begin{align}
    \displaystyle\int_{\omega_1 - \Delta \omega/2}^{\omega_1 + \Delta \omega/2} S_{\rm sig}(\omega) {d\omega}= \pi^2 (\gagg \eta_{10} B_p)^2 \left( {Q_1 \over \omega_1 }\right) V_{\rm cav} \, ,
\end{align}
using the normalization of (\ref{eqn:psdnormalization}). To {obtain the} sensitivity, we compute the total noise power within the frequency bin into which the axion signal falls. For the optimistic parameters presented in~\cite{Berlin2020, Berlin2021}, the dominant noise source at the relevant frequency is thermal noise with total bin power 

\begin{align}
     \displaystyle\int_{\omega_1 - \Delta \omega/2}^{\omega_1 + \Delta \omega/2} \left( {Q_1 \over Q_{\rm int}} \right) {4\pi T_{\rm cav} (\omega \omega_1/Q_1)^2 \over (\omega^2 - \omega_1^2)^2 +(\omega \omega_1/Q_1)^2}{d\omega} = 4\pi T_{\rm cav} \left( {Q_1 \over Q_{\rm int}} \right) \Delta \omega.
\end{align}

\section{DMRadio-GUT Projection}

To estimate the projected sensitivity for DMRadio-GUT, we follow the formalism in~\cite{ABRACADABRA}. The detector consists of a large toroidal magnet and an external pickup loop inductively coupled to a SQUID magnetometer that measures the magnetic flux through the toroid. In the presence of a static magnetic field, the axion field generates an effective charge density, $\rho_{a,\rm eff} = -\gagg \boldsymbol{\nabla} a \cdot {\bfB}$ (not to be confused with the axion energy density), and current density, ${\bf J}_{a, \rm eff} = \gagg \dot{a} \bfB$. In axion DM experiments, the charge density is ignored as it is gradient-suppressed. For relativistic sources, such as pulsar-sourced axions, however, $|\boldsymbol{\nabla} a| \sim \dot{a}$, and the {charge and current densities} are comparable. However, the effect of the charge density is suppressed relative to that of the current density by a factor of $\Omega L_p \ll 1$, where $\Omega$ is the frequency of the axion signal and $L_p$ is the inductance of the pickup loop. The magnetic flux through the pickup loop due to the   oscillating axion field is 
\begin{align}
    \Phi_{\rm pickup} &= \gagg B_{\rm max} \sqrt{2 \rhops} V_B \cos(\Omega t) \, ,
\end{align}
 where $B_{\rm max}$ is the peak magnetic flux in the toroid, $\rhops$ is the axion density in the detector,  and $V_B$ is the effective volume containing the magnetic field. The flux through the inductively-coupled SQUID is~\cite{ABRACADABRA} 
 \begin{align}
     \Phi_{\rm SQUID} &= {\alpha \over 2} \sqrt{L \over L_p} \Phi_{\rm pickup} \, ,
 \end{align}
where $\alpha \approx 0.7$ and $L$ is the inductance of the SQUID, which we take to be $L \approx 1$ nH. 

An axion DM search could employ either broadband or resonant readout, with the former being preferred at low frequency. The reason is that for a resonant search, the need to scan over DM mass significantly reduces the interrogation time in each frequency bin. In contrast, the signal frequency of pulsar-sourced axions is known to very high precision, set by the precision of timing measurements of pulsars. Thus, we focus on resonant readout with an $LC$ circuit. The bandwidth of the readout circuit is $\Delta \omega_{\rm RO} = \omega/Q_0$, where $Q_0 = 1/(\omega R C)$ is the quality factor of the circuit and $C$~($R$) is the capacitance~(resistance) of the capacitor, respectively. 

We can model the system as an RLC circuit, with the pickup loop placed at the center of the toroid playing the role of the inductor. On resonance, the resistance of the circuit is related to the intrinsic quality factor through $R = \Omega L_p/Q_0$. The axion field drives an AC voltage in the readout circuit with PSD~\cite{Benabou:2022qpv}
\begin{align}
    S^a_{VV}(\omega) = \pi \Omega^2 \gagg^2 B_{\rm max}^2 c_{\rm pu}^2 V_B^2  \rhops \delta (\omega - \Omega) \,,
\end{align}
where $c_{\rm pu}$ quantifies the fraction of the axion-sourced magnetic field that couples to the resonant circuit. We have assumed that the axion signal is sufficiently narrow that its spectrum is well-approximated by a delta function.  
In reality, the axion signal will be spread over the  finite bandwidth of the circuit $\Delta \omega = \Omega/Q_0$. On resonance, the induced current PSD is $S^a_{II} = S^a_{VV}/R^2$, where $R$ is {the resistance defined above}. One way to parameterize the noise contributions is with a ``noise temperature,'' $k_B T_N(\omega) = N(\omega) \hbar \omega$, where $k_B$ is the Boltzmann constant and $\hbar$ is the reduced Planck constant. $N(\omega)$ receives contributions from thermal noise, amplifier noise, and $1/f$ noise. The noise PSD may be written as follows~\cite{Benabou:2022qpv}
\begin{align}\label{eqn:siinoise}
    S^{\rm noise}_{II}(\omega)  = {Q_0 \over \pi \Omega L_p} (2 k_B T_{\rm LC} + \eta_A \hbar \omega) \, ,
\end{align}
where $T_{\rm LC}$ is the temperature of the LC circuit and $\eta_A$ describes how far the amplifier is from operating at the standard quantum limit~(SQL), with $\eta_A = 1$ corresponding to the SQL ($1/2$ from quantum-limited amplifier noise and 1/2 from zero-point thermal fluctuations). For a pulsar-sourced axion search with $\Omega/(2\pi) \sim 30$ Hz, the contribution of $1/f$ noise cannot be neglected. In keeping with the convention of (\ref{eqn:siinoise}), we absorb the contribution of $1/f$ noise into $\eta_A$. The flux PSD of $1/f$ noise is given by $S_\Phi(\omega) = A_\Phi^2 (\omega/ 2\pi \times {\rm Hz})^{\xi}$, where $\xi \approx -1$ 
and, for the low temperatures considered here, $A_\Phi^2 \lesssim (10^{-6} \Phi_0)^2/$Hz, where $\Phi_0 = h/2e$ is the magnetic flux quantum, with $h$ being the Planck constant and $e$ being the magnitude of the electron charge~\cite{2013PhRvL.110n7002A}. On resonance, the contribution of $1/f$ noise to $\eta_{A}$ is approximately $10^3$. Since $T_{\rm LC} \gg \eta_A \Omega$, {the dominant noise source is thermal}. The signal and thermal noise at frequency $\omega_a = \Omega$ are
\begin{align}
    P_{\rm sig} = Q_0 {\Omega \Phi_{\rm pickup}^2 \over 2 L_p}\, , \quad P_{\rm noise} = k_B T_{\rm LC} \times {2\pi \over T_{\rm int} } \, ,
\end{align}
where $T_{\rm int}$ is the integration time. We compute the sensitivity by setting $P_{\rm sig} = 8.48 \times P_{\rm noise}$ (as in the main text) and use the following fiducial parameters roughly in line with those of DMRadio-GUT~\cite{Brouwer2022}: $B_{\rm max} = 16$ T, $c_{\rm pu} = 0.1$, $V_B = 10 \ {\rm m}^3$, $L_p = \pi R = \pi \times 2.2 \ {\rm m}$, $T = 10 \ {\rm mK}$, $Q_0 = 2 \times 10^7$, and $T_{\rm int} = 1$ year. 

\clearpage
\section{CASPEr Projection}

This section discusses detection of pulsar-sourced axions through their coupling to fermions. For non-relativistic nuclei, the axion field acts as an effective magnetic field with interaction Hamiltonian $H_{\rm int} = \gamma_{_N} {\bf B}_a \cdot {\bf S}_N$, where $\gamma_{_N}$ is the gyromagnetic ratio of the nucleon, ${\bf B}_a = - g_{aNN} \boldsymbol{\nabla} a/\gamma_{_N}$ is the effective axion magnetic field, and ${\bf S}_N$ is the nucleon spin operator. Several experiments, including CASPEr~\cite{Casper2013} and co-magnetometer experiments~\cite{Lee2023, Bloch2022}, use nuclear magnetic resonance techniques to look for axion DM. This section computes the sensitivity of a CASPEr-wind-type experiment to pulsar-sourced axions. The starting point is a macroscopic sample of nuclear spins aligned using a strong, static magnetic field, ${\bf B}_E = B_E \hat{z}$. We assume the apparatus is aligned such that ${\bf B}_a = B_{a,0} \hat{x} \cos(\Omega t - k_a x)$, where $k_a^2 = \Omega^2 - m_a^2$. The axion magnetic field causes the spins to precess about $\hat{z}$, leading to an oscillating transverse magnetization of the sample. This effect is resonantly enhanced when the axion frequency matches the Larmor frequency, $\omega_0 = \gamma_{_N} B_E$. The width of the resonance is $\Delta \omega \simeq 1/T_2$, where $T_2$ is the transverse spin relaxation time. The evolution of the transverse magnetization, ${M_x}$, is described by the Bloch equations, which give
\begin{align}
    \ddot{M}_x + {2 \over T_2} \dot{M}_x + \omega_0^2 M_x = \gamma_{_N} M_0 \omega_0 B_{a,0} \cos{\Omega t} \, ,
\end{align}
where $M_0$ is the initial magnetization of the system and $B_{a,0} = - \gan k_a \sqrt{2\rho_a}/(\Omega\gamma_{_N})$. We have taken the system to be located at $x=0$,  assume that the size of the apparatus is much less than the axion wavelength, and  work in the limit where $T_2 \ll T_{\rm int} \ll \tau_a$, with $T_{\rm int}$ the integration time and $\tau_a$ the coherence time of the signal. The coherence time of the axion signal is set by the spin-down parameters of the pulsar being observed, $P/\dot{P}$, which is in general much longer than a year (taken to be the fiducial integration time). To compute the sensitivity, we envision taking $N$ measurements with interval $\Delta t = T_{\rm int}/N$. In frequency space, this corresponds to bins with frequency $\omega_n = 2\pi n/T_{\rm int}$. Since $\tau_a \gg T_{\rm int}$, the signal lies entirely within one bin with {PSD}
\begin{align}
    P_a &= \gan^2 \rho_a k_a^2 M_0^2 {T_2^2 T_{\rm int} \over 2 \Omega^2} \, .
\end{align}

The axion-driven transverse magnetic field is measured using a SQUID magnetometer. The main sources of noise come from: thermal effects, spin projection noise, and quantum noise from the SQUID. As in~\cite{Dror2022}, we assume the system temperature is sufficiently low that thermal noise is subdominant. Spin projection noise arises due to the intrinsic uncertainty in spin measurement along a given direction.
The {PSDs} of spin projection and magnetometer noise on resonance are, respectively,~\cite{Dror2022}
\begin{align}
    P_{\rm SP} = {\gamma_{_N}^2 n J T_2\over 2 V} \, , \quad P_{\rm SQUID} = {1 \over A_{\rm eff}^2} {(10^{-6} \Phi_0)^2 \over {\rm Hz}} \, ,
\end{align}
where $n$ and $V$ are the number density and volume of the sample, respectively, $J$ is the nuclear spin,  and $A_{\rm eff}$ is the effective area of the SQUID pickup loop.

\color{black}

\section{Polar Cap Model Astrophysical Uncertainties}\label{sec:PG-models-sm}

The projections for the polar cap~(PC) model reflect particular choices of the PC geometry and accelerating electric field, $\epar$. In the main text, we model the geometry of the PC vacuum gap as a cylinder of radius $\rpc$ and height $h$. This is the minimal geometry arising from a purely dipolar field. The PC geometry of pulsars with considerable multipolar fields may deviate significantly. For example, in a quadrudipole magnetosphere~\cite{1982ApJ...254..713B}, one of the PCs has an annular shape that wraps around the star and has area larger than the area of the canonical dipolar PC~\cite{Gralla_2017}. Additionally, the strength of multipolar magnetic fields (and the corresponding electric fields) may be considerably stronger than the dipole field near the surface (see, e.g.,~\cite{Kalapotharakos_2021}). Thus the choice of a dipole magnetic field with antipodal cylindrical PCs is a conservative choice. In a pure dipole geometry, the radius of the cylinder is well-defined, as discussed in the main text. Uncertainties arise from the strength of the electric field in the gap and the height of the PC. Below, we consider two extreme limits of the electric field and height; the values adopted for the projections in the main text are pessimistic for both quantities. 

The accelerating electric field in the gap is determined self-consistently from Gauss' law in the co-rotating frame $\nabla \cdot \vec{E} = (\rho - \rho_{\rm GJ})$, where $\rho$ is the charge density present in the gap and $\rho_{\rm GJ}$ the Goldreich-Julian charge density. When sufficient charge is present, the gap is fully screened, and there is no electric field in the co-rotating frame (the co-rotating electric field $\vec{E}_{\rm cor} = - (\vec{\Omega} \times \vec{r})\times\vec{B}$ yields $\vec{E} \cdot \vec{B} = 0$). The filling of the gap with plasma prior to the ignition of pair production depends on the relationship of the stellar surface temperature and the binding energy of electrons and ions in the upper crust~\cite{RudermanSutherland1975, 1979ApJ...231..854A, Usov1995, 1998ApJ...508..328H, harding2007pulsarhighenergyemissionpolar}. If temperature is sufficiently low that charges remain confined to the surface, the electric field may be as high as the vacuum value, $\epar = \Omega \rns B_0$, and if the temperature is high enough to supply charges to the gap, the electric field is space-charge limited and $\epar \simeq \Omega z B_0$, where $z$ is the height above the neutron star surface~\cite{harding2007pulsarhighenergyemissionpolar}. For both the vacuum gap and the space-charge limited gap, $\epar$ is only non-zero up to some height above the NS surface, beyond which the electric field is screened. A customary choice for the height is $h \simeq \rpc$ (see, e.g.,~\cite{Philippov2020, Tolman:2022unu} and Sec. 2 in~\cite{Philippov2022Review}). The height of the gap will ultimately be regulated by the production of pairs. As discussed in the main text, pair production in pulsar PCs occurs through the emission and subsequent absorption of high-energy photons. Therefore, the minimum height of the gap can be taken to be the mean-free-path of a pair-producing photon, $h = h_{\rm Rud}$, computed in~\cite{RudermanSutherland1975} and quoted in (\ref{eqn:hgap}) in the main text. For a rapidly-rotating pulsar like the Crab, $h_{\rm Rud}/\rpc \approx 10^{-2}$, although the ratio may be more modest for middle-aged pulsars.  

\begin{table}
    \newcolumntype{M}[1]{>{\centering\arraybackslash}m{#1}}
    \newcolumntype{N}{@{}m{0pt}@{}}
    \centering
    \begin{tabular}{|M{0.3cm}|M{0.3cm}|M{3cm}|M{3cm}|N}
        \hline
        \multicolumn{2}{|c|}{} & \multicolumn{2}{c|}{ Polar gap height, $h$ }\\ 
        \cline{3-4}
        \multicolumn{2}{|c|}{} & optimistic & pessimistic \\ 
        \hline
        \multirow{4}{*}{\rotatebox[origin=c]{90}{Polar Gap electric field, $E$}}
        & \multirow{2}{*}{\rotatebox[origin=c]{90}{optimistic}}
        & \vspace{5mm} $h = r_{pc}$ & \vspace{5mm}$ h = h_\text{Rud}$ \\ [15pt]
        & & $E_{\parallel} = E_{\parallel, \rm vac} = \Omega R_\text{NS} B_0$
        & $E_{\parallel} = E_{\parallel, \rm vac} = \Omega R_\text{NS} B_0$ \\ [10pt]
        \cline{2-4}
         & \multirow{2}{*}{\rotatebox[origin=c]{90}{pessimistic}} 
         & \vspace{0.2in}$h = r_{pc}$ & \vspace{0.2in}$ h = h_\text{Rud}$ \\ [15pt]
        & & $E_{\parallel} = E_{\parallel, \rm SCL} = 2\Omega h B_0 = 2\Omega r_{pc} B_0 $ & $E_{\parallel} = E_{\parallel, \rm SCL} = 2\Omega h B_0 = 2\Omega h_\text{Rud} B_0 $ \\ [10pt]
        \hline
    \end{tabular}
    \caption{\color{black} The different PC models used to estimate axion-emission power in PC-only scenarios, including the pessimistic and optimistic cases for electric field strength and PC height, resulting in four different models.  For a detailed description, see Section~\ref{sec:PG-models-sm}. }
    \label{tab:PG_models}
\end{table}

\color{black}

In the main text, we found that the axion power from the PC is $P_a \propto E^2 h^2$. In the table above, we adopt two choices for the electric field: the (optimistic) vacuum value $E_{\parallel, \rm vac} = \Omega \rns B_0$ and the (pessimistic) space-charge limited value, $E_{\parallel, \rm SCL} = 2\Omega h B_0$.  We also adopt two choices for the height: $h = \rpc$~(optimistic) and $h = h_{\rm Rud}$~(pessimistic). The projections shown in Fig.~\ref{fig:results} are based on the most pessimistic choices of $\epar$ and $h$. Figure~\ref{tab:PG_models} shows how the projected sensitivities of DMRadio-GUT, Dark~SRF, and CASPEr are modified when different modeling choices are adopted. For all but the most pessimistic modeling choices, the projected sensitivity of the Dark-SRF experiment could improve upon laboratory constraints (e.g., CAST) for $m_a \lesssim 10^{-13}$ eV, but is unlikely to be competitive with astrophysical limits in this mass range. 

\begin{figure}
    \centering
    \includegraphics[width = \textwidth]{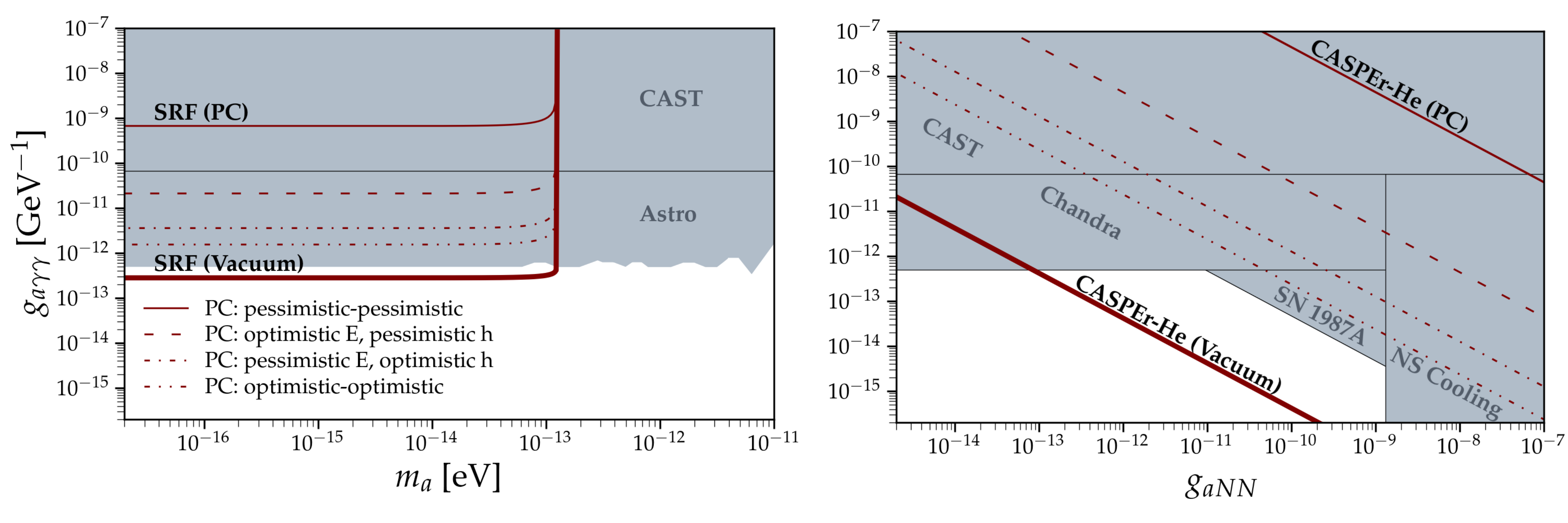}
    \caption{\color{black} Similar to Fig.~\ref{fig:results}, with different PC models. \textbf{Left:} Projected sensitivity to the axion-photon coupling for the SRF experiment. \textbf{Right:} Projected sensitivity for the CASPEr-wind~(He) experiment. The sensitivity difference between pessimistic and optimistic PC models for DMRadio-GUT and CASPEr-wind~(Xe), not shown here, will be the same as in the experiments presented. The four PC models are summarized in Tab.~\ref{tab:PG_models}.}
    \label{fig:uncertainty_PG_models}
\end{figure}

\color{black}

 We conclude this section by commenting on the feasibility of determining PC properties directly from observations of electromagnetic emission from the PC.\footnote{For the Crab pulsar, the main peak of the radio emission is theorized to come from the Y-point near the light cylinder, rather than the PC~\cite{Philippov2019}. Radio observations suggest that the full rotation period contains seven radio components, suggesting many possible sites of radio emission. In fact, Eilek \& Hankins~\cite{Eilek_2016} suggest that one of the components (the ``low-frequency component'') may be coming from the PC region. The existence of many radio components is consistent with the picture in~\cite{Bransgrove:2022afn}. Thus, this observation suggests additional sites of non-stationary pair creation, and possibly axion emission, which may lead to significantly more power than predicted in the PC model. Identifying sites of pair creation using particle-in-cell simulations is the subject of ongoing work. } For example, PC radio emission is intricately linked with pair production and non-stationary screening of the gap~\cite{Philippov:2020jxu}. The radio luminosity from this process is estimated to be~\cite{Tolman:2022unu}

\begin{align}
    L_{\rm rad} &= {\bar{E}_\star^2 \sin^2\theta (\eta \pi \rpc^2) \over 4\pi } \,,
\end{align}
where $\theta$ is the angle of propagation between the produced radio wave and the background magnetic field, $\eta$ is the fraction of pair producing field lines in the PC, and $\bar{E}_\star$ is defined as the electric field at the transition from the early (non-linear) screening phase to the late (linear) stage. Therefore, it is the amplitude of the electric field at this stage of the limit cycle that produces radio emission. In contrast, the axion luminosity is dominated by the early stage of the gap, before sufficient screening has taken place. Around the transition from non-linear screening to linear screening, axion emission is suppressed due to both the smaller amplitude of $\epar$ and its oscillations, which tend to average to a smaller value~\cite{Tolman:2022unu}. Therefore, in this model, PC parameters relevant for axion production cannot be directly inferred from observations of the radio emission. Self-consistent modeling of the pulsar limit cycle, including the initial electric field prior to screening, was presented in~\cite{okawa2024modelpairproductionlimit}. This model predicts a maximum electric field in the gap $E = (10^4$--$10^5) m_e \omega_p$, where $\omega_p = \sqrt{e^2 n_e/m_e}$ is the plasma frequency associated with the local Goldreich-Julian number density. Applied to the Crab, this electric field corresponds, in a space-charge limited gap, to a gap height $h = (15$--$150)$ m, which is a factor of $\sim (2$--$20)$ larger than the value adopted in the main text (\ref{eqn:hgap}).

\end{document}